\documentclass[iop,apj,appendixfloats]{emulateapj}
\usepackage{apjfonts}

\gdef\qg{qCMG}
\gdef\sg{sCMG}
\gdef\kms{km\,s$^{-1}$}
\gdef\msun{$M_{\odot}$}
\gdef\ha{H$\alpha$}
\gdef\nii{[N{\sc ii}]}
\gdef\oiii{[O{\sc iii}]}
\gdef\niiha{[N{\sc ii}]/H$\alpha$}

\lefthead{van Dokkum et al.}
\righthead{}
\slugcomment{Accepted for publication in the Astrophysical Journal}
\begin{document}

\title{Forming Compact Massive Galaxies}

\author{Pieter G.\ van Dokkum\altaffilmark{1},
Erica June Nelson\altaffilmark{1},
Marijn Franx\altaffilmark{2},
Pascal Oesch\altaffilmark{1},
Ivelina Momcheva\altaffilmark{1},
Gabriel Brammer\altaffilmark{3},
Natascha M.\ F\"orster Schreiber\altaffilmark{4},
Rosalind E.\ Skelton\altaffilmark{5},
Katherine E.\ Whitaker\altaffilmark{6},
Arjen van der Wel\altaffilmark{7},
Rachel Bezanson\altaffilmark{8},
Mattia Fumagalli\altaffilmark{2},
Garth D.\ Illingworth\altaffilmark{9},
Mariska Kriek\altaffilmark{10},
Joel Leja\altaffilmark{1},
Stijn Wuyts\altaffilmark{4}} 

\altaffiltext{1}
{Department of Astronomy, Yale University, New Haven, CT 06511, USA}
\altaffiltext{2}
{Leiden Observatory, Leiden University, 2300-RA Leiden, The
Netherlands}
\altaffiltext{3}
{Space Telescope Science Institute, Baltimore, MD 21218, USA}
\altaffiltext{4}
{Max-Planck-Institut f\"ur Extraterrestrische Physik,
Giessenbachstrasse, D-85748 Garching, Germany}
\altaffiltext{5}
{South African Astronomical Observatory, PO Box 9, Observatory,
Cape Town 7935, South Africa}
\altaffiltext{6}
{Astrophysics Science Division, Goddard Space Center, Greenbelt,
MD 20771, USA}
\altaffiltext{7}
{Max Planck Institute for Astronomy (MPIA), K\"onigstuhl 17,
D-69117 Heidelberg, Germany}
\altaffiltext{8}
{Steward Observatory, University of Arizona, 933 North Cherry
Avenue, Tucson, AZ 85721, USA}
\altaffiltext{9}
{UCO/Lick Observatory, University of California, Santa Cruz, CA 95064}
\altaffiltext{10}
{Department of Astronomy, University of California, Berkeley,
CA 94720, USA}

\begin{abstract}
In this paper we study a key phase in the formation of massive galaxies:
the transition of star forming galaxies into massive ($M_{\rm stars}
\sim 10^{11}$\,\msun), compact ($r_e \sim 1$\,kpc) quiescent galaxies,
which takes place from $z\sim 3$ to $z\sim 1.5$. We use HST grism
redshifts and extensive photometry in all five 3D-HST/CANDELS fields,
more than doubling the area used previously for such studies, and
combine these data with Keck MOSFIRE and NIRSPEC spectroscopy.
We first confirm that a population of massive,
compact, {\em star forming} galaxies exists at $z\gtrsim 2$, using
$K$-band spectroscopy of 25 of these objects
at $2.0<z<2.5$. They
have a median \niiha\ ratio of 0.6, are highly
obscured with SFR(tot)/SFR(\ha{})\,$\sim 10$, and have a large
range of observed line widths. We infer from
the kinematics and spatial distribution
of H$\alpha$ that
the galaxies have rotating disks of ionized gas that are a factor of
$\sim 2$ more
extended than the stellar distribution. By combining
measurements of individual galaxies, we find that the kinematics
are consistent with
a nearly Keplerian fall-off from $V_{\rm rot}
\sim 500$\,\kms\ at 1\,kpc to $V_{\rm rot}\sim 250$\,\kms\ at
7\,kpc, and that the total mass out to this
radius is dominated by the dense stellar component.
Next, we study the size and mass evolution
of the progenitors of compact massive
galaxies. Even though individual galaxies
may have had complex histories
with periods of compaction and mergers,
we show that the {\em population} of progenitors likely
followed
a simple inside-out growth track
in the size-mass plane of $\Delta \log r_e \sim 0.3
\Delta \log M_{\rm stars}$. This mode of growth gradually increases
the stellar mass within a fixed physical radius, and
galaxies quench when they reach a stellar density or
velocity dispersion threshold.
As shown in other studies,
the mode of growth changes after quenching, as dry
mergers take the galaxies on a relatively steep
track in the size-mass plane.
\end{abstract}


\section{Introduction}

Many studies have shown that massive
galaxies with low star formation rates were remarkably compact
at $z\gtrsim 2$ (e.g., {Daddi} {et~al.} 2005; {Trujillo} {et~al.} 2006; {van Dokkum} {et~al.} 2008; Damjanov et al.\ 2011;
Conselice 2014). At fixed stellar mass of
$M_{\rm stars} \approx 10^{11}$\,\msun,
quiescent galaxies 
are a factor of $\sim 4$ smaller at
$z=2$ than at $z=0$ (e.g., {van der Wel} {et~al.} 2014b).
As the stellar mass of the galaxies also evolves, the inferred
size growth of individual galaxies is even larger
({van Dokkum} {et~al.} 2010; {Patel} {et~al.} 2013). 
It is unlikely that all massive galaxies in
the present-day Universe had a compact progenitor
({van Dokkum} {et~al.} 2008, 2014;
{Franx} {et~al.} 2008; {Newman} {et~al.} 2012; {Poggianti} {et~al.} 2013; {Belli}, {Newman}, \& {Ellis} 2014a);
however, the vast majority of
compact, massive galaxies that are observed at $z=2$ ended
up in the center of a much larger galaxy today
({Belli} {et~al.} 2014a; {van Dokkum} {et~al.} 2014).
Their size growth after $z=2$ is probably dominated
by minor mergers: such mergers are expected, and
other mechanisms cannot easily produce the observed
$\dot{r}_{\rm e}/
\dot{M}_{\rm stars} \approx 2$ scaling between size growth
and mass growth
({Bezanson} {et~al.} 2009; {Naab}, {Johansson}, \& {Ostriker} 2009; {Hopkins} {et~al.} 2010; Trujillo et al.\ 2011;
{Hilz}, {Naab}, \& {Ostriker} 2013).

It is not yet clear how these
massive, extremely compact  galaxies were
formed, and this question has significance well beyond the
somewhat narrow context of the size evolution of quiescent
galaxies.  The dense centers of
massive galaxies today
are home to the most massive black holes in the Universe
({Magorrian} {et~al.} 1998); have an enrichment history that
is very
different from that of the Milky Way ({Worthey}, {Faber}, \&  {Gonzalez} 1992); and
probably had a bottom-heavy stellar initial mass function
(IMF) ({Conroy} \& {van Dokkum} 2012). All these characteristics
are the product of processes that took place in the
star forming progenitors of $z\sim 2$ massive quiescent galaxies.
Furthermore, stars in very 
dense regions represent only a very
small fraction ($\sim 0.1$\,\%) of the stellar mass in
the Universe today, but their contribution rises sharply
with redshift: depending on the IMF, stars inside dense
cores with $M^{\rm stars}_{r<1\,{\rm kpc}}>3
\times 10^{10}$\,\msun\ 
may contribute 10\,\% -- 20\,\% of the stellar mass
density at $z>2$ ({van Dokkum} {et~al.} 2014).

The formation of compact massive galaxies requires large
amounts of gas to be funneled in a region that is
only 1--2 kpc in diameter, while preventing significant
star formation at larger radii. 
Galaxy formation models have been able to reproduce the
broad characteristics of compact massive galaxies, either
by mergers that are accompanied by a strong central
star burst (e.g., {Hopkins} {et~al.} 2009b; {Wuyts} {et~al.} 2010; {Wellons} {et~al.} 2015),
by in-situ formation
from highly efficient gas cooling ({Naab} {et~al.} 2009; {Wellons} {et~al.} 2015),
or by contraction (``compaction'') of star forming gas disks
({Dekel} \& {Burkert} 2014; {Zolotov} {et~al.} 2015).
These scenarios have testable predictions:
for example, if compact
massive galaxies formed in mergers
then they may be expected to show tidal
features. Furthermore, the star formation rates of galaxies, and
their evolution in the size-mass plane, can be compared
to observations.

Observationally, the challenge is to identify these star forming
progenitors of compact massive galaxies. Once they are found
they can be studied, to measure the physical conditions inside
them and to test proposed mechanisms for their
formation
(see {Barro} {et~al.} 2013, 2014b; {Nelson} {et~al.} 2014; {Williams} {et~al.} 2014, 2015, for examples of such studies).
The main observational complication is that
typical quiescent galaxies at $z\gtrsim 2$ are structurally
very different from typical star forming galaxies 
(see, e.g., {Franx} {et~al.} 2008). At fixed mass, star forming
galaxies are larger, have a lower {Sersic} (1968) index and,
as a result, a much lower central density
(e.g., {Franx} {et~al.} 2008; {Kriek} {et~al.} 2009a; {van der Wel} {et~al.} 2014b; {van Dokkum} {et~al.} 2014). It may
be that a subset of the star forming galaxies decrease their
size through mergers or ``compaction'', but it would be
difficult to pinpoint which among the many
large, star forming
galaxies are destined to go through these phases.
A similar problem arises when linking compact,
quiescent descendants at $z=2$ to (lower mass) star forming
galaxies at much higher redshift ({Williams} {et~al.} 2014, 2015): although there may be
progenitors of massive quiescent galaxies
among small, blue, low mass star forming galaxies
at $z>3$, most of those galaxies will likely follow
different paths.

{Barro} {et~al.} (2013, 2014b) and {Nelson} {et~al.} (2014) use
a relatively model-independent and straightforward way to
identify plausible progenitors: they select massive
star forming galaxies at $z\gtrsim 2$ with the same
small sizes as quiescent galaxies. 
These objects form the compact tail of the size
distribution of star forming galaxies: for every massive
star forming galaxy at $z=2-2.5$ that is compact, there are several
that are not (see Sect.\ \ref{candidates.sec}, and
{van der Wel} {et~al.} 2014b).
It seems plausible that star forming
galaxies with the same structure as quiescent galaxies
are the direct ancestors of these galaxies, and there may
be physical reasons why the most compact star forming galaxies
are the most likely to shut off: many proposed quenching
and maintenance
mechanisms operate most effectively when a significant bulge
(and associated black hole) has formed ({Croton} {et~al.} 2006; {Hopkins} {et~al.} 2008; {Johansson}, {Naab}, \&  {Ostriker} 2009; {Conroy}, {van Dokkum}, \&  {Kravtsov} 2015). 


In this paper we build on previous studies by identifying
a sample of massive, compact, star forming galaxies
at $z=2-2.5$ in the 3D-HST survey
({van Dokkum} {et~al.} 2011; {Brammer} {et~al.} 2012b; {Skelton} {et~al.} 2014).
We study all five 3D-HST/CANDELS fields in a homogeneous
way, providing improved measurements of the number density
of candidate compact galaxies in formation. 
We present extensive Keck spectroscopy of a subset
of these candidates, and measure redshifts, emission
line widths, and emission line ratios.
The H$\alpha$ line profile and spatial extent is used to probe the
potential beyond the stellar effective radius, allowing us to
reconstruct the average rotation curve of this class of objects.
In the second part of the paper
we discuss a framework for the formation and evolution of massive
galaxies that places the results of the Keck spectroscopy in context.
We show that, even though individual galaxies likely have
complex formation histories, the evolution of the {\em population} of
massive galaxies can be described with a simple model in which galaxies
follow parallel tracks in the size-mass plane.
For consistency with previous studies we
assume $\Omega_{\rm m}=0.3$, $\Omega_{\Lambda}=0.7$,
and $H_0=70$\,\kms\,Mpc$^{-1}$.

\section{Compact Massive Star Forming Galaxies}

\subsection{Catalogs and Derived Parameters}
\label{catalogs.sec}

We use data from the 3D-HST project ({van Dokkum} {et~al.} 2011; {Brammer} {et~al.} 2012b)
to identify candidate
compact massive galaxies. The 3D-HST catalogs
({Skelton} {et~al.} 2014) provide multi-band photometry
for objects in the five extra-galactic fields of the
CANDELS survey ({Grogin} {et~al.} 2011; {Koekemoer} {et~al.} 2011). Objects
were selected using
a signal-to-noise (S/N)
optimized combination of the WFC3 $J_{125}$,
$JH_{140}$, and $H_{160}$ images. The catalogs encompass
nearly all publicly available data in the CANDELS fields,
including deep IRAC data, as well as medium-band imaging
in the  optical and the near-IR.
Stars were excluded, as well as
objects that have {\tt{use\_phot=0}}
(see {Skelton} {et~al.} 2014).

The imaging data are combined with 3D-HST WFC3 G141 grism
spectroscopy, which -- together with data from 
program GO-11600 -- covers $\approx 80$\,\% of the CANDELS
photometric area (see {Brammer} {et~al.} 2012b). The analysis
of the combined photometric and spectroscopic dataset will be
described in detail in I.\ Momcheva et al., in preparation.
Briefly, the photometric data from
{Skelton} {et~al.} (2014) and the 2D grism data were fit simultaneously
with a modified version of the EAZY code ({Brammer}, {van Dokkum}, \& {Coppi} 2008) to measure
redshifts, rest-frame colors, and the strengths of emission lines
({Brammer} {et~al.} 2012a). If there are no significant
emission or absorption features
in the grism spectrum, or if no grism spectrum is available,
the fit is similar to a standard
photometric redshift analysis. In version 4.1.4 of our
data release spectra
are extracted only to $H_{160}<24$ (and obviously only
in the area covered by the grism observations).

In addition to the Skelton et al.\ photometric information
and the grism spectroscopy we use
Spitzer MIPS 24\,$\mu$m data to estimate total IR luminosities
and star formation rates, as described in {Whitaker} {et~al.} (2012, 2014). These IR luminosities
and star formation rates are
consistent (within a factor of $\sim 2$)
with those derived from the full mid- and far-IR
SEDs, at least for the IR-luminous galaxies that have reliable
far-IR photometry
(see, e.g., {Muzzin} {et~al.} 2010; {Elbaz} {et~al.} 2011; {Wuyts} {et~al.} 2011; {Utomo} {et~al.} 2014).

Structural parameters of galaxies in the Skelton et al.\
catalogs were measured
by {van der Wel} {et~al.} (2014b), using the methodology described in
{van der Wel} {et~al.} (2012). Sizes, total luminosities, and ellipticities
were measured from the WFC3 imaging using the GALAPAGOS
implementation ({Barden} {et~al.} 2012) of GALFIT ({Peng} {et~al.} 2002).
In Sect.\ 7.2 we show with
a stacking analysis that the structural
parameters in the van der Wel et al.\ (2014b)
catalogs are reliable for the compact, massive galaxies
studied in this paper.
The catalog contains a small number of ``catastrophic''
failures. To identify these, we compared the total galaxy
fluxes from the GALFIT fit to the total fluxes in the
Skelton et al.\ catalogs. Galaxies were excluded from
the analysis if the absolute difference between these two
measurements exceeds 0.5 magnitudes. In this paper we
use circularized half-light radii throughout, defined as
\begin{equation}
\log r_{\rm e} = \log r_{\rm e,a} + 0.5\log q,
\end{equation}
with $r_{\rm e,a}$ the half-light radius along the major
axis and $q\equiv b/a$ the axis ratio of the galaxy. The sizes
are determined from data in the
$H_{160}$ band, which corresponds
to rest-frame $g$ at $z=2.3$.

Finally, stellar masses were determined from fits of
stellar population synthesis models to the 0.3\,$\mu$m --
8\,$\mu$m photometry, as described in {Skelton} {et~al.} (2014).
The fits were done with the FAST code ({Kriek} {et~al.} 2009b),
using a {Chabrier} (2003) IMF,
the {Calzetti} {et~al.} (2000) attenuation law, and
exponentially-declining star formation histories. These
parameters were chosen for consistency with previous studies;
small changes such as using ``delayed $\tau$''
models do not change the masses significantly. In this
paper we do not use the best-fitting
star formation rates, ages, or extinction from these fits,
as they tend to be less robust than the stellar masses
(see, e.g., {Kriek} {et~al.} 2009b; {Muzzin} {et~al.} 2009a).
A small (typically $\sim 5$\,\%)
correction was applied to each galaxy to make its half-light
radius and stellar mass self-consistent:
\begin{equation}
\log M_{\rm stars} = \log M_{\rm stars,FAST} + \log (L_{\rm G}/L_{\rm tot}),
\end{equation}
with $L_{\rm G}$ the total $H$ band luminosity as implied by
the GALFIT fit and $L_{\rm tot}$ the total $H$ band luminosity
in the Skelton et al.\ catalog (see {Taylor} {et~al.} 2010a; {van Dokkum} {et~al.} 2014).

\subsection{Selection of Star Forming Galaxies}
\label{uvj.sec}

In this paper we use the rest-frame colors of galaxies
to separate (candidate) star forming galaxies from quiescent
galaxies. As shown by {Labb{\'e}} {et~al.} (2005), {Wuyts} {et~al.} (2007),
{Whitaker} {et~al.} (2011), and many others, galaxies occupy
distinct regions in the space spanned by the rest-frame
$U-V$ and $V-J$ colors, depending on their specific
star formation rate. The reason is that dust
and age have a subtly different effect on the spectral
energy distributions (SEDs) of galaxies: galaxies that
are young and dusty are red in both $U-V$ and $V-J$,
whereas galaxies that are old and dust-free are red in $U-V$ but
(relatively) blue in $V-J$. With high quality redshifts
and photometry it has been demonstrated that there is
a gap between the
(age-)sequence of quiescent galaxies and the
(dust-)sequence of star forming galaxies in the $UVJ$ plane
({Whitaker} {et~al.} 2011; {Brammer} {et~al.} 2011), leading to a
relatively unambiguous separation of the two galaxy
classes.

\begin{figure}[hbtp]
\epsfxsize=8.5cm
\epsffile{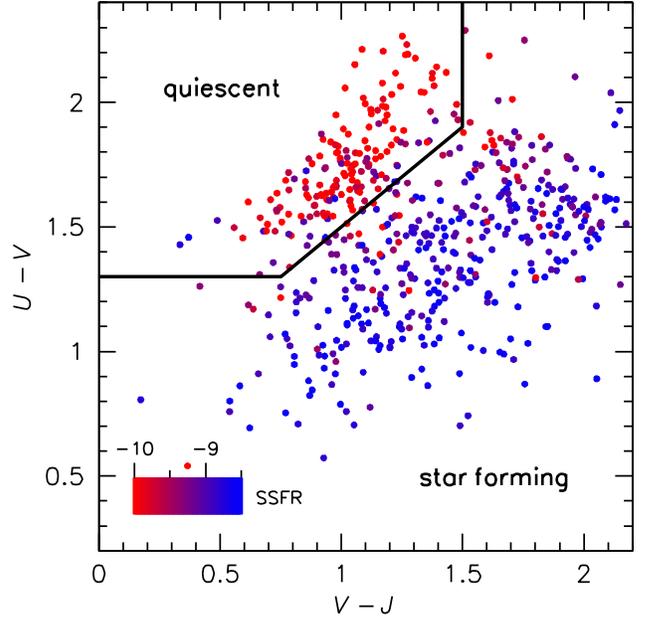}
\caption{\small
Distribution of galaxies with $\log M_{\rm stars}>10.6$\,\msun\
and $2.0<z<2.5$ 
in the $UVJ$ plane. The galaxies are color-coded by
the logarithm of their
specific star formation rate, SSFR\,=\,SFR/$M_{\rm stars}$.
The star formation rates are derived from the UV+IR emission,
with the IR emission determined from the Spitzer/MIPS flux.
In this paper ``star forming galaxies''
refers to all objects outside
of the $UVJ$ quiescent box.
\label{uvj.fig}}
\end{figure}

The distribution of galaxies with $\log M_{\rm stars}>10.6$
and $2.0<z<2.5$ in the $UVJ$ plane is shown in
Fig.\ \ref{uvj.fig}. The quiescent box is indicated with
the black lines; galaxies inside this box satisfy the
equations
\begin{eqnarray}
\label{uvj.eq}
V-J& < &1.5, \nonumber\\
U-V&>& 1.3, \nonumber \\
U-V & >& 0.8 (V-J) + 0.7.
\end{eqnarray}
Galaxies are color-coded by their specific star formation
rates, defined as SSFR\,=\,SFR/$M_{\rm stars}$, with SFR
the star formation rate derived from their UV+IR
emission (see {Whitaker} {et~al.} 2014, and references therein).
As can be seen in Fig.\ 1
the $UVJ$ selection corresponds very well to a selection
on specific star formation rate. This was expected from previous
studies (e.g., {Wuyts} {et~al.} 2011); nevertheless,
the correspondence is striking as the MIPS 24\,$\mu$m
measurements (which
dominate the star formation rates in this stellar mass range)
are entirely independent from the $U-V$ and $V-J$ colors.

We note that a subset of
quiescent galaxies has high SSFRs in Fig.\ \ref{uvj.fig};
these are galaxies whose rest-frame optical/near-IR SEDs show
no signs of star formation even though they have high
MIPS 24\,$\mu$m fluxes. These galaxies are difficult to interpret:
they may be quiescent galaxies with an active nucleus,
or their star formation is so
obscured that the young stars do not contribute significantly
to the SED. Fumagalli et al.\ (2014) show that the optical/near-IR
SEDs
of these galaxies are very similar to the ones that have no MIPS
detection.
Approximately 20\,\% of galaxies in the {Barro} {et~al.} (2013)
sample fall in this category.

Of 582 galaxies with $\log M_{\rm stars}>10.6$ and $2.0<z<2.5$,
185 (32\,\%)
are quiescent and 397 (68\,\%) are star forming.
The total area of the five fields is 896\,arcmin$^2$,
and the number densities of massive quiescent galaxies
and massive star forming galaxies are $1.2\times
10^{-4}$\,Mpc$^{-3}$ and $2.7\times 10^{-4}$\,Mpc$^{-3}$
respectively. These numbers are consistent with previous
measurements from other datasets
(e.g., {Marchesini} {et~al.} 2009; {Brammer} {et~al.} 2011; {Muzzin} {et~al.} 2013).

\subsection{Selection of Compact Massive Star Forming Galaxies}
\label{candidates.sec}

\begin{figure}[hbtp]
\epsfxsize=8.5cm
\epsffile{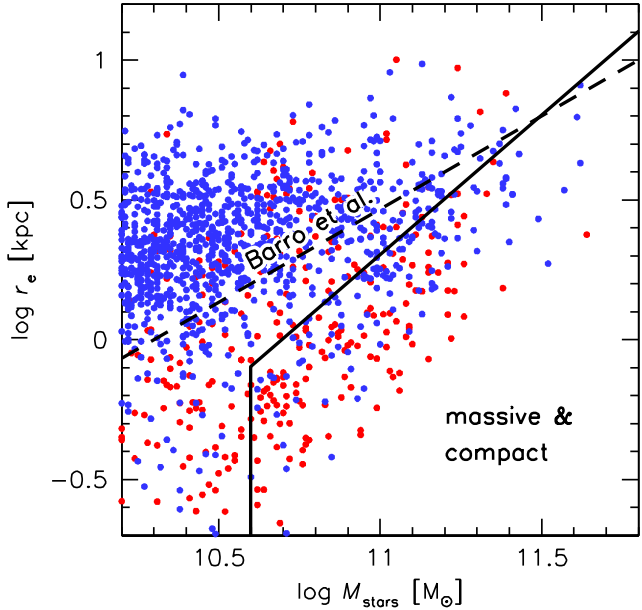}
\caption{\small
Size-mass relation for galaxies with $2.0<z<2.5$.
Sizes are circularized half-light
radii. Red symbols are $UVJ$-selected
quiescent galaxies, blue symbols are star forming galaxies. The solid
lines shows our selection criteria for compact, massive galaxies:
$\log M_{\rm stars}>10.6$ and $\log r_{\rm e}< \log M_{\rm stars}
-10.7$. This criterion is more restrictive than that used by
{Barro} {et~al.} (2013, 2014b) (dashed line); we did not use
the Barro et al.\ criterion as 60\,\% of star forming
galaxies with $\log M_{\rm stars}>10.8$ fall below the dashed line,
and their median size is significantly larger than that of massive
quiescent
galaxies.
\label{masssize.fig}}
\end{figure}

The size-mass relation for galaxies in the 3D-HST survey
with $2.0<z<2.5$ 
is shown in Fig.\ \ref{masssize.fig}. Quiescent and
star forming galaxies, identified using Eq.\ \ref{uvj.eq},
are indicated with red and blue points
respectively. As is well known, star forming galaxies are
larger than quiescent galaxies at fixed mass
(e.g., {Franx} {et~al.} 2008; {Williams} {et~al.} 2010; {van der Wel} {et~al.} 2014b). Note that the galaxy
distribution in Fig.\ \ref{masssize.fig}
is displaced with respect to that in
Fig.\ 5 of {van der Wel} {et~al.} (2014b), as we use
circularized half-light radii and van der Wel et al.\ use half-light
radii along the major axis.

Compact massive galaxies (CMGs) are in the lower right portion
of the size-mass diagram. {Barro} {et~al.} (2013) use the criterion
$\log r_{\rm e} < (\log M_{\rm stars} - 10.3)/1.5$ to isolate compact
galaxies (dashed line in Fig.\ \ref{masssize.fig}). However, at
masses of $\sim 10^{11}$\,\msun\ this
selection does not produce a sample of compact star forming galaxies that is
directly comparable to compact quiescent galaxies. The
median size of quiescent galaxies with $\log M_{\rm stars}>10.8$ that
satisfy the Barro et al.\ compactness criterion
is $r_{\rm e}=1.3$\,kpc. The median size of star forming galaxies
with $\log M_{\rm stars}>10.8$ that satisfy this
criterion is $2.2$\,kpc. For comparison, the median size
of the full sample of star forming galaxies with $\log M_{\rm stars}>10.8$
is $2.8$\,kpc. That is, at high masses,
the Barro et al.\ criterion selects
star forming galaxies whose sizes are closer to those
of the full sample of star forming galaxies than to those of
compact quiescent galaxies.
The reason is that the Barro et al.\ ``compactness''
criterion is not very restrictive at high masses, as it
selects 60\,\% of all star forming galaxies that have
$\log M_{\rm stars}>10.8$. 

As our goal is to select plausible progenitors of massive, compact
quiescent galaxies we adopt a slightly more restrictive criterion:
\begin{equation}
\label{selcompact.eq}
\log r_{\rm e}< \log M_{\rm stars}-10.7,
\end{equation}
with $M_{\rm stars}$ in units of M$_{\odot}$ and $r_{\rm e}$ in units
of kpc. This limit
is indicated by the solid diagonal line in Fig.\
\ref{masssize.fig}. 
Thirty-nine percent of star forming galaxies with
$\log M_{\rm stars}>10.8$ satisfy this criterion and their median
size is $r_{\rm e}=1.8$\,kpc.
As we discuss below, the slope of unity of our compactness criterion
can be readily interpreted in terms of
a physical parameter, namely the velocity dispersion.
The slope of $1/1.5=0.67$ used by Barro et al.\ (2013) was
chosen to
be consistent with the slope of the size-mass relation of quiescent
galaxies as found by Newman et al.\ (2012). We note that van
der Wel et al.\ (2014b) find a slightly steeper slope than
Newman et al.\ (2012) at $z\sim 2.3$ ($0.76\pm 0.04$
versus $0.69\pm 0.17$).

In addition to their compactness criterion
Barro et al.\ apply a mass limit of $\log M_{\rm stars}>10$. This
relatively low limit is also used for their comparison samples of
quiescent galaxies and spatially-extended star forming galaxies.
However, very few galaxies that have
$M_{\rm stars}\sim 10^{10}$\,\msun\ at $z=2$ will grow into $M_{\rm stars}
\sim 10^{11}$\,\msun\ galaxies by $z=0$ (e.g., {van Dokkum} {et~al.} 2010; {Leja}, {van Dokkum}, \&  {Franx} 2013a; {Behroozi} {et~al.} 2013).
We therefore apply a mass limit that is higher by a factor of 4:
$\log M_{\rm stars}>10.6$. This selection produces
homogeneous samples of {\em massive} compact
galaxies. 
Another consideration when choosing this mass limit
is that sizes are
uncertain when the effective radius is significantly
smaller than
the pixel size (the drizzled pixel size is
$0\farcs 06$, corresponding to 0.5\,kpc at $z=2$).

In the remainder of the paper we will use ``CMG'', for ``Compact Massive
Galaxy'', to denote objects with $\log M_{\rm stars}>10.6$ and
$\log r_{\rm e}<\log M_{\rm stars}-10.7$. 
Based on their location in
the $UVJ$ diagram we distinguish ``qCMG'', for quiescent compact
massive galaxy, and ``sCMG'', for star forming compact galaxy.
There are 112 sCMGs at $2.0<z<2.5$
in the five 3D-HST/CANDELS fields.
Five of these have effective radii $r_{\rm e}<0.5$\,kpc; when
calculating dynamical masses and expected velocity dispersions of
these galaxies we use 0.5\,kpc instead of their best-fitting radius.
It should be noted that many of the star forming progenitors of
$2<z<2.5$ qCMGs are expected to be at higher redshift than
$z=2.5$; we discuss the evolution of sCMGs and qCMGs in Sections
7 and 8.

\subsection{Expected Galaxy-Integrated Velocity Dispersions and
Number Densities}
\label{predsig.sec}

We quantify the compactness of galaxies by their expected
galaxy-integrated velocity
dispersion, as this quantity follows directly from our size-mass
selection and can be compared to
observations (see Sect.\ \ref{sigcomp.sec}).
For simplicity, we
use the following relation:
\begin{equation}
\label{sigpred.eq}
\log \sigma_{\rm pred} = 0.5\left(\log M_{\rm stars} - \log r_{\rm e} - 5.9
\right),
\end{equation}
with $\sigma_{\rm pred}$ the predicted velocity dispersion in \kms, 
$M_{\rm stars}$ in units of M$_{\odot}$, and $r_{\rm e}$ in units of kpc
({Franx} {et~al.} 2008; {van Dokkum}, {Kriek}, \&  {Franx} 2009). This relation has been shown to
reasonably predict
the observed
stellar velocity dispersions of both quiescent galaxies and
star forming galaxies, at least in the regime where this has
been tested: out to
$z\sim 0.7$ 
for massive star forming
galaxies ({Taylor} {et~al.} 2010a; {Bezanson}, {Franx}, \& {van  Dokkum} 2015) and
out to $z\sim 2$ for massive quiescent galaxies
({Bezanson} {et~al.} 2013; {van de Sande} {et~al.} 2013; {Belli} {et~al.} 2014a).

\begin{figure}[hbtp]
\epsfxsize=8.5cm
\epsffile{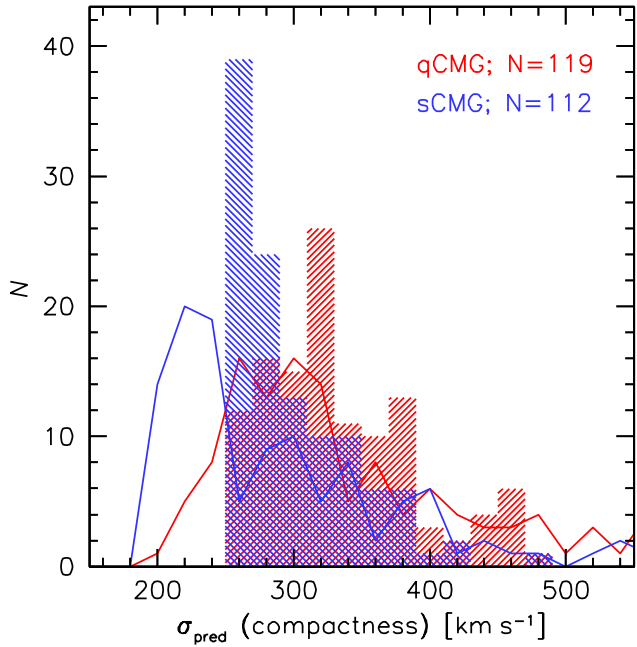}
\caption{\small
Distribution of expected galaxy-integrated velocity dispersions
at $2.0<z<2.5$, for quiescent compact massive galaxies (qCMGs; red)
and for star forming compact massive galaxies (sCMGs; blue).
Histograms use a simple relation of the form $\sigma^2 \propto M_{\rm stars}
/r_{\rm e}$. Our compactness criterion corresponds to
$\sigma_{\rm pred}>250$\,\kms. Lines use an expression that
takes the Sersic index of the galaxies into account.
sCMGs have a median
predicted dispersion of $284$\,\kms.
\label{sigpred.fig}}
\end{figure}

Our compactness criterion (Eq.\ \ref{selcompact.eq}) corresponds to
$\log \sigma_{\rm pred} > 2.40$, or $\sigma_{\rm pred}>250$\,\kms.
The distributions of predicted dispersions of sCMGs and qCMGs
are shown by the
histograms in Fig.\
\ref{sigpred.fig}. The median expected dispersions of the two populations
are similar but not identical:
$\sigma_{\rm pred}=324$\,\kms\ for quiescent galaxies and
$\sigma_{\rm pred}=284$\,\kms\ for star forming galaxies. The reason
for this difference is that the size distribution of quiescent galaxies
is different from that of star forming galaxies. For star forming
galaxies we select the tail of the distribution, with the largest
number of galaxies close to the compactness cutoff, whereas for
quiescent galaxies we select the bulk of the population
(see {van der Wel} {et~al.} 2014b, for a discussion of the form of the size
distributions of quiescent and star forming galaxies).
Phrased differently, irrespective of the exact compactness criterion,
the smallest galaxies tend to be quiescent. We will return to this
in Sect.\ \ref{simplemod.sec},
where we define a ``quenching line'' just inside the
compact massive galaxy box.

As shown in {Taylor} {et~al.} (2010a), the residuals between expected and
observed dispersions correlate with the Sersic index.
The lines in Fig.\ \ref{sigpred.fig} show the distributions when the
Sersic index of the galaxies is taken into account, using
\begin{equation}
\label{sigpredn.eq}
\log \sigma_{\rm pred} = 0.5\left(\log G + \log \beta(n) + \log M_{\rm stars} - \log r_{\rm e}\right),
\end{equation}
with
\begin{equation}
\beta(n) =  8.87-0.831 n + 0.0241 n^2
\end{equation}
({Cappellari} {et~al.} 2006). Here $n$ is the Sersic index and
$G=4.31\times 10^{-6}$ when $M_{\rm stars}$ is in units of
\msun, $r_e$ is in kpc, and $\sigma_{\rm pred}$ is in km\,s$^{-1}$.
\sg{}s have a slightly smaller median
Sersic index ($\langle n\rangle = 2.4$) than \qg{}s
($\langle n\rangle=2.9$). For quiescent galaxies the line and
histogram are nearly the same, but for star forming galaxies the
Sersic-dependent dispersions are on average $\approx 10$\,\% lower than
those calculated with Eq.\ \ref{sigpred.eq}.

The number density of \qg{}s and \sg{}s is
the same, $0.8\times 10^{-4}$\,Mpc$^{-3}$ (for
reference, the number density of the full population of
quiescent galaxies with $\log M_{\rm stars}>10.6$ is $1.2\times
10^{-4}$\,Mpc$^{-3}$; see Sect.\ \ref{uvj.sec}). This result is
consistent with previous studies that noted the overlap of the
compact tail of star forming galaxies and the bulk of the quiescent
population ({Barro} {et~al.} 2013;
van der Wel et al.\ 2014b). We therefore confirm that
a population of star forming galaxies can be identified at $2.0<z<2.5$
that has a median mass, median size, and number density similar to the
population of massive quiescent galaxies at the same redshifts.
If all these compact star forming galaxies quench in the near future,
the number density of massive quiescent galaxies will increase by
70\,\%, and the number density of
\qg{}s will double.

\begin{deluxetable}{lcccc}
\tabletypesize{\footnotesize}
\tablewidth{0pt}
\tablecaption{Coordinates
of Confirmed Star Forming Compact Massive Galaxies}
\tablehead{
\colhead{id\tablenotemark{a}} &
\colhead{RA} &
\colhead{DEC} &
\colhead{$R_{606}$} &
\colhead{$H_{160}$}
}
\startdata
AEGIS\_9163 & 14$^{\rm h}$21$^{\rm m}$03\fs{}68 & \phn53\arcdeg{}04\arcmin{}37\farcs3 & 25.8 & 23.2 \\ 
AEGIS\_26952 & 14$^{\rm h}$20$^{\rm m}$40\fs{}81 & \phn53\arcdeg{}04\arcmin{}51\farcs9 & 25.2 & 22.2 \\ 
AEGIS\_41114 & 14$^{\rm h}$18$^{\rm m}$32\fs{}92 & \phn52\arcdeg{}46\arcmin{}06\farcs7 & 25.1 & 22.7 \\ 
COSMOS\_163 & 10$^{\rm h}$00$^{\rm m}$25\fs{}01 & \phn\phn{}2\arcdeg{}10\arcmin{}44\farcs1 & 25.9 & 23.2 \\ 
COSMOS\_1014 & 10$^{\rm h}$00$^{\rm m}$35\fs{}92 & \phn\phn{}2\arcdeg{}11\arcmin{}27\farcs8 & 23.1 & 21.5 \\ 
COSMOS\_11363 & 10$^{\rm h}$00$^{\rm m}$28\fs{}71 & \phn\phn{}2\arcdeg{}17\arcmin{}45\farcs4 & 24.2 & 21.3 \\ 
COSMOS\_12020 & 10$^{\rm h}$00$^{\rm m}$17\fs{}91 & \phn\phn{}2\arcdeg{}18\arcmin{}07\farcs2 & 25.8 & 22.0 \\ 
COSMOS\_22995 & 10$^{\rm h}$00$^{\rm m}$17\fs{}15 & \phn\phn{}2\arcdeg{}24\arcmin{}52\farcs3 & 24.6 & 22.1 \\ 
COSMOS\_27289 & 10$^{\rm h}$00$^{\rm m}$41\fs{}58 & \phn\phn{}2\arcdeg{}27\arcmin{}51\farcs5 & \nodata & 22.1 \\ 
GOODS-N\_774 & 12$^{\rm h}$36$^{\rm m}$27\fs{}73 & \phn62\arcdeg{}07\arcmin{}12\farcs8 & 27.1 & 23.0 \\ 
GOODS-N\_6215\tablenotemark{b} & 12$^{\rm h}$36$^{\rm m}$06\fs{}86 & \phn62\arcdeg{}10\arcmin{}21\farcs4 & 25.2 & 21.5 \\ 
GOODS-N\_13616\tablenotemark{b} & 12$^{\rm h}$36$^{\rm m}$06\fs{}33 & \phn62\arcdeg{}12\arcmin{}32\farcs9 & 25.9 & 22.8 \\ 
GOODS-N\_14283\tablenotemark{b} & 12$^{\rm h}$37$^{\rm m}$02\fs{}60 & \phn62\arcdeg{}12\arcmin{}44\farcs0 & 25.0 & 22.9 \\ 
GOODS-N\_22548\tablenotemark{b} & 12$^{\rm h}$37$^{\rm m}$00\fs{}46 & \phn62\arcdeg{}15\arcmin{}08\farcs9 & 25.5 & 22.5 \\ 
GOODS-S\_5981 & \phn{}3$^{\rm h}$32$^{\rm m}$14\fs{}55 & $-27$\arcdeg{}52\arcmin{}56\farcs5 & 24.9 & 22.4 \\ 
GOODS-S\_30274 & \phn{}3$^{\rm h}$32$^{\rm m}$31\fs{}46 & $-27$\arcdeg{}46\arcmin{}23\farcs2 & 23.5 & 21.3 \\ 
GOODS-S\_37745 & \phn{}3$^{\rm h}$32$^{\rm m}$43\fs{}88 & $-27$\arcdeg{}44\arcmin{}05\farcs7 & 24.1 & 22.0 \\ 
GOODS-S\_45068\tablenotemark{b} & \phn{}3$^{\rm h}$32$^{\rm m}$33\fs{}02 & $-27$\arcdeg{}42\arcmin{}00\farcs4 & 25.0 & 22.5 \\ 
GOODS-S\_45188 & \phn{}3$^{\rm h}$32$^{\rm m}$15\fs{}18 & $-27$\arcdeg{}41\arcmin{}58\farcs7 & 25.4 & 22.9 \\ 
UDS\_16442 & \phn{}2$^{\rm h}$17$^{\rm m}$20\fs{}80 & \phn$-5$\arcdeg{}13\arcmin{}16\farcs0 & 27.4 & 23.4 \\ 
UDS\_25893 & \phn{}2$^{\rm h}$18$^{\rm m}$02\fs{}97 & \phn$-5$\arcdeg{}11\arcmin{}21\farcs3 & \nodata & 23.1 \\ 
UDS\_26012 & \phn{}2$^{\rm h}$17$^{\rm m}$03\fs{}66 & \phn$-5$\arcdeg{}11\arcmin{}22\farcs2 & 25.4 & 22.4 \\ 
UDS\_33334 & \phn{}2$^{\rm h}$16$^{\rm m}$55\fs{}01 & \phn$-5$\arcdeg{}09\arcmin{}52\farcs8 & 26.2 & 23.3 \\ 
UDS\_35673 & \phn{}2$^{\rm h}$17$^{\rm m}$05\fs{}33 & \phn$-5$\arcdeg{}09\arcmin{}25\farcs7 & 25.1 & 22.4 \\ 
UDS\_42571 & \phn{}2$^{\rm h}$17$^{\rm m}$43\fs{}95 & \phn$-5$\arcdeg{}07\arcmin{}51\farcs3 & 27.0 & 22.8 
\enddata
\tablenotetext{a}{Id number in Skelton et al.\ (2014).}
\tablenotetext{b}{Confirmation from {Barro} {et~al.} (2014b); RA, DEC,
$R_{606}$ and $H_{160}$ from {Skelton} {et~al.} (2014).}
\end{deluxetable}

\section{Near-IR Spectroscopy}

We observed candidate sCMGs
with the
near-IR spectrographs MOSFIRE ({McLean} {et~al.} 2012) and NIRSPEC
({McLean} {et~al.} 1998)
on Keck in 2014 and 2015. The resulting spectra
provide spectroscopic redshifts
(measured from \ha\ and \nii\ at $2.0<z<2.7$), which can be
used to verify that a population of sCMGs exists at these redshifts.
Furthermore,  the
spectroscopic observations provide
galaxy-integrated kinematics of
the ionized gas: if compact
star forming galaxies are in the process of forming the stars
that are later in compact quiescent galaxies,
their gas kinematics should be
similar to the stellar kinematics of quiescent galaxies.
In addition to redshifts and kinematics the spectra
provide star formation rates and  strong
line ratios; these are important for
understanding the physical processes that take place in these galaxies,
although their interpretation is often not unique.

\subsection{MOSFIRE}
\label{mosfire.sec}

The MOSFIRE spectra were obtained in three separate observing runs:
January 11, 12 2014; April 18, 23, 25 2014; and Dec 12, 13, 15 2014.
The January run suffered from clouds and poor seeing; conditions
were generally good
during the other two runs. Compact, massive star forming galaxies
were not always the main targets, and were not always selected using
the criteria of Sect.\ 2.2. One target from the April run,
a galaxy at $z=7.730$, is described in {Oesch} {et~al.} (2015).
The December run gave higher
priority to galaxies at $3.0<z<3.6$ than to galaxies at lower redshift.
In this paper we will limit the discussion to star forming
galaxies at $2<z<2.5$
that satisfy the criteria of Sect.\ \ref{candidates.sec}.

The observations were all taken in the $K$-band, using
a standard ABAB dither pattern. 
The exposure times varied from $\sim 1$\,hr to $\sim 4$\,hrs,
depending on conditions and the requirements imposed by the primary
targets in the masks. One of the slits in each mask
was devoted to a relatively bright, relatively blue star.
This has four important functions: the S/N ratio of the star is used
to weight individual exposures in the reduction; the $y-$position
of the star is used to correct the data for small vertical
drifts of the mask relative to the sky (see {Kriek} {et~al.} 2015); the
extracted spectrum is used to identify regions of strong sky
absorption; and the width of the 2D stellar spectrum in the spatial
direction provides us with a model of the point spread
function (PSF) that is otherwise very difficult
to construct (see Sect.\ \ref{morphline.sec}).

The data reduction used the standard MOSFIRE pipeline
DRP,\footnote{https://code.google.com/p/mosfire/} with small
modifications (see {Oesch} {et~al.} 2015). Individual sequences
were reduced and shifted to a common reference{} frame before
stacking.
One-dimensional spectra were obtained from the 2D spectra by summing
rows, as dictated by the observed spatial extent of the galaxies.
For each mask an empirical noise spectrum was created by removing all
rows with signal, and determining the width of the pixel
distribution of the remaining rows for each pixel in the wavelength
direction. The width was measured by removing the lowest and highest
16\,\% of values, and is therefore equivalent to the $\pm 1 \sigma$
width of a Gaussian. For each individual galaxy in a mask
the noise spectrum was multiplied by the square
root of the number of rows that was summed to create the 1D spectrum
of that object.

\subsection{NIRSPEC}

The NIRSPEC data were obtained in two runs, January 10, 13, 14 2014 and
January 25, 26 2015. Conditions were poor in the 2014 run and the
only object in our final sample that came from it is GOODS-N\_774,
which was published in
{Nelson} {et~al.} (2014). Conditions in 2015 were excellent, with
the seeing
ranging from $0\farcs 3 - 0\farcs 6$ during both nights. 
The selection for the NIRSPEC runs was very similar to that described
in Sect.\ \ref{candidates.sec}; within
these criteria priority was generally given to galaxies
with higher star formation rates (and with good blind
offset stars; see below).

\begin{figure*}[hbtp]
\begin{center}
\epsfxsize=17cm
\epsffile{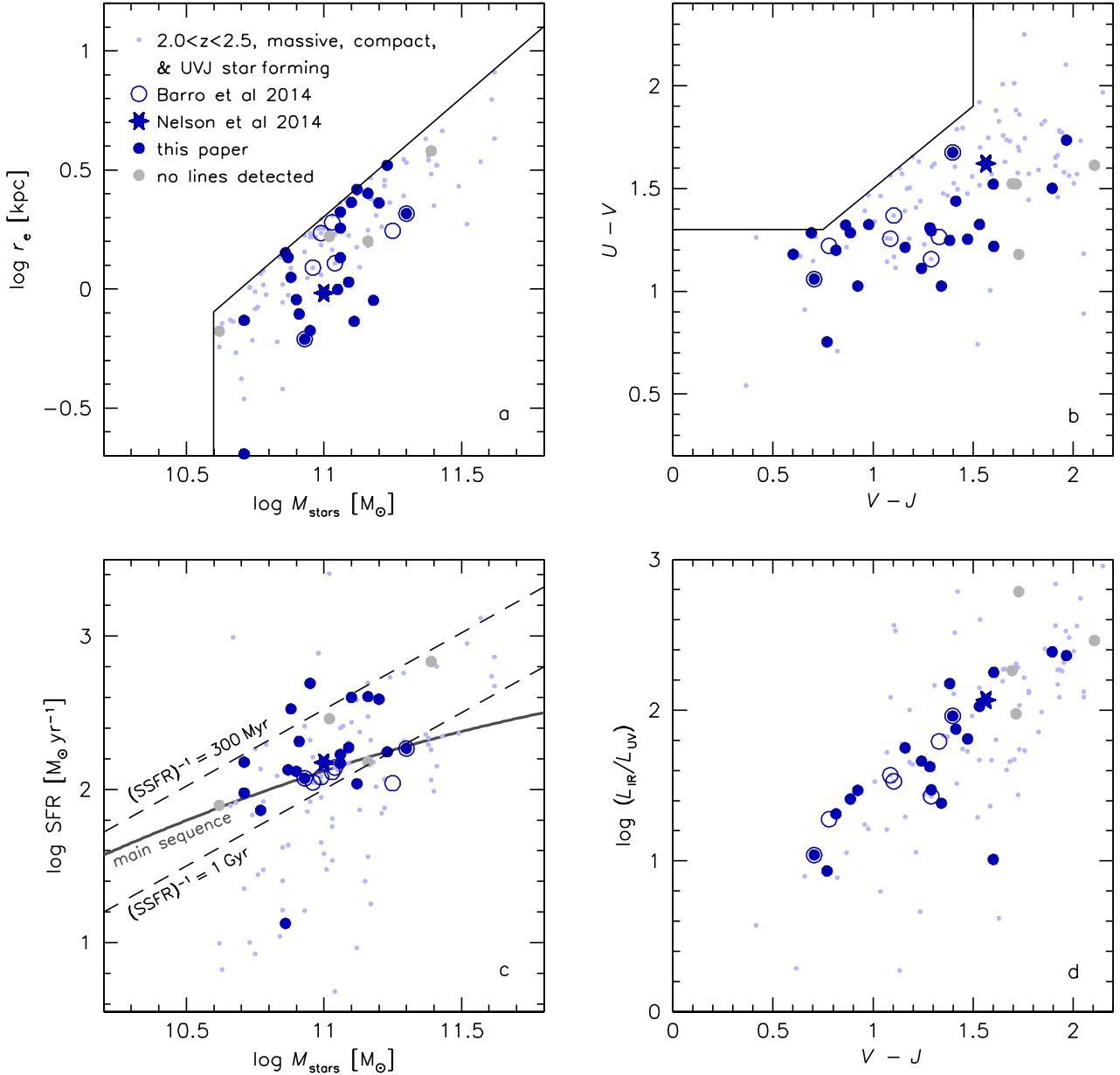}
\end{center}
\vspace{-0.3cm}
\caption{\small
Comparison of objects with near-IR spectra to the parent population
of compact, massive star forming galaxies at $2.0<z<2.5$. 
Panels show the size-mass relation (a), the $UVJ$ diagram (b), the
star formation -- mass relation (with
the {Whitaker} {et~al.} (2014) ``main sequence'' indicated) (c), and
the relation between $L_{\rm IR}/L_{\rm UV}$ to the rest-frame
$V-J$ color (d). Solid blue symbols are objects in the sample described
here. Open symbols are galaxies from
{Barro} {et~al.} (2014b) that fall in our selection
box. Grey points are observed galaxies whose
spectrum did not show any clear features.
\label{select.fig}}
\end{figure*}

We followed standard observing procedures for NIRSPEC
spectroscopy of faint targets
(see, e.g., {Erb} {et~al.} 2003; {van Dokkum} {et~al.} 2004).
Target aquisition was done with blind offsets from nearby stars,
as the galaxies are not detected in the SCAM slit-viewing
camera.  The N6 filter was used for GOODS-N\_774; all data in the
2015 run were taken with the N7 filter.
A typical observing sequence consisted of four 900\,s exposures in an ABBA
pattern with $7\arcsec$ offsets between nods. The data were continuously
inspected as objects sometimes drift out of the slit.

The data reduction followed standard procedures for near-IR, single
slit data (see, e.g., {van Dokkum} {et~al.} 2004). 
The data were initially
reduced in pairs, using the sky of the A frame for the B frame and vice
versa. This method yields relatively clean, photon noise-dominated
spectra, at the expense of reducing the S/N in the final frames by $\sqrt{2}$
(see, e.g., {Kriek} {et~al.} 2015). Wavelength calibration was done using
sky lines, which were also used to determine the spectral resolution
of the data (see Sect.\ \ref{fitting.sec}). The slit is not
long enough
to obtain an accurate noise spectrum from empty regions; therefore, we
calculate the noise spectrum from the sky spectrum and the noise in
the darks. An analysis of the residuals from fits to the emission
lines shows that this is sufficient for our purposes (see
Sect.\ \ref{fitting.sec}).

\begin{figure*}[hbtp]
\begin{center}
\epsfxsize=17cm
\epsffile{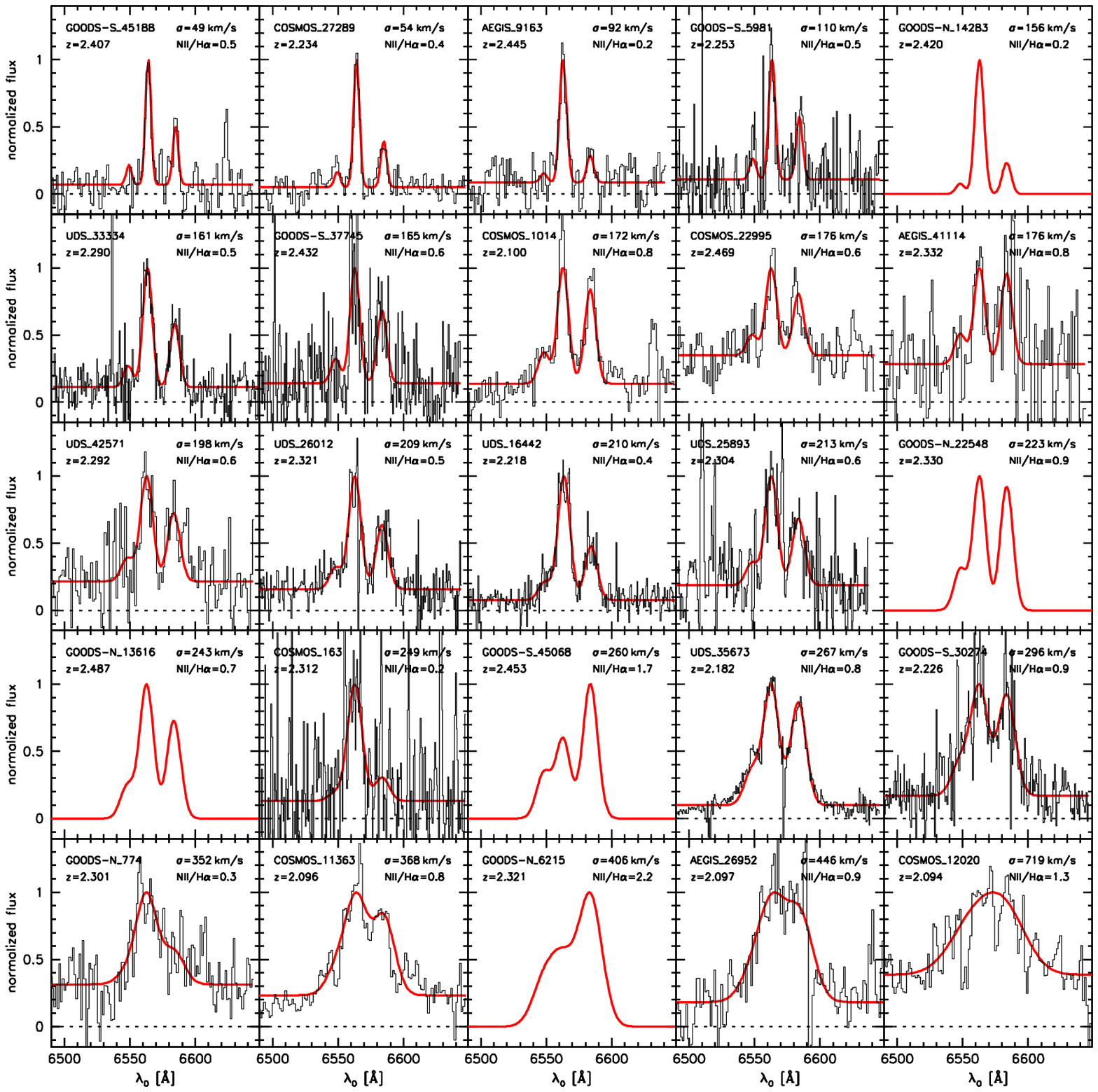}
\end{center}
\vspace{-0.3cm}
\caption{\small
Spectra of the 20 \sg{}s in our sample with $2.0<z<2.5$.
Red lines show best-fitting models, as determined
with the {\tt emcee} code ({Foreman-Mackey} {et~al.} 2013). We also show
the best-fitting models of
five galaxies from {Barro} {et~al.} (2014b) that satisfy our
selection criteria (red lines without data); these objects
are included in our analysis.
The galaxies are ordered by their observed line
widths, which range from $\sim 50$\,\kms\ to $\sim 700$\,\kms.
\label{spectra.fig}}
\end{figure*}

\begin{figure*}[hbtp]
\begin{center}
\epsfxsize=17cm
\epsffile{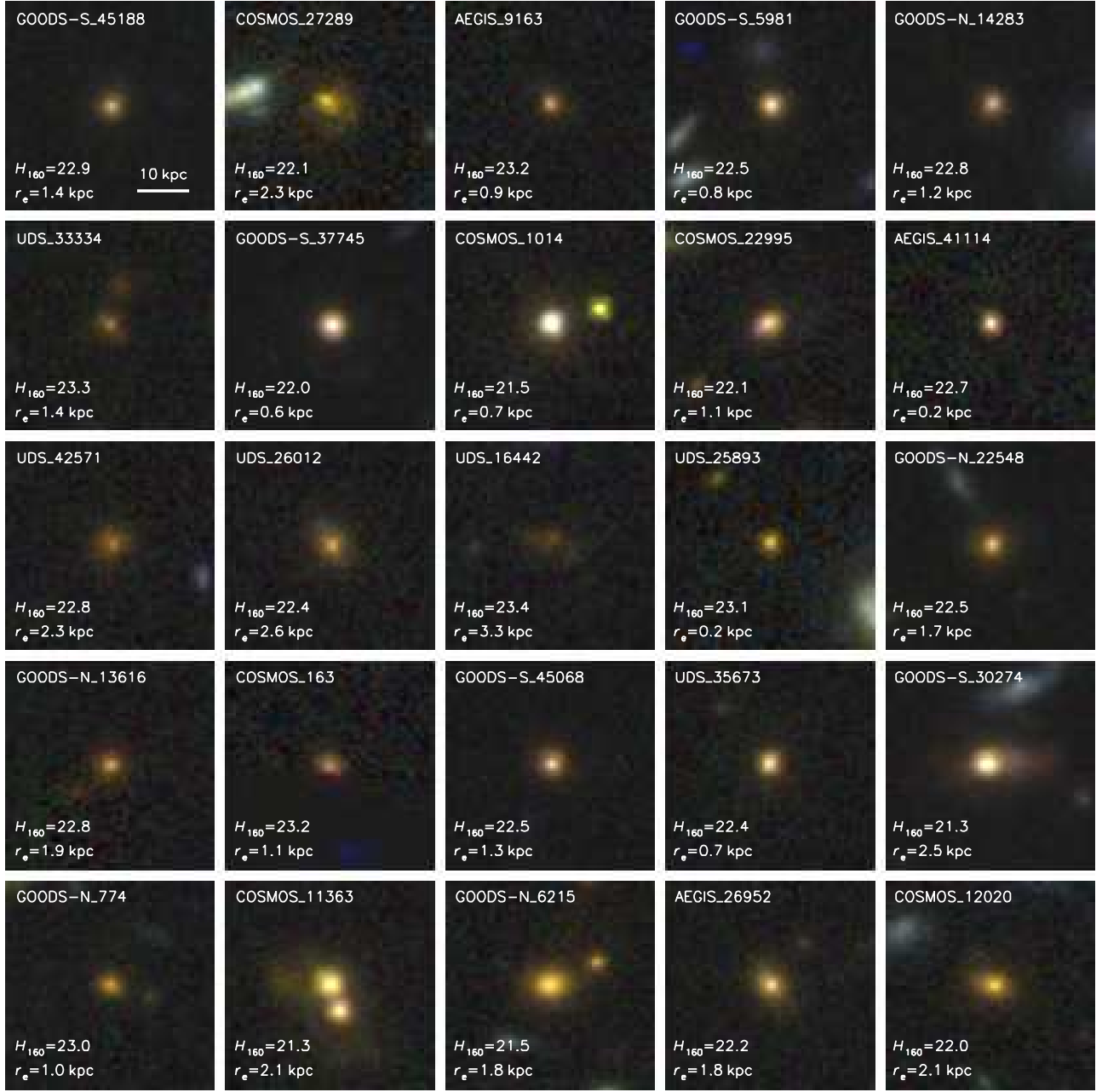}
\end{center}
\vspace{-0.3cm}
\caption{\small
HST images of the galaxies of Fig.\ \ref{spectra.fig}, created from
the WFC3 $H_{160}$, $J_{125}$ and summed ACS $V_{606} +
I_{814}$ images. Each image is $4\farcs 8 \times 4\farcs8$,
corresponding to approximately 40\,kpc\,$\times\,40$\,kpc.
The $H_{160}$ magnitudes and circularized effective radii are
listed in the images. Note that the galaxies were selected
to be compact in mass, and are not necessarily compact
in light. There is generally
little evidence for spiral arms, star forming
clumps, or other structure.
Two galaxies show
evidence for past (GOODS-S\_30274) and
ongoing (COSMOS\_11363) mergers. The galaxies are ordered by
their H$\alpha$ velocity dispersion, as in Fig.\ \ref{spectra.fig}.
There is no clear relation between HST morphology and H$\alpha$
velocity dispersion in this sample.
\label{hstim.fig}}
\end{figure*}

\begin{figure*}[hbtp]
\begin{center}
\epsfxsize=17cm
\epsffile{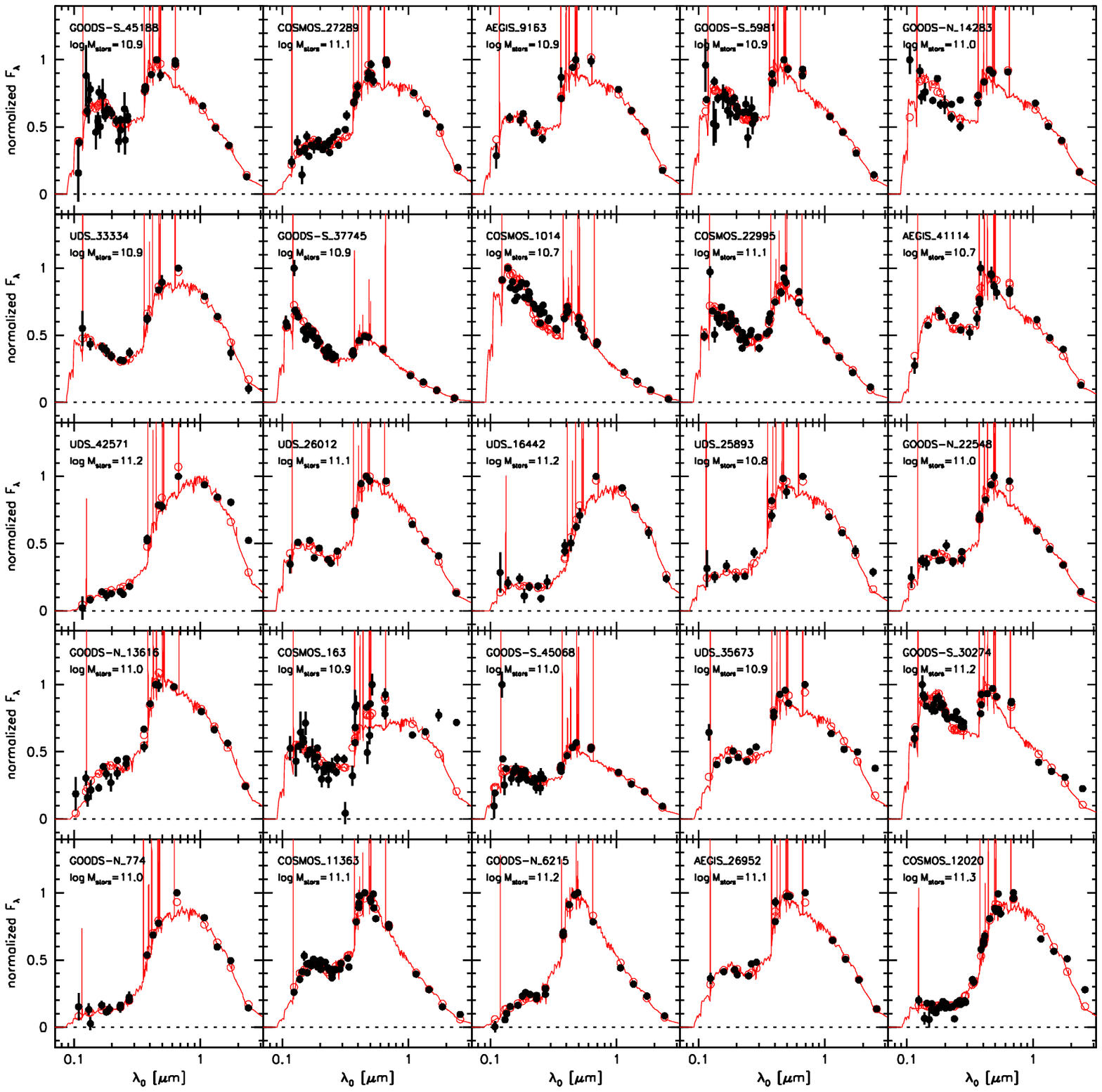}
\end{center}
\vspace{-0.3cm}
\caption{\small
Restframe UV to near-IR spectral  energy distributions of
the galaxies of Fig.\ \ref{spectra.fig}. The red
spectra are the best-fitting
EAZY ({Brammer} {et~al.} 2008) models; open red circles show the model fluxes
in the observed filters.
The SEDs show a large variety, ranging from
blue, relatively unobscured emission (COSMOS\_1014) to very red SEDs with
high inferred dust content (e.g., UDS\_42571 and GOODS-N\_774). As in Fig.\
\ref{hstim.fig}, there is no obvious relation between the SEDs of
the galaxies and the measured velocity dispersions of their ionized gas.
\label{seds.fig}}
\end{figure*}

\subsection{Results and Comparison to Parent Sample}
\label{parent.sec}

We identify the redshifted \ha\ and \nii\
emission lines in 20 out of 24 compact, massive star forming
galaxies with expected redshifts in the range $2.0<z<2.5$.
This success rate of 86\,\% is
encouraging,\footnote{Somewhat amazing really, particularly
when considering that only a
handful of these objects had a previously measured secure
redshift from the
ground or the grism.}
but it should be noted that our selection at the telescope was somewhat
subjective, particularly in the NIRSPEC runs. As an example, if there
were two plausible targets and one showed a hint of an H$\alpha$ 
contribution to the broad band flux we would generally give it preference.
Additionally, there are five non-overlapping
galaxies in {Barro} {et~al.} (2014b) that
satisfy our criteria (see Sect.\ 3.5); the
total sample of  massive compact star forming galaxies
with \ha\ measurements is therefore 25 (Table 1).

The properties of the galaxies in the spectroscopic sample are compared
to the parent sample in Fig.\ \ref{select.fig}. The median size and 
mass are $r_{\rm e}=1.3$\,kpc and $M_{\rm stars}=1.0\times 10^{11}$\,\msun\
respectively, close to the medians of the parent sample. The spread is
somewhat smaller; 24 out of 25 galaxies are in the mass
range $10.7< \log M_{\rm stars}<11.3$. The galaxies have bluer $U-V$
colors and slightly
higher UV+IR star formation rates than the parent sample.
This is by selection: galaxies with specific star formation rates
${\rm SSFR}<10^{-9}$\,yr$^{-1}$ were given lower priority. Despite
the lack of galaxies with low star formation rates in the spectroscopic
sample, the median SSFR is only 0.1\,dex higher than that of the
parent sample ($\log {\rm SSFR}=-8.8$\,yr$^{-1}$ compared to 
$\log {\rm SSFR}=-8.9$\,yr$^{-1}$ for the parent sample).
Both medians are close
to the {Whitaker} {et~al.} (2014) main sequence for this redshift
(dark grey line in Fig.\ \ref{select.fig}c).
Panel d of Fig.\ \ref{select.fig} shows the dust content of the
galaxies, as parameterized by both the ratio of the IR and UV
luminosities and the rest-frame $V-J$ color. Galaxies in the upper
right part of this panel are very dusty, with the re-radiated IR
emission exceeding the UV emission by a factor of $>100$.
The median $L_{\rm IR}/L_{\rm UV}$ ratio of the parent sample
is $\langle L_{\rm IR}/L_{\rm UV}\rangle = 64$. 
The median ratio for the galaxies in the spectroscopic sample is
slightly lower, at 42. We only have a few spectroscopic objects in
this part of the diagram, and all four  spectroscopic failures are
located here. We infer that the most likely explanation for the
failures is
that the H$\alpha$
emission in these galaxies is too obscured for a detection in
our current observations.

The Keck spectra of the 20 galaxies that we observed are shown
in Fig.\ \ref{spectra.fig}.
The galaxies are ordered by the measured velocity
dispersion (see below). 
We include the
five objects from {Barro} {et~al.} (2014b) that satisfy our selection
criteria;
as we cannot show the spectra of these objects
in Fig.\ \ref{spectra.fig}, we instead
show models that are based on their published best-fitting
parameters. 

Figures \ref{hstim.fig} and \ref{seds.fig} show the HST images and the
rest-frame UV -- near-IR spectral energy distributions (SEDs) of the
25 galaxies of Fig.\ \ref{spectra.fig}. The $H_{160}$ images
are shown separately at high dynamic range in Appendix A.
The SEDs range from relatively unobscured
(COSMOS\_1014)
to extremely dusty (e.g., GOODS-N\_774). Some have excess emission in
the IRAC bands (UDS\_42571; see, e.g., {Mentuch} {et~al.} 2009)
Two galaxies show clear signs of merging:
COSMOS\_11363 is an ongoing merger between two
compact massive galaxies that
are only $0\farcs 6$ apart, and GOODS-S\_30274
is probably a merger remnant (see Sect.\ \ref{sb.sec}).
Interestingly, there is no
clear relation between the measured velocity dispersion and either
the morphology
or the SED.
Phrased differently, it is not possible to predict the H$\alpha$
line width based on the information shown in Figs.\ \ref{hstim.fig}
and \ref{seds.fig}.

\subsection{Redshifts, Fluxes, Line Widths, and Line Ratios}
\subsubsection{Fitting}
\label{fitting.sec}
The spectra were fitted with a model that has the redshift,
the continuum level, the
\nii\ and H$\alpha$ line fluxes, and the line width as free parameters.
The instrumental resolution
is explicitly taken into account.
The model has the following form:
\begin{equation}
M(\lambda) = L(\lambda) \ast R(\lambda) + C,
\end{equation}
with $L(\lambda)$ the model for the line emission, $R(\lambda)$ the
instrumental resolution, $C$ the continuum level, and
$\ast$ denoting convolution.
The instrumental resolution is modeled with a Gaussian:
\begin{equation}
R(\lambda) = \frac{\Delta \lambda}{\sqrt{2\pi}\sigma_{\rm instr}}\exp\left(-0.5\left(\frac{\lambda-\lambda_{\rm cen}}
{\sigma_{\rm instr}}\right)^2\right),
\end{equation}
with $\sigma_{\rm instr}$ measured from sky lines in the vicinity of
the redshifted H$\alpha$ line, $\Delta \lambda$ the pixel size
in \AA, and $\lambda_{\rm cen}$ the
center of the fitting range. Expressed as a velocity, the resolution of
the MOSFIRE spectra is $\approx 35$\,\kms, and the resolution of the
NIRSPEC data is $\approx 80$\,\kms. 
The lines are parameterized as follows:
\begin{equation}
\label{lines.eq}
L(\lambda) = f_{{\rm H}\alpha} L_{6563}(\lambda) +
f_{[{\rm N}{\textsc{ii}}]} \left( L_{6584}(\lambda) + \frac{1}{3}L_{6548(\lambda)},
\right)
\end{equation}
with
\begin{equation}
L_{\lambda_0}(\lambda) = \frac{\Delta\lambda}{\sqrt{2\pi}\sigma}\exp\left(-0.5\left(\frac{\lambda-(1+z)\lambda_0}
{\sigma}\right)^2\right).
\end{equation}
Here $f$ is the line strength, $\sigma$ is the galaxy-integrated
line-of-sight velocity dispersion,
$\lambda_0$ is the rest-frame wavelength of the line
(with $\lambda_0=6562.8$ and $\lambda_0=6548.1$, $6583.6$
for H$\alpha$ and the two [N{\sc ii}] lines respectively),
and $z$ is the redshift.

Some galaxies show evidence for multiple velocity components
(e.g., COSMOS\_1014).
We do not attempt to separately fit broad and narrow velocity
components to these galaxies
(as was done by, e.g., {F{\"o}rster Schreiber} {et~al.} 2014). As discussed
later, broad components could indicate the presence of
winds but could also indicate rapidly rotating gas at
small radii in the galaxies. In the absense of high spatial
resolution data, it is difficult to distinguish these
possibilities; we therefore simply interpret the
H$\alpha$-luminosity-weighted velocity profiles in this paper.
It should be noted that the formal uncertainties underestimate
the error in the velocity dispersion if the velocity
distribution is not Gaussian. This is particularly important
for galaxies with a high S/N ratio, such as COSMOS\_12020.

The {\tt emcee} MCMC algorithm ({Foreman-Mackey} {et~al.} 2013) was used to fit this model
to the galaxy spectra. The fit was done over the wavelength region
$(1+z)\lambda_{6548}-200 < \lambda <
(1+z)\lambda_{6584}+200$; the results are not dependent on the choice
of fitting region as long as the continuum is reasonably well covered.
Priors are top hats with boundaries that comfortably encompass the
fitting results. That is, the Bayesian aspects of {\tt emcee} were
essentially turned off.
We used 100 walkers and generated 500 chains
in each fit. Burn-in was typically fast, but we removed the first
200 chains when calculating errors. For each fit
parameter the best fit is defined as the
median of the 300 remaining samples. Errors were determined from the
16$^{\rm th}$ and 84$^{\rm th}$ percentiles (see {Foreman-Mackey} {et~al.} 2013, for details).
The best fit models are shown by red lines in Fig.\
\ref{spectra.fig}. Residuals from the fits are shown in
Fig.\ \ref{spectra_noise.fig}. As discussed in Appendix
\ref{noise.sec} the residuals are consistent with the
expected noise in almost all cases.

\subsubsection{Calibration}
The redshifts and velocity dispersions follow directly from the MCMC fit,
but the line fluxes, equivalent widths, and line ratios need to be calibrated or corrected.
The continuum
is detected for every galaxy, which makes it possible to calculate
equivalent widths directly from the spectra. The equivalent
widths, in turn, enable us to calibrate the
line fluxes using the known $K$-band magnitudes of the galaxies.
The equivalent width
of H$\alpha$ in the observed frame is given by
\begin{equation}
{\rm EW}_{{\rm H}\alpha} = \Delta\lambda \frac{f_{{\rm H}\alpha}}{C}
+{\rm EW}_{{\rm H}\alpha,\,{\rm abs}}(1+z).
\end{equation}
The second term is a correction for the underlying stellar continuum
absorption, which has a non-negligible effect on the measured
equivalent widths and line ratios in our sample.
We adopt ${\rm EW}_{{\rm H}\alpha,\,{\rm abs}} =3$\,\AA\
({Moustakas} \& {Kennicutt} 2006; {Alonso-Herrero} {et~al.} 2010). 
The relation between rest-frame equivalent width
and the observed equivalent width is
${\rm EW}_{{\rm H}\alpha}^0 = {\rm EW}_{{\rm H}\alpha}/(1+z).$
The mean rest-frame equivalent width in our sample is
$\langle {\rm EW}_{{\rm H}\alpha}^0\rangle = 71$\,\AA,
consistent with the general population of (detected)
massive star forming galaxies at these redshifts
({Fumagalli} {et~al.} 2012).
The \niiha\ ratio, corrected for absorption, is
\begin{equation}
\frac{[{\rm N}\,{\textsc{ii}}]}{{\rm H}\alpha} = 
\frac{f_{[{\rm N}{\textsc{ii}}]}}{f_{{\rm H}\alpha}} \times
\frac{{\rm EW}_{{\rm H}\alpha}^0 - {\rm EW}_{{\rm H}\alpha,\,{\rm abs}}}
{{\rm EW}_{{\rm H}\alpha}^0},
\end{equation}
with $f$ taken from the MCMC fit. Note that we use positive values for
both absorption equivalent widths and emission equivalent widths
in these expressions, as
``absorption'' here is more accurately described
as ``emission that is filling in the underlying
absorption line''.


The line flux is calculated from the observed
equivalent width and the $K$ magnitude using
\begin{equation}
F_{{\rm H}\alpha} = 1.02\times 10^{-15}
\times \frac{{\rm EW}_{{\rm H}\alpha}}{2730} \times
10^{(K_s-22)/-2.5},
\end{equation}
with $K_s$ the AB magnitude of the object and
$F$ in units of ergs\,s$^{-1}$\,cm$^{-2}$. This expression ignores
small differences between the filters used in each field as well as
the detailed shape of the
continuum within the $K_s$ filter. We verified that the
transmission at the observed wavelenghts of the
lines is within $\sim 5$\,\% of the central transmission
of the filter in all cases.
Finally, the line luminosity is calculated using
\begin{equation}
L_{{\rm H}\alpha} =1.20\times10^{50}\times D^2 F_{{\rm H}\alpha},
\end{equation}
with $D$ the luminosity distance in Mpc and $L$ in ergs\,s$^{-1}$.
The results for all galaxies are listed in Table 2. The
error bars reflect the (propagated) MCMC errors;
no additional calibration
uncertainty was included in the error budget.

\subsection{Comparison to Barro et al.}

There are seven galaxies in the {Barro} {et~al.} (2014b) sample that satisfy our
more restrictive selection criteria.
Two of these seven galaxies, COSMOS\_12020 and GOODS-S\_37745,
are also in our sample: COSMOS\_12020 was observed with NIRSPEC and
GOODS-S\_37745 with MOSFIRE. For COSMOS\_12020 we find $\sigma =
719^{+30}_{-14}$\,\kms\ 
and \niiha\,$= 1.39 \pm 0.23$, whereas Barro et al.\ have $\sigma
= 352 \pm 213$\,\kms\ and \niiha\,$= 0.25\pm 0.25$. The kinematics
of this galaxy are
very complex, and a Gaussian is a poor fit (see Fig.\ \ref{spectra.fig},
and Sect.\ \ref{complex.sec}); this probably explains the differences
between the two measurements and the large uncertainty
in the Barro et al.\ values. As noted in Sect.\ 3.4.1 the
formal uncertainty in our measurement of this galaxy is smaller than
the true uncertainty, as it does not take deviations from a
Gaussian into account. Given that a Gaussian is clearly a poor
fit, the velocity dispersion of this galaxy is not well determined.
For GOODS-S\_37745 we find $\sigma = 163^{+27}_{-24}$\,\kms\ and
\niiha\,$=0.65 \pm 0.23$, compared to $\sigma = 197\pm 37$\,\kms\
and \niiha\,$=0.77 \pm 0.30$ in {Barro} {et~al.} (2014b).  These values are in
agreement within the (relatively large) $1\sigma$ uncertainties.

For the two galaxies that overlap we use our own measurements.
The other five  galaxies from Barro et al.\ are added to our sample
(see Tables 1 and 2). We
do not have measurements of the line flux or spatial extent of the
emission line gas for these objects, but they are included in the
analysis whenever only the redshift,
velocity dispersion, or line ratio is needed. They are shown in Fig.\
\ref{spectra.fig} by their best-fitting models. The total number of
sCMGs at $2.0<z<2.5$ that are studied in this paper is 25.

\begin{deluxetable*}{lccccccccccccc}
\tabletypesize{\footnotesize}
\tablewidth{0pt}
\tablecaption{Properties of Star Forming Compact Massive
Galaxies\tablenotemark{a}}
\tablehead{
\colhead{id\tablenotemark{b}} &
\colhead{$z$} &
\colhead{$M_{\rm stars}$} &
\colhead{$r_{\rm e}$} &
\colhead{$n$} &
\colhead{$q$} &
\colhead{SFR\tablenotemark{c}} &
\colhead{$\log\frac{L_{\rm IR}}{L_{\rm UV}}$} &
\colhead{X-ray} &
\colhead{instr} &
\colhead{$F_{{\rm H}\alpha}$} &
\colhead{EW$^0_{{\rm H}\alpha}$} &
\colhead{$\sigma$} &
\colhead{[N{\sc ii}]/H$\alpha$} \\
\colhead{} &
\colhead{} &
\colhead{$10^{11}$\,\msun} & 
\colhead{kpc} & 
\colhead{} &
\colhead{} &
\colhead{\msun\,yr$^{-1}$} &
\colhead{} &
\colhead{} &
\colhead{} &
\colhead{$10^{-17}$\,ergs\,s$^{-1}$\,cm$^{-2}$} &
\colhead{\AA}  &
\colhead{km\,s$^{-1}$} &
\colhead{}
}
\startdata
AEGIS\_9163 & 2.445 & 0.8 & 0.9 & 5.4 & 0.72 & 131 & 1.81 &  & NIRS & $\phn{}6.6_{-1.3}^{+1.3}$ & $\phn{}74_{-12}^{+12}$ & $\phn{}92_{-18}^{+18}$  & $0.21_{-0.07}^{+0.07}$ \\ 
AEGIS\_26952 & 2.097 & 1.1 & 1.8 & 3.6 & 0.64 & 148 & 1.62 & yes & NIRS & $18.4_{-3.2}^{+3.1}$ & $\phn{}95_{-13}^{+13}$ & $446_{-54}^{+89}$  & $0.90_{-0.14}^{+0.14}$ \\ 
AEGIS\_41114 & 2.332 & 0.5 & 0.2 & 8.0 & 0.62 & 95 & 1.38 &  & NIRS & $\phn{}3.2_{-0.7}^{+0.7}$ & $\phn{}30_{-6}^{+6}$ & $176_{-37}^{+46}$  & $0.85_{-0.24}^{+0.24}$ \\ 
COSMOS\_163 & 2.312 & 0.8 & 1.1 & 2.5 & 0.60 & 336 & 2.25 & yes & MOSF & $\phn{}7.4_{-2.1}^{+2.3}$ & $\phn{}95_{-26}^{+27}$ & $249_{-34}^{+43}$  & $0.19_{-0.12}^{+0.12}$ \\ 
COSMOS\_1014 & 2.100 & 0.5 & 0.7 & 8.0 & 0.79 & 150 & 0.93 &  & NIRS & $18.5_{-2.5}^{+2.5}$ & $\phn{}70_{-6}^{+6}$ & $172_{-13}^{+13}$  & $0.77_{-0.08}^{+0.08}$ \\ 
COSMOS\_11363 & 2.096 & 1.1 & 2.1 & 5.2 & 0.76 & 169 & 1.31 & yes & NIRS & $22.3_{-2.5}^{+2.4}$ & $\phn{}65_{-3}^{+3}$ & $368_{-20}^{+32}$  & $0.78_{-0.04}^{+0.04}$ \\ 
COSMOS\_12020 & 2.094 & 2.0 & 2.1 & 5.7 & 0.57 & 185 & 1.96 & yes & NIRS & $\phn{}8.3_{-1.4}^{+1.4}$ & $\phn{}34_{-5}^{+5}$ & $719_{-32}^{+14}$  & $1.26_{-0.21}^{+0.21}$ \\ 
COSMOS\_22995 & 2.469 & 1.2 & 1.1 & 2.8 & 0.67 & 188 & 1.41 & yes & NIRS & $\phn{}5.1_{-0.7}^{+0.7}$ & $\phn{}23_{-2}^{+2}$ & $176_{-18}^{+19}$  & $0.61_{-0.09}^{+0.09}$ \\ 
COSMOS\_27289 & 2.234 & 1.3 & 2.3 & 3.3 & 0.81 & 398 & 2.02 &  & NIRS & $25.5_{-3.9}^{+4.0}$ & $106_{-12}^{+13}$ & $\phn{}54_{-13}^{+11}$  & $0.36_{-0.04}^{+0.04}$ \\ 
GOODS-N\_774 & 2.301 & 1.0 & 1.0 & 2.9 & 0.59 & 150 & 2.07 &  & NIRS & $\phn{}5.7_{-0.8}^{+0.8}$ & $\phn{}45_{-4}^{+4}$ & $352_{-30}^{+36}$  & $0.34_{-0.07}^{+0.07}$ \\ 
GOODS-N\_6215 & 2.321 & 1.8 & 1.8 & 2.6 & 0.72 & 110 & 1.28 & yes & MOSF\tablenotemark{d} & \nodata & \nodata & $406_{-69}^{+69}$  & $2.17_{-0.28}^{+0.28}$ \\ 
GOODS-N\_13616 & 2.487 & 1.1 & 1.9 & 5.6 & 0.97 & 130 & 1.79 &  & MOSF\tablenotemark{d} & \nodata & \nodata & $243_{-30}^{+30}$  & $0.73_{-0.18}^{+0.18}$ \\ 
GOODS-N\_14283 & 2.420 & 0.9 & 1.2 & 2.7 & 0.86 & 111 & 1.43 & yes & MOSF\tablenotemark{d} & \nodata & \nodata & $156_{-27}^{+27}$  & $0.23_{-0.39}^{+0.39}$ \\ 
GOODS-N\_22548 & 2.330 & 1.0 & 1.7 & 5.9 & 0.78 & 120 & 1.53 & yes & MOSF\tablenotemark{d} & \nodata & \nodata & $223_{-56}^{+56}$  & $0.92_{-0.30}^{+0.30}$ \\ 
GOODS-S\_5981 & 2.253 & 0.8 & 0.8 & 4.4 & 0.85 & 206 & 1.75 &  & MOSF & $\phn{}8.0_{-1.9}^{+1.9}$ & $\phn{}54_{-11}^{+11}$ & $110_{-15}^{+18}$  & $0.49_{-0.12}^{+0.12}$ \\ 
GOODS-S\_30274 & 2.226 & 1.4 & 2.5 & 8.0 & 0.46 & 404 & 1.47 & yes & MOSF & $32.6_{-4.6}^{+4.6}$ & $\phn{}81_{-8}^{+8}$ & $296_{-17}^{+19}$  & $0.90_{-0.08}^{+0.08}$ \\ 
GOODS-S\_37745 & 2.432 & 0.9 & 0.6 & 3.6 & 0.94 & 118 & 1.04 &  & MOSF & $11.7_{-3.0}^{+3.2}$ & $\phn{}59_{-14}^{+15}$ & $165_{-21}^{+27}$  & $0.60_{-0.16}^{+0.16}$ \\ 
GOODS-S\_45068 & 2.453 & 1.1 & 1.3 & 4.9 & 0.97 & 139 & 1.57 &  & MOSF\tablenotemark{d} & \nodata & \nodata & $260_{-18}^{+18}$  & $1.70_{-0.78}^{+0.78}$ \\ 
GOODS-S\_45188 & 2.407 & 0.7 & 1.4 & 4.3 & 0.90 & 134 & 1.66 & yes & NIRS & $\phn{}8.0_{-1.6}^{+1.5}$ & $\phn{}72_{-12}^{+12}$ & $\phn{}49_{-18}^{+17}$  & $0.46_{-0.08}^{+0.08}$ \\ 
UDS\_16442 & 2.218 & 1.7 & 3.3 & 1.6 & 0.52 & 176 & 2.36 &  & MOSF & $15.9_{-2.1}^{+2.2}$ & $145_{-12}^{+14}$ & $210_{-6}^{+6}$  & $0.43_{-0.04}^{+0.04}$ \\ 
UDS\_25893 & 2.304 & 0.6 & 0.2 & 8.0 & 0.92 & 73 & 1.88 & yes & MOSF & $\phn{}4.9_{-4.1}^{+4.3}$ & $\phn{}54_{-44}^{+46}$ & $213_{-23}^{+20}$  & $0.58_{-0.28}^{+0.28}$ \\ 
UDS\_26012 & 2.321 & 1.3 & 2.6 & 3.5 & 0.73 & 109 & 1.47 &  & MOSF & $11.3_{-1.4}^{+1.4}$ & $\phn{}65_{-5}^{+5}$ & $209_{-8}^{+7}$  & $0.54_{-0.05}^{+0.05}$ \\ 
UDS\_33334 & 2.290 & 0.7 & 1.4 & 2.4 & 0.56 & 13 & 1.01 &  & MOSF & $\phn{}6.6_{-3.7}^{+4.8}$ & $\phn{}74_{-41}^{+54}$ & $161_{-16}^{+9}$  & $0.51_{-0.31}^{+0.31}$ \\ 
UDS\_35673 & 2.182 & 0.9 & 0.7 & 6.4 & 0.75 & 492 & 2.18 &  & MOSF & $24.7_{-2.6}^{+2.7}$ & $136_{-5}^{+5}$ & $267_{-3}^{+4}$  & $0.84_{-0.02}^{+0.02}$ \\ 
UDS\_42571 & 2.292 & 1.6 & 2.3 & 1.9 & 0.82 & 388 & 2.39 & yes & NIRS & $\phn{}7.1_{-1.1}^{+1.1}$ & $\phn{}46_{-6}^{+6}$ & $198_{-20}^{+22}$  & $0.60_{-0.10}^{+0.10}$ 
\enddata
\tablenotetext{a}{Uncertainties do not include possible effects of
non-Gaussian velocity distributions.}
\tablenotetext{b}{Id number in Skelton et al.\ (2014).}
\tablenotetext{c}{Star formation rate from UV+IR emission.}
\tablenotetext{d}{Velocity dispersion and \niiha\ from {Barro} {et~al.} (2014b).}
\end{deluxetable*}

\section{Interpretation of the Line Ratios and Luminosities}

\subsection{Line Ratios}

\begin{figure}[hbtp]
\epsfxsize=8.5cm
\epsffile{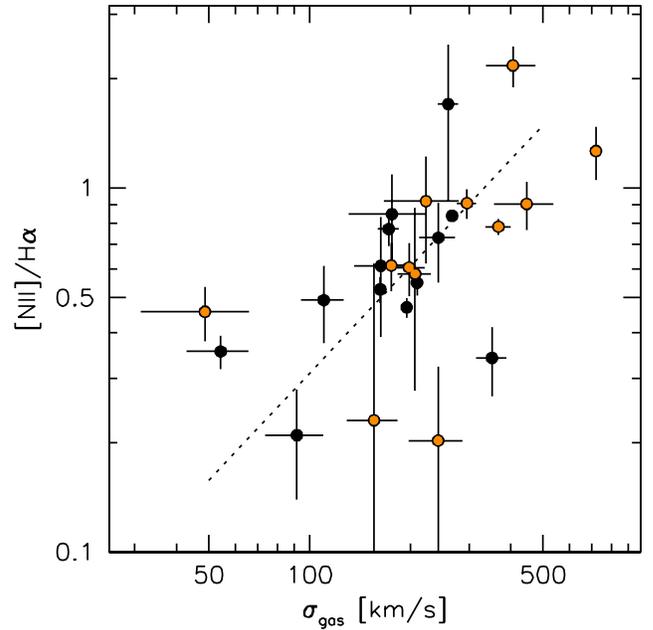}
\caption{\small
Relation between \niiha\ ratio and H$\alpha$ velocity dispersion
for the 25 \sg{}s.
There is a significant
correlation, such that galaxies with higher velocity dispersions
have higher \niiha\ ratios.
Orange symbols
are galaxies with X-ray-identified AGN. The four galaxies with
the highest observed dispersions are all X-ray AGN, as are
five of the six galaxies with the highest \niiha\ ratios. The black point
with \niiha$=0.3$ and $\sigma=352$\,\kms\  is GOODS-N\_774, which was
previously
published in {Nelson} {et~al.} (2014).
\label{lrsig.fig}}
\end{figure}

Considering that the 25 \sg{}s of Fig.\ \ref{spectra.fig}
were selected in a very restricted region of parameter space,
their emission lines show a surprisingly large range of properties.
The velocity dispersions range from 50\,\kms\
to $>500$\,\kms, the \niiha\ ratios from 0.2 to $>2$, and
the H$\alpha$ line luminosities from $1.3\times 10^{42}$\,L$_{\odot}$
to $1.2\times 10^{43}$\,L$_{\odot}$. Two of these parameters,
the \niiha\ ratio and the velocity dispersion, show
a significant correlation: as shown in Fig.\ \ref{lrsig.fig},
galaxies with the highest velocity
dispersions tend to have the highest line ratios.
The correlation has
a formal significance of $>99$\,\%.  The broken line is the best fit
relation, which has the form
\begin{equation}
\log \frac{[{\rm N}{\textsc{ii}}]}{{\rm H}\alpha}  = (-0.51 \pm 0.08) + (1.0
\pm 0.2) \log \left( \frac{\sigma_{\rm gas}}{100}\right).
\end{equation}
The canonical high-metallicity saturation
value for \niiha\ in low redshift star forming galaxies is $\sim 0.4$
(e.g., {Baldwin}, {Phillips}, \&  {Terlevich} 1981; {Denicol{\' o}}, {Terlevich}, \&  {Terlevich} 2002; {Pettini} \& {Pagel} 2004; {Kewley} {et~al.} 2013).
Although this limit is observed to be higher at $z\gtrsim 2$
(e.g., {Brinchmann}, {Pettini}, \&  {Charlot} 2008; {Steidel} {et~al.} 2014; {Shapley} {et~al.} 2015),
values of \niiha$\,> 1$ are extreme at any redshift
(see, e.g., {Leja} {et~al.} 2013b; {Shapley} {et~al.} 2015).
A likely explanation for the highest $\sigma$, highest
\niiha\ galaxies
in Fig.\ \ref{lrsig.fig} is that
shocks ({Dopita} \& {Sutherland} 1995) and/or emission from AGNs
({Kewley} {et~al.} 2013) are responsible
for the line ratios.

This is supported by the X-ray luminosities of the objects, obtained
from all public catalogs in the CANDELS fields.\footnote{The
catalogs 
were searched using the tools of the NASA High Energy Astrophysics
Science Archive Research Center (http://heasarc.gsfc.nasa.gov/).
We note, however, that
the X-ray coverage in the CANDELS fields is not uniform.}
Twelve of the
25 \sg{}s (48\,\%) have $L_X>10^{43}$\,ergs\,s$^{-1}$ and are classified
as AGN. The X-ray luminosities
range from $L_X = 1.4\times 10^{43}$\,ergs\,s$^{-1}$ for
GOODS-S\_30274 to $L_X=
6\times 10^{44}$\,ergs\,s$^{-1}$ for COSMOS-11363. This high AGN fraction
is consistent with previous studies of massive star forming
galaxies at these redshifts
(e.g., {Papovich} {et~al.} 2006; {Daddi} {et~al.} 2007; {Kriek} {et~al.} 2007; {Barro} {et~al.} 2013; {F{\"o}rster Schreiber} {et~al.} 2014).
The four galaxies with the highest velocity dispersions are all classified
as X-ray AGN.\footnote{The correlation between
\niiha\ and $\sigma$ is no longer significant when these
four objects are removed.}
Their kinematics are complex (see Fig.\ \ref{spectra.fig}),
and their \niiha\ ratios range from 0.8 to 2.2.
It is likely that
the observed emission line properties of these galaxies are
affected by the presence of the AGN, either directly through
emission
from the broad line region or indirectly through AGN-driven
winds (see {F{\"o}rster Schreiber} {et~al.} 2014; {Genzel} {et~al.} 2014a).

However, it is not clear whether AGNs or winds {\em dominate}
the observed, galaxy-integrated kinematics, even for
these four objects -- 
and whether the presence of a central point source influenced their
selection as apparently compact, apparently massive galaxies.
As shown in Fig.\ \ref{seds.fig} the UV -- near-IR SEDs of
all galaxies are well fit
by stars-only models.
Most galaxies have strong Balmer breaks
(including the most powerful X-ray source in the sample,
COSMOS-11363), and as discussed
in {Kriek} {et~al.} (2007) and later studies
(e.g., {Marsan} {et~al.} 2015) this
strongly constrains the contribution of continuum emission
from an AGN
at $\lambda_{\rm rest}\sim 4000$\,\AA.
As we show below and in the following section, the properties
of most of the galaxies can be understood in a model where AGN
are present but do
not dominate the kinematics, line ratios, line luminosities,
or morphology.

\subsection{Star Formation Rates}
\label{sfr.sec}

The H$\alpha$ luminosities can be converted to star formation rates
if it is assumed
that the H$\alpha$ emission largely originates in H{\sc ii} regions.
By comparing these star formation rates to those derived from the
UV and the bolometric UV+IR luminosities we can assess whether this
assumption is reasonable, and also constrain the amount of
obscuration in the galaxies. The H$\alpha$ star formation rates
were determined
using the {Kennicutt} (1998) relation, converted to a {Chabrier} (2003)
IMF.\footnote{For consistency with previous studies we use a
{Chabrier} (2003) IMF as the default, even though these galaxies
may have a more bottom-heavy IMF (see, e.g., Conroy \& van
Dokkum 2012).} 
The UV luminosities come from the best-fitting {Brammer} {et~al.} (2008)
models at $\lambda_{\rm rest}=2500$\,\AA,
and the IR luminosities are converted Spitzer/MIPS 24\,$\mu$m
fluxes (see {Whitaker} {et~al.} 2012 and Sect.\ 2.1).

\begin{figure}[hbtp]
\epsfxsize=8.5cm
\epsffile{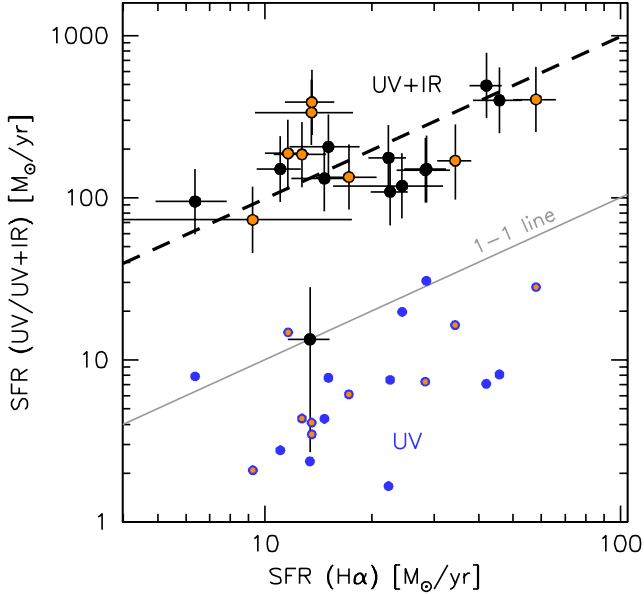}
\caption{\small
Relation between the star formation rate derived
from H$\alpha$ and the star formation rate derived from the
UV (blue points) and UV+IR (black points with errorbars). X-ray AGN
are indicated with orange centers. The H$\alpha$ star formation
rates fall in between the UV and UV+IR ones, as expected from
the effects of dust extinction. The obscuration toward H$\alpha$
is a factor of 10, with a scatter of only a factor of 2.
The X-ray sources are indistinguishable from the
other galaxies.
\label{sfrsfr.fig}}
\end{figure}

The relation between the UV/UV+IR star formation rates and the H$\alpha$
star formation rate is shown in Fig.\ \ref{sfrsfr.fig}. Only the 20
galaxies from our own spectroscopy are considered here, as we do not have
self-consistent measurements of $L$(H$\alpha$) for the five objects
from {Barro} {et~al.} (2014b).
The H$\alpha$
star formation rates range from 6\,\msun/yr -- 58\,\msun/yr. They
correlate with the UV star formation rates (98\,\% significance)
and with the
UV+IR star formation rates, which are dominated by the IR
(96\,\% significance). The mean offset between SFR(H$\alpha$) and
SFR(UV) is $0.47\pm 0.06$ dex, with an rms scatter of 0.22 dex. The offset
between SFR(H$\alpha$) and SFR(UV+IR) is $-1.00\pm 0.09$
dex, with a scatter of 0.27 dex. The implication is that the H$\alpha$
emission misses $\sim 90$\,\% of the star formation, and the UV
misses $\sim 97$\,\%. The ratios between
the three indicators are broadly consistent with expectations from a
{Calzetti} {et~al.} (2000) reddening curve, if there is $\sim 50$\,\%
more dust toward nebular emission line gas than toward the UV
continuum.\footnote{We refer to other studies for more detailed
analysis of the attenuation toward H\,{\sc ii} regions
(e.g., Price et al.\ 2014, Reddy et al.\ 2015).}

The X-ray AGNs are indicated by orange points in Fig.\ \ref{sfrsfr.fig}.
Remarkably, they are indistinguishable from the other objects: they
span the same range in H$\alpha$ luminosity, and they follow the same
relations with the UV and UV+IR luminosities. The offsets between the
AGN and non-AGN are consistent with zero. This suggests, but does
not prove, that the H$\alpha$, UV, and IR luminosities of most 
galaxies are dominated by star formation. 

\begin{figure*}[hbtp]
\begin{center}
\epsfxsize=17cm
\epsffile{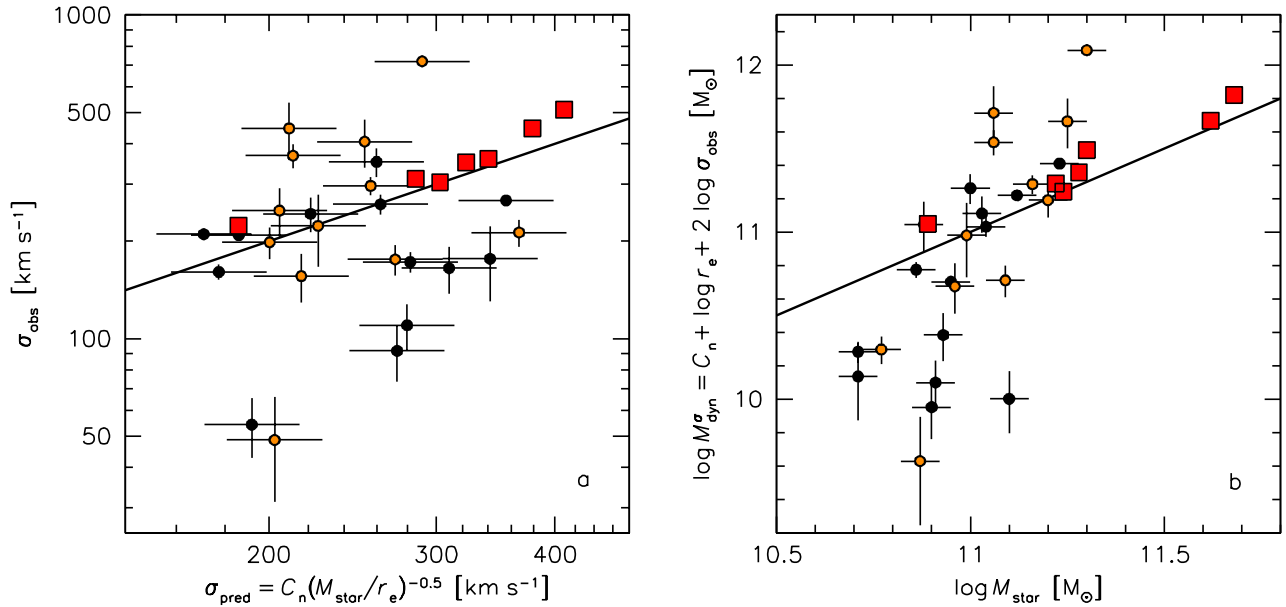}
\end{center}
\vspace{-0.3cm}
\caption{\small
a) Comparison of observed and predicted velocity dispersions. The
predicted dispersions are calculated from the stellar mass, the half-light
radius, and the Sersic index. Red squares are quiescent galaxies at $2<z<2.5$
from {van Dokkum} {et~al.} (2009), {van de Sande} {et~al.} (2013), and {Belli} {et~al.} (2014b). Points
with errorbars are the 25 \sg{}s; orange centers indicate galaxies with
X-ray AGN.  b) Comparison between dynamical mass and stellar mass.
The galaxies show a very large range, and 
the dynamical masses often appear to be lower than the
stellar masses. The gas in \sg{}s does not have the same
distribution and/or kinematics as the stars in \qg{}s.
\label{sigsig.fig}}
\end{figure*}

\section{Interpretation of the Velocity Dispersions}
\label{mass.sec}

\subsection{Are the Gas Dynamics Similar to the Stellar Dynamics of
Compact Quiescent Galaxies?}
\label{gasstars.sec}

The velocity dispersions we measure come from Gaussian fits to the
galaxy-integrated, luminosity-weighted H$\alpha$ line profile
and are equivalent to the second moment
of the velocity distribution of the gas. They should not be
confused with the rotation-corrected  gas
dispersions within spatially-resolved disks,
such as discussed by, e.g., {Kassin} {et~al.} (2012) and
{F{\"o}rster Schreiber} {et~al.} (2014). The measured dispersions
are a complex function of the dynamics and gas distribution in the
galaxies:
\begin{equation}
\label{siggen.eq}
\sigma_{\rm gas}^2 \sim \alpha^2 V_{\rm rot}^2 \sin^2(i) + \sigma_{\rm ISM}^2
+w^2(i)\sigma_{\rm wind}^2,
\end{equation}
with $\alpha \sim 0.8$ ({Franx} 1993; {Rix} {et~al.} 1997; {Weiner} {et~al.} 2006; see also Appendix C),
$i$ the inclination of the galaxy ($i=0\arcdeg$ is face-on, and
$i=90\arcdeg$ is edge-on), $\sigma_{\rm ISM}$ the galaxy-integrated dispersion within
the gas clouds, and $w(i)
\sigma_{\rm wind}$ an inclination-dependent term that takes
non-gravitational motions into account.
A further complication is that Eq.\ \ref{siggen.eq} is
the result of an integral over the area of the galaxy that falls
within the slit, 
weighted by the spatially-varying luminosity of the H$\alpha$ line.

We first assume that the gas in the \sg{}s ``behaves'' in a similar way
as the stars in \qg{}s. That is, we assume that the stars in \qg{}s
were formed directly out of the (detected) gas in \sg{}s, such that they
have the same distribution and kinematics. This has been assumed
in previous studies of the kinematics of compact massive 
star forming galaxies (Nelson et al.\ 2014; Barro et al.\ 2014b)
and it may be reasonable
if many compact, massive
quiescent galaxies are direct descendants of the \sg{}s.
As
discussed in Sect.\ \ref{predsig.sec} the stellar velocity dispersions
of quiescent galaxies
can be predicted from their stellar masses and effective
radii (e.g., {Taylor} {et~al.} 2010a; {Bezanson} {et~al.} 2011; {Belli} {et~al.} 2014b).
Figure \ref{sigsig.fig}a
shows the relation between $\sigma_{\rm gas}$
and the predicted velocity dispersion. The predicted dispersions
were calculated using the Sersic-dependent relation Eq.\ \ref{sigpredn.eq}.

There is no significant correlation between $\sigma_{\rm gas}$
and $\sigma_{\rm pred}$, for either the full sample or the
sample with the X-ray AGN removed. The rms
scatter in $\sigma_{\rm gas}/\sigma_{\rm
pred}$ is 0.26 dex. Given that we are ignoring the effects of non-gravitational
motions, it is striking that many galaxies have {\em lower} velocity
dispersions than the expectations. The mean offset is $-0.08$ dex for
the full sample, and $-0.16$ dex when the AGN are excluded.
These results stand
in sharp contrast to the stellar velocity dispersions of quiescent
galaxies. Red squares are seven galaxies with $2<z<2.5$ and measured
$\sigma_{\rm stars}$, $r_e$, $n$, and $M_{\rm stars}$ from
{van Dokkum} {et~al.} (2009), {van de Sande} {et~al.} (2013), and {Belli} {et~al.} (2014b). They have
a mean offset in $\sigma_{\rm stars}/\sigma_{\rm pred}$
of $+0.05$ dex and an rms scatter of only $0.03$ dex.

As dynamical mass is proportional to $\sigma^2$ the offsets of the \sg{}s
are even more dramatic in Fig.\ \ref{sigsig.fig}b, which shows the relation
between dynamical mass and stellar mass. Here dynamical mass was calculated
using
\begin{equation}
M^{\sigma}_{\rm dyn} = \frac{\beta(n) \sigma_{\rm obs}^2 r_e}{G},
\end{equation}
as derived by {Cappellari} {et~al.} (2006) and following
studies of quiescent galaxies
at high redshift (e.g., {van de Sande} {et~al.} 2013). For \sg{}s $\sigma_{\rm obs} =
\sigma_{\rm gas}$ and for quiescent galaxies $\sigma_{\rm obs} =
\sigma_{\rm stars}$. Note that,
given our definition of $\sigma_{\rm pred}$
(see Eq.\ \ref{sigpredn.eq}), panels a and b of Fig.\ \ref{sigsig.fig}
are two different ways of presenting the same information.
The mean mass offset of the \sg{}s is $-0.16$ dex for the full
sample, and $-0.32$ dex for the sample with the AGN removed.
That is, the dynamical masses of the non-AGN galaxies are on average
a factor of two lower than the stellar masses. Several of the galaxies
have apparent dynamical masses that are a factor of $\gtrsim 10$ lower than
their stellar masses.
Again, the quiescent galaxies show a tight relation in Fig.\
\ref{sigsig.fig}b, with a mean offset of $+0.1$ dex. 

We conclude that the gas dynamics of \sg{}s are {\em not} similar to
the stellar dynamics of quiescent galaxies in the same
mass and redshift range. The stellar masses and
sizes are not useful indicators of the observed gas
velocity dispersions; in fact, the observed \niiha\ ratio is a better
predictor of the observed H$\alpha$ linewidth
of a galaxy than its compactness is. There are many ways to
{\em increase} the
velocity dispersion of a galaxy so it falls above the lines of equality
in the two panels of Fig.\ \ref{sigsig.fig}: the broad line region
of an AGN, AGN-induced winds, and supernova-driven winds can all lead
to broad H$\alpha$ lines
(e.g., {Westmoquette} {et~al.} 2009; {Le Tiran} {et~al.} 2011; {F{\"o}rster Schreiber} {et~al.} 2014; {Banerji} {et~al.} 2015).
This is likely the case for several galaxies in the sample: the four
galaxies with the largest dynamical masses are all X-ray AGN
with
\niiha\ ratios in the range $0.8 - 2.2$.
However, it is difficult to {\em decrease} the observed dispersion.
Setting aside the possibility that the stellar masses of some
galaxies could be in error by a factor of $\sim 10$, this
is only possible if the detected ionized gas
is \sg{}s is distributed very differently from the stars
in quiescent galaxies. As we show below,
there is strong evidence that this is indeed the case.

\subsection{Evidence for Rotating Gas Disks}
\label{rot.sec}

\begin{figure}[hbtp]
\epsfxsize=8.5cm
\epsffile{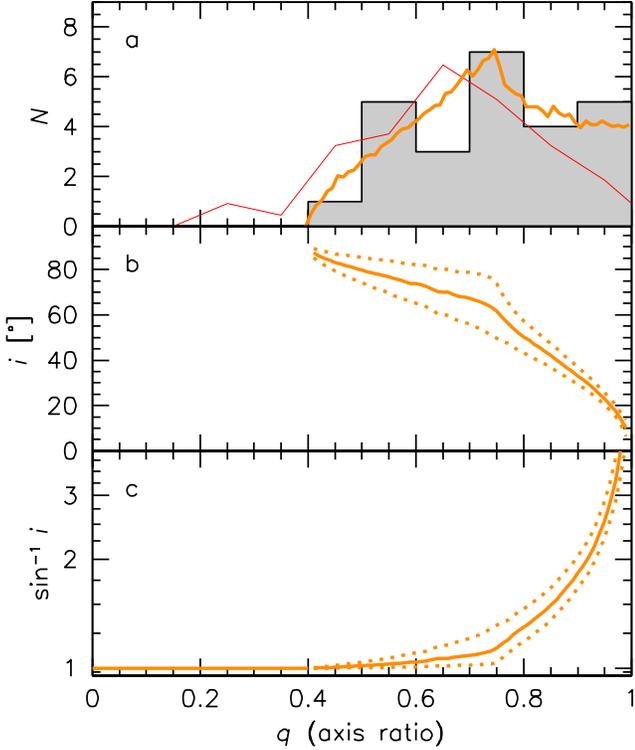}
\caption{\small
(a) Distribution of axis ratios among the 25 \sg{}s at $2<z<2.5$.
The distribution is not uniform, and is inconsistent
with thin disks
under random viewing angles. The axis ratio distribution
of \qg{}s in the same redshift range is shown in red.
The orange line is a model for
randomly oriented oblate objects  with intrinsic thickness $q_0
\equiv C/A=0.4-0.75$.
(b) The relation between median
inclination and observed axis ratio in this model.
Dotted lines indicate the $\pm 1\sigma$ spread. (c) Inclination
correction as a function of observed axis ratio.
\label{incl.fig}}
\end{figure}

A possible interpretation of the large range
of velocity dispersions is that the
dynamics are dominated by rotation, and we are  seeing
disks under a large
range of viewing angles. In Fig.\ \ref{incl.fig}a we show the distribution
of projected axis ratios $q = b/a$ in our sample, as determined from the
$H_{160}$ data (see {van der Wel} {et~al.} 2014b).  The axis ratios of the 25 galaxies
are inconsistent with a uniform distribution, which would be
expected for thin, randomly oriented
disks. We find no galaxies
with $q<0.4$ and the distribution peaks at $q\sim 0.75$. The
distribution is consistent with that observed for \qg{}s,
shown by the red line in
Fig.\ \ref{incl.fig}a: according to the
Kolgomorov-Smirnov test the probability that both samples were
drawn from the same distribution of axis ratios is 27\,\%.
The distributions are also consistent with results for the
general population of massive galaxies at $z\sim 2$
({Chang} {et~al.} 2013; {van der Wel} {et~al.} 2014a).
We note that we do not detect a significant wavelength
dependence of the mean axis ratio of the 25 sCMGs: we find
$\langle q \rangle = 0.76 \pm 0.03$ in $J_{125}$ and
$\langle q \rangle = 0.74 \pm 0.03$ in $H_{160}$.

\begin{figure}[hbtp]
\epsfxsize=8.5cm
\epsffile{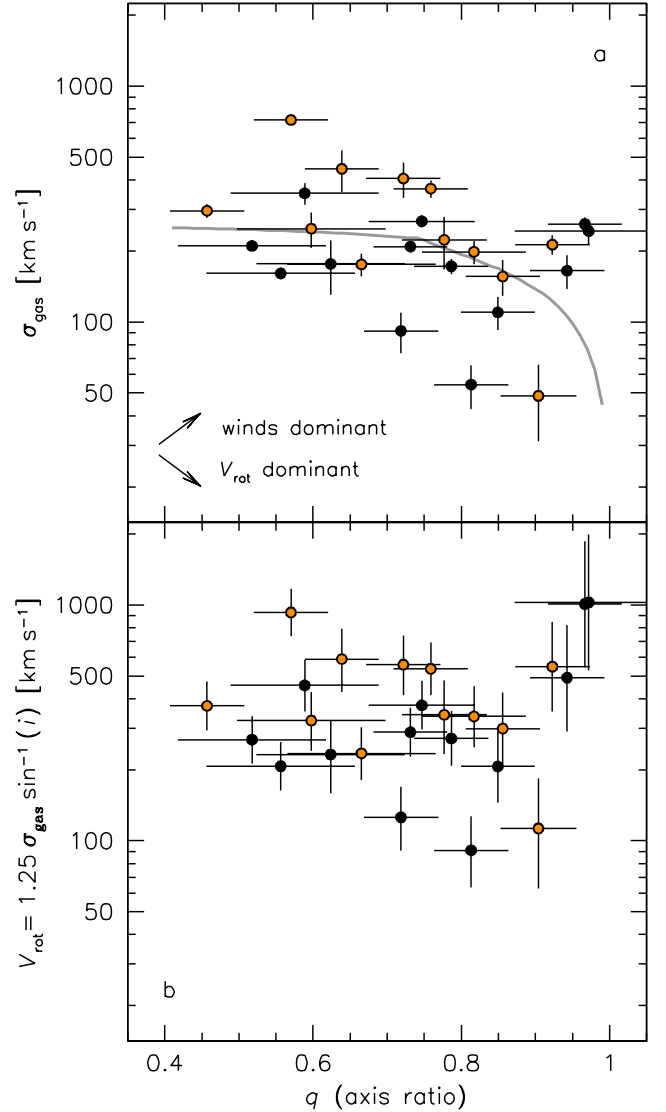}
\caption{\small
(a) Observed relation between the H$\alpha$ velocity dispersion and
the $H_{160}$ axis ratio. Orange centers indicate X-ray AGN.
There is a significant (anti-)correlation between $\sigma_{\rm gas}$
and $q$,
as expected if there is a significant contribution from rotation
to $\sigma_{\rm gas}$ and the H$\alpha$ disks are aligned with the
stellar distribution. The grey line indicates the
expected trend for rotating disks (Fig.\ \ref{incl.fig}c).
(b) Inferred rotation velocity versus axis ratio.
The rotation velocities were corrected for inclination using the
observed axis ratios (see text). The median 
rotation velocity is 338\,\kms\
for the full sample and 271\,\kms\ when AGN are excluded. 
\label{sigq.fig}}
\end{figure}

Even though the stars are not in thin disks, the gas can be.
If the gas is in rotationally-supported disks that are aligned with
the stellar distribution, the measured velocity dispersions are
expected to show an anti-correlation with the observed axis ratios
of the galaxies. As shown in Fig.\ \ref{sigq.fig}a we see precisely this
effect: 
there is an anti-correlation, with a correlation coefficient of $-0.38$
and a significance of 94\,\%. This is strong evidence that
the gas is in disks and that the measured dispersions
are dominated by gravitational motions.\footnote{The correlation
between $\sigma / \sqrt{M}$ and $q$ has slightly less scatter,
and equal significance.}
This anti-correlation is {\em not} consistent with 
M82-style galactic winds: outflows that are perpendicular to
the disk lead to the highest observed velocities (and
hence integrated velocity dispersions) when the disk
is viewed face-on.

Going back to Eq.\ \ref{siggen.eq}, we now assume that $\sigma_{\rm ISM}$
and $\sigma_{\rm wind}$ can be neglected, so that
\begin{equation}
\label{correcti.eq}
V_{\rm rot} = \frac{\sigma_{\rm gas}}{\alpha \sin^{-1} (i)}.
\end{equation}
To derive rotation velocities we need to determine the relation
between inclination and axis ratio in our sample.
We  constructed a model with long, intermediate,
and short axes $A$, $B$, and $C$ that reproduces the
observed axis ratio distribution for random viewing angles. The orange
line in Fig.\ \ref{incl.fig}a shows the predicted distribution of
$q$ for thick disks -- or oblate spheroids -- with $A/B=1$ and $q_0 \equiv
C/A$ uniformly distributed between $q_0 = 0.40$ and $q_0=0.75$.
This model is an excellent fit\footnote{It is well known that
the axis ratio distribution by itself is insufficient to constrain
all three axes $A$, $B$, and $C$ (see, e.g., Franx et al.\ 1991).
Although there is some evidence that the stellar distribution of compact
$z\sim 2$ galaxies is oblate or disk-like rather than triaxial
(e.g., van der Wel et al.\ 2014a;
Zolotov et al.\ 2015), in our paper this is an
assumption, not a result.}
to the observed distribution of
$q$.
It should be emphasized that this is a model for the intrinsic shapes
of the {\em stellar} distribution, not for the gas distribution: the gas
is likely in thinner disks,\footnote{Although the gas disks likely have
lower $C/A$ than the stellar distribution, they are probably
not as thin as disks in the local Universe (see, e.g.,
Cresci et al.\ 2009).} and all we assume is that the
gas disks of the galaxies are aligned with their
stellar distributions.

For galaxies with intrinsic thickness $q_0$
the relation between the inclination and the observed axis ratio is
given by
\begin{equation}
\cos^2(i) = \frac{q^2 - q_0^2}{1-q_0^2}.
\end{equation}
As $q_0$ is not a constant in our model the relation between $i$ and
$q$ is not single-valued.
The solid line in Fig.\ \ref{incl.fig}b shows the median relation, and
the broken lines indicate
the $1\sigma$ scatter.
Figure\ \ref{incl.fig}c shows the inclination
correction $\sin^{-1} (i)$ as a function of $q$.

The inclination-corrected rotation velocities are shown in Fig.\
\ref{sigq.fig}b. They are derived from the gas velocity dispersions
and the observed axis ratios of the galaxies
using the average
relation in \ref{incl.fig}c and assuming $\alpha = 0.8 \pm 0.2$
(see {Rix} {et~al.} 1997; {Weiner} {et~al.} 2006). In Appendix C
we show that this value is a reasonable approximation for the
geometries of both the mass and the ionized gas
that we derive in this paper.
The large uncertainty reflects
the fact that the conversion of dispersion to rotation velocity
depends on the spatial distribution of the gas, and the
underlying velocity field (see Appendix C).
Data of much higher spatial resolution
and S/N ratio
are needed to measure $\alpha$ directly for these extremely
compact galaxies.\footnote{For completeness, we note the interesting
possibility that
the two peaks in the spectra may not be H$\alpha$
and \nii\ but two narrow peaks in a ``double-horned'' \ha\ profile
that happen to have exactly the separation of H$\alpha$ and \nii\
$\lambda$6584. This may happen when the H$\alpha$ emission originates
from a narrow ring rather than a disk.
In most cases that interpretation can readily be ruled out,
from the spatially-resolved line profile (see Sect.\
\ref{morphline.sec}) or from the detection of the \nii\ $\lambda$6548
line, but in a few cases (e.g., AEGIS\_41114) it is difficult
to exclude
this possibility without observing other emission lines.}
The uncertainty in $\alpha$
and 50\,\% of the (logarithmic) inclination
correction were added in quadrature to the error budget. The median
rotation velocity for the full sample is $\langle V_{\rm rot}\rangle
= 339$\,\kms. Excluding the X-ray AGN we find $\langle V_{\rm rot}\rangle
= 271$\,\kms. 

\begin{figure}[hbtp]
\epsfxsize=8.5cm
\epsffile{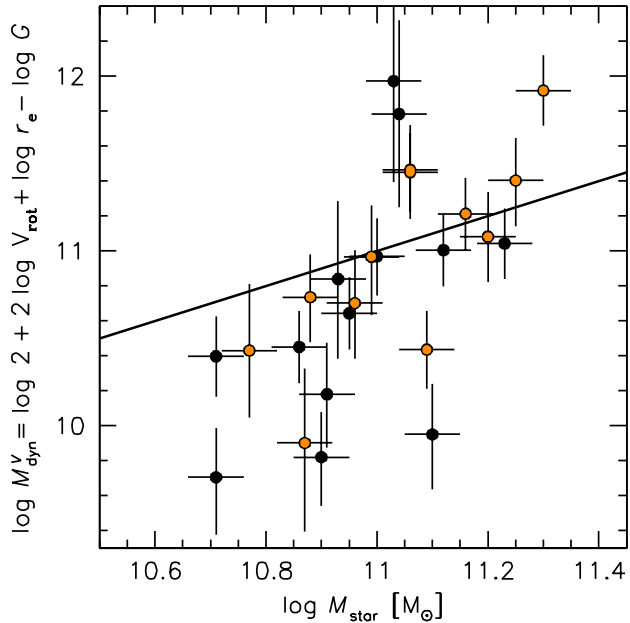}
\caption{\small
Relation between dynamical mass and stellar mass, with dynamical masses
calculated from the inclination-corrected rotation velocities and the
stellar half-light radii. Most galaxies fall below the line of equality.
\label{vdyn.fig}}
\end{figure}

If it is assumed that $r_e$ is not only the half-light radius in the
$H_{160}$ band but also the half-light radius of the H$\alpha$ emission,
we can define the dynamical mass as
\begin{equation}
\label{mvrot.eq}
M^{V}_{\rm dyn} \equiv 2 \frac{V_{\rm rot}^2 r_e}{G}.
\end{equation}
This is not a true total mass but simply twice the enclosed mass within
the half-light radius. In Fig.\ \ref{vdyn.fig} this dynamical mass
is compared to the stellar mass.
Although the inclination corrections have lessened the offsets
of the most extreme outliers, it is clear that orientation
effects are not sufficient to explain the relatively low
velocities that have been measured for a large fraction of the sample.
The mean offset for the whole sample is $-0.19$ dex, and the scatter is
$0.55$ dex. In the next Section we show that variation
in the spatial extent
of the ionized gas with respect to that of the stars
is a likely source of both the offset and scatter
in Fig.\ \ref{vdyn.fig}.

\section{Spatially-Extended Gas Disks}

\subsection{Inferred Sizes of Gas Disks}
\label{sigcomp.sec}

In the previous Section we showed that many galaxies have
galaxy-integrated velocity
dispersions that are much smaller than expected from their
stellar masses and sizes. As demonstrated in Sect.\ \ref{rot.sec}
this is partly caused
by the $\sin (i)$ reduction of the velocity of rotating disks.
However, even after correcting the observed
dispersions to inclination-corrected rotation velocities 
the dynamical masses are typically lower than the stellar masses,
particularly for galaxies that do not host an X-ray AGN.

So far we have assumed that the spatial extent of the gas is similar
to that of the stars, that is, $r_{\rm gas} \sim r_{\rm stars} \equiv
r_{\rm e}$, where $r_{\rm gas}$ is the half-light radius of the
measured H$\alpha$
distribution.\footnote{That is, the distribution of
the ionized gas, with no extinction corrections applied.
Measuring the  true ``$r_{\rm gas}$'' requires
molecular line measurements with high spatial resolution.}
There is no a priori reason why this should be the case;
e.g., in the models of {Zolotov} {et~al.} (2015) compact galaxies
often  have rings of gas and young stars around their dense
centers, which originate from ongoing
accretion from the halo. Furthermore, as shown earlier $\sim 90$\,\%
of the star formation in \sg{}s is obscured, and the extinction is expected
to be particularly high toward the central regions
(e.g., {Gilli} {et~al.} 2014; {Nelson} {et~al.} 2014).
The distribution of
detected H$\alpha$ emission may therefore be less centrally
concentrated than the distribution of star formation.

The radius of the gas disks can be inferred from $V_{\rm rot}$
if we assume that the observed velocity is the circular
velocity of the stellar body at the radius of the gas.
The gas radius then depends on $V_{\rm rot}$,
the stellar mass, and the structural parameters of the galaxy:
\begin{equation}
\label{rgas.eq}
r_{\rm gas} \sim \frac{G}{V_{\rm rot}^2} f(r_{\rm gas})M_{\rm stars},
\end{equation}
with $V_{\rm rot}$ the inclination-corrected rotation velocity and
$f(r_{\rm gas})$ a function that depends on the mass distribution
of the galaxies:
\begin{equation}
f(r_{\rm gas}) = \frac{\int_0^{r_{\rm gas}} I(r) 2 \pi r\,dr}
{\int_0^{\infty} I(r) 2 \pi r\,dr}.
\end{equation}
Here $I(r)$ is the best-fitting Sersic profile to the light distribution.
For $r_{\rm gas} = r_{\rm stars}$ ($= r_{\rm e}$),
$f(r_{\rm gas})=0.5$
and Eq.\ \ref{rgas.eq} is equivalent to
Eq.\ \ref{mvrot.eq} with $M_{\rm dyn} = M_{\rm stars}$.
These expressions ignore the fact that the 2D radii are not identical
to the 3D radii, assume that the stellar mass distribution can
be approximated by the $H_{160}$ luminosity distribution,
and assume that the
contributions of gas and dark matter to the total mass can be neglected
on the scales that are probed by the H$\alpha$ emission.

\begin{figure}[hbtp]
\epsfxsize=8.5cm
\epsffile{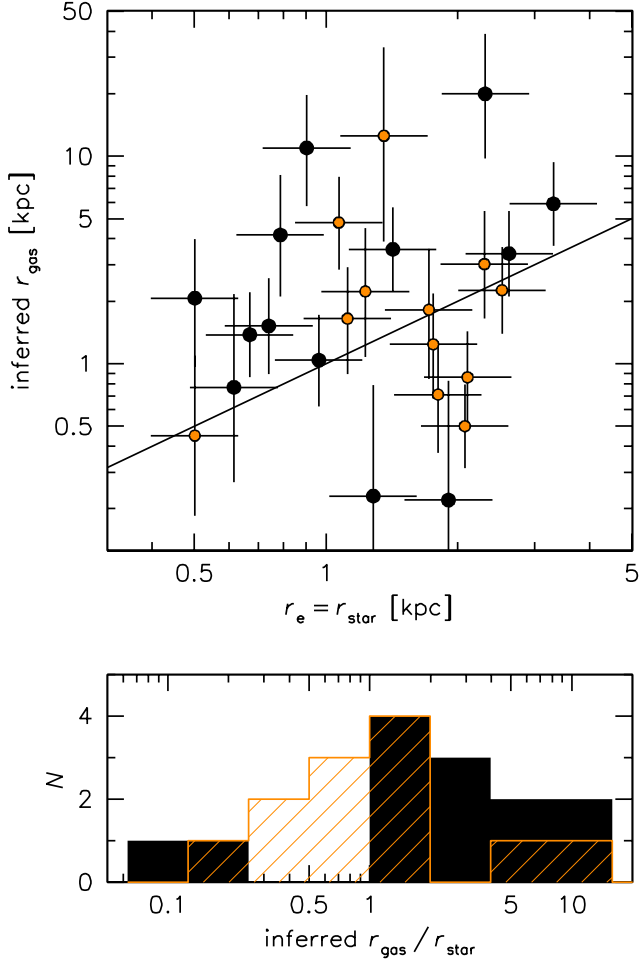}
\caption{\small
Relation between inferred radius of the gas distribution and the
stellar half-light radius. Orange points indicate galaxies with
X-ray AGN. The gas radii were determined from the
stellar masses and the inclination-corrected rotation velocities.
There is a large scatter, reflecting the large scatter in
Fig.\ 13.
The ratio between the gas size and the stellar size is shown in
the bottom panel. Non-AGN (black) and AGN (orange) are shown
separately. The galaxies with
AGN have, on average, compact inferred gas distributions.
For the non-AGN (black histogram) the average spatial
extent of the gas is $\sim 2.3 \times$ larger than that
of the stars.
\label{sizepred.fig}}
\end{figure}

Solving Eq.\ \ref{rgas.eq} numerically, we find that the inferred gas
disk sizes range from $\sim 0.2$\,kpc to
$\gtrsim 10$\,kpc.\footnote{We note that there are two
solutions to Eq.\ \ref{rgas.eq}, as the gas could
in principle also be located in the inner $\lesssim 50$\,pc
where the rotation curve is still rising (see, e.g., Fig. 18).
This is unlikely given that the galaxies have, by selection,
star forming SEDs with a spatial extent of $\sim 1$\,kpc.
Furthermore,
as we show later, the large radius solutions are corroborated by
the measured spatial extent of the H$\alpha$ emission.}
This large
range is not surprising, as it is effectively an interpretation of
the large variation that is seen in Fig.\  \ref{vdyn.fig}. 
Figure \ref{sizepred.fig} shows the relation between inferred 
$r_{\rm gas}$ and the stellar effective radius. The gas radii are typically
larger than the stellar radii, particularly for the galaxies that do not
have an X-ray AGN (black points). The ratio between the gas radius
and the stellar radius is shown explicitly in the bottom panel of
Fig.\ \ref{sizepred.fig}. The mean ratio, calculated with the biweight
statistic (Beers et al.\ 1990),
is $\log r_{\rm gas} - \log r_{\rm stars} = 0.18 \pm 0.10$
for the full sample.
Excluding galaxies with an AGN, we find
$\log r_{\rm gas} - \log r_{\rm stars} = 0.37 \pm 0.14$. That is, the
gas disks are a factor of $\sim 2.3$ more extended than the stellar
distribution.  This is strictly a lower limit, as it is assumed that
only stars contribute to the stellar mass, the galaxies have a 
relatively ``light''
{Chabrier} (2003) IMF, and there are no
contributions from non-gravitational motions to the measured velocity
dispersions.

\begin{figure*}[hbtp]
\begin{center}
\epsfxsize=17cm
\epsffile{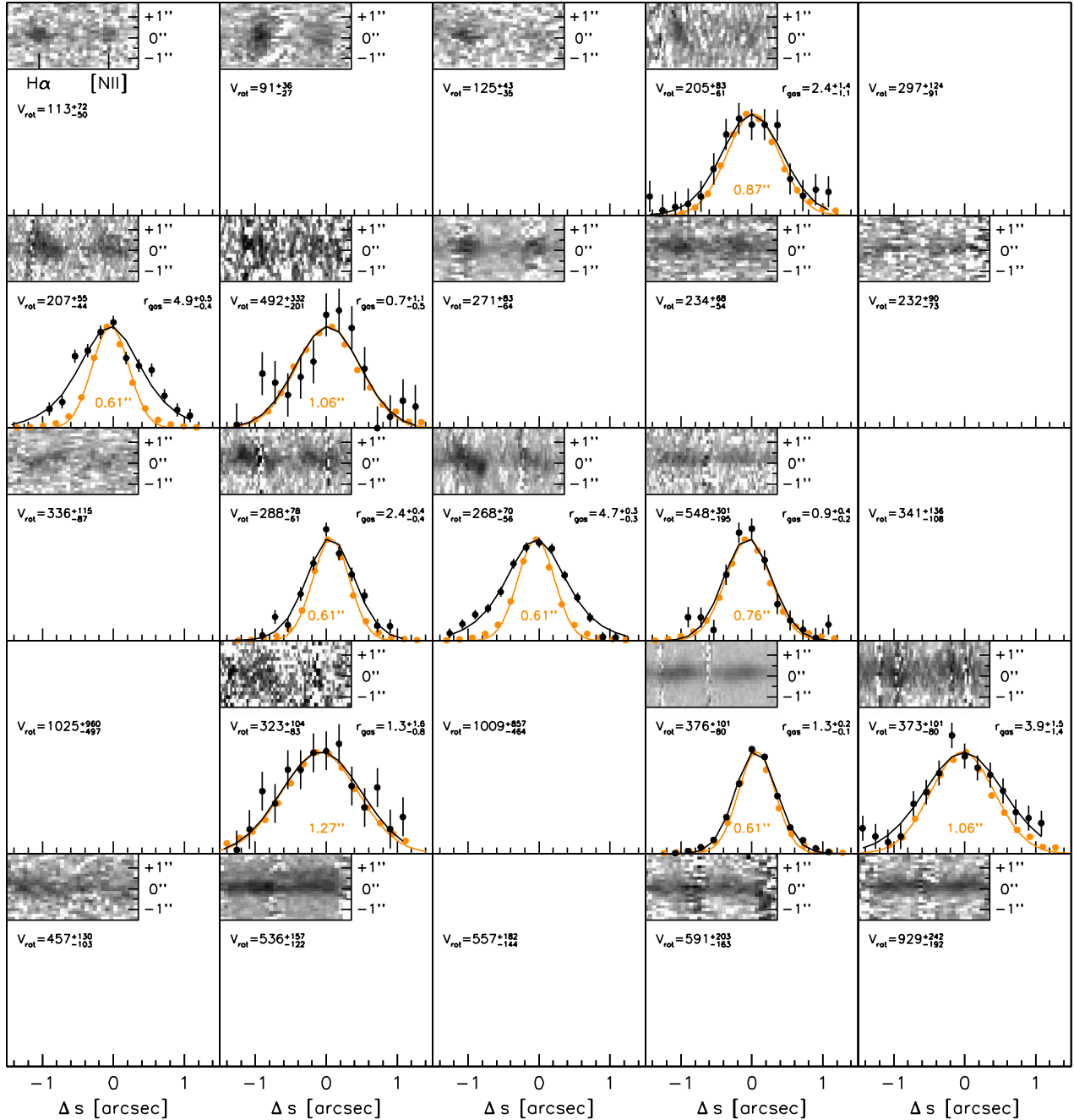}
\end{center}
\vspace{-0.3cm}
\caption{\small
Two-dimensional MOSFIRE and NIRSPEC
spectra centered on the redshifted
H$\alpha$ and \nii\ lines. The galaxies are ordered by their observed
galaxy-integrated
velocity dispersion, as in Figs.\ \ref{spectra.fig}, \ref{hstim.fig},
and \ref{seds.fig}. The inclination-corrected rotation velocity $V_{\rm rot}$
(in \kms)
is indicated in each panel. At least 1/3 of the
galaxies show velocity gradients,
demonstrating that their ionized gas distributions are spatially resolved
in these ground-based, seeing-limited data. For the nine galaxies observed
with MOSFIRE the spatial extent of the gas can be measured, using the
profile of a star (orange). Black curves are the best fitting
exponential profiles convolved with the PSF.
The measured half-light radii of the H$\alpha$
emission ($r_{\rm gas}$, in kpc) are indicated.
\label{2dspec.fig}}
\end{figure*}

\subsection{Measured Sizes of Gas Disks}
\label{morphline.sec}

We can test directly whether the \sg{}s are embedded in large gas disks
by examining the observed spatial extent of the emission lines. Even though
the galaxies were selected to be extremely compact, the inferred
spatial extent of the emission line gas is so large that it should (just)
be detectable in ground-based, seeing-limited observations.
The 2D spectra are shown in Fig.\ \ref{2dspec.fig}; they cover a
rest-frame wavelength range from 6551\,\AA\ to 6596\,\AA\ and a spatial
extent along the slit of $\pm 1\farcs 5$. The five empty panels are the
\sg{}s from {Barro} {et~al.} (2014b).

Remarkably, about 1/3 of the galaxies show velocity gradients. They
are most prominent in UDS\_33334, UDS\_26012, and UDS\_16442, but also
visible in GOODS-S\_5981, UDS\_42571, and UDS\_35673. 
The seeing ranged from $0\farcs 6$ to $\gtrsim 1\farcs 0$,
and the stellar
half-light radii of the galaxies are typically $0\farcs 1$.
Therefore, the
fact that we spatially resolve the H$\alpha$ emission 
immediately demonstrates that the ionized gas extends to larger
radii than the stars in these galaxies.
We emphasize here that we do {\em not} attempt to measure
rotation curves directly from these velocity gradients,
as this can only
be done reliably when the sizes of galaxies are similar to, or
larger than, the spatial resolution of the
data (see, e.g., Vogt
et al.\ 1996; Miller et al.\ 2011; Newman et al.\ 2013).

For the nine galaxies that were observed with MOSFIRE we can measure the
spatial extent of the H$\alpha$ emission. As discussed in Sect.\
\ref{mosfire.sec} a bright star was included in
all MOSFIRE masks, and the
profile of this star in the spatial direction
can be used to approximate the PSF. We extracted spatial profiles of the
combined \ha\ and \nii\ emission for the nine galaxies by averaging the
data in the wavelength direction. Each column was
weighted by the inverse of the noise (which is dominated by sky emission
lines); we did not weight by the signal
as this would bias the profile towards the central regions. The
spatial profiles are shown in Fig.\ \ref{2dspec.fig} (black points with
errorbars). Each panel also shows the profile of the star that was
observed in that particular mask (orange points); the FWHM of this profile
is also indicated.

The profiles were fit by a model to determine the half-light
radii of the ionized gas. The model has the form
\begin{equation}
\label{proffit.eq}
M(r) = \Sigma(r) \ast P(r),
\end{equation}
with $r$ the position along the slit,
$\Sigma(r)$ the model for the one-dimensional surface brightness
profile of \ha\ along the slit,  $P(r)$ a Gaussian fit
to the profile of the star, and $\ast$ denoting a
convolution. The Gaussian fits to the stellar profiles are shown by
orange lines in Fig.\ \ref{2dspec.fig}.
Parameterizing $P(r)$ with the sum of two Gaussians
does not improve the fit to the stellar profile or change the results.
It is not
possible to constrain the functional form of the surface brightness
profile with our data. Instead, we assume that the H$\alpha$ is
in an
exponential disk (see {Nelson} {et~al.} 2013):
\begin{equation}
\Sigma(r) = \Sigma(0)
\exp \left(- \frac{1.678 |r - r_{\rm cen}|}{r_{\rm gas}}\right).
\end{equation}
Here 
$\Sigma(0)$ is a normalization factor, $r_{\rm cen}$ is the
center of the profile,
and $r_{\rm gas}$ is the half-light radius of the
ionized gas.

We fitted this model to the data using the {\tt emcee} code, as described
for the fits in the wavelength direction in Sect.\ \ref{fitting.sec}. Again,
the priors are top hats with bounds that do not constrain the fits or the
errorbars. Rather than $r_{\rm gas}$ itself we fit
$\log r_{\rm gas}$: the error distribution of $r_{\rm gas}$ is highly
asymmetric, which means that the peak of the distribution of samples
does not coincide with its 50$^{\rm th}$ percentile. The distribution
of the $\log r_{\rm gas}$ samples is symmetric. The resulting
measured gas
radii, converted to kpc, are listed in the panels of Fig.\
\ref{2dspec.fig}.
For seven out of nine galaxies the value of $r_{\rm gas}$
is different from zero with $>2\sigma$ significance.

A geometric correction needs to be applied to the measured values
of $r_{\rm gas}$ to account for the fact that the slit is typically
not aligned with the major axis of the gas disk. This correction
depends on the orientation of the slit and on the inclination
of the gas disk:
\begin{equation}
r_{\rm gas}^{\rm c} \approx \left[
\cos^2\,(i) + \cos^2\,({\rm PA}_{\rm slit} - {\rm PA}_{\rm gal})
\left( 1 - \cos^2\,(i)\right)\right]^{-0.5} r_{\rm gas}
\end{equation}
with $i$ the inclination (as derived in Sect.\ \ref{rot.sec}),
${\rm PA}_{\rm slit}$ the position angle of the slitmask,
and ${\rm PA}_{\rm gal}$ the orientation
of the galaxy on the sky (as determined with GALFIT).
Note that the corrected $r_{\rm gas}^{\rm c}$
is measured along the major axis (and is not a circularized
radius), consistent with our interpretation
that the gas is in thin, rotating disks.
The median correction is small at 9\,\%. For GOODS-S\_30274
we use the median correction of the other eight galaxies, as its
PA mostly reflects the orientation of its tidal tail.
We use the corrected
radii when comparing the measured
radii to predicted radii and when deriving the rotation curve
of the galaxies in Sect.\ \ref{rotcurve.sec}.

For three galaxies, UDS\_35673, GOODS-S\_30274, and GOODS-N\_6215,
we obtained an independent measurement of the extent
of the emission line gas from their WFC3/G141 grism spectra. These are
the only galaxies in the sample of 25 that have grism spectra covering
the redshifted \oiii\ $\lambda 4959,5007$ lines and 
a detection of these lines with $>5\sigma$
significance. As shown in {Nelson} {et~al.} (2012) emission lines in grism
spectra are images of the galaxy in the light of that line, providing
direct information on the distribution of ionized gas at $0\farcs 14$
resolution. The interpretation of the \oiii\ lines is
complicated by the fact that the two lines are very close together
on the detector. We fit the lines simultaneously
with GALFIT ({Peng} {et~al.} 2002), keeping their relative location and flux
ratio fixed and using a PSF generated with Tiny Tim ({Krist} 1995).
Two of the three galaxies (UDS\_35673 and GOODS-S\_30274)
are also in the MOSFIRE sample. The best-fit G141 \oiii\
radii of these objects
are $1.6\pm 0.3$\,kpc and $5.1 \pm 1.5$\,kpc, in excellent agreement
with the MOSFIRE \ha\ values ($1.3^{+0.2}_{-0.1}$\,kpc and
$3.9^{+1.5}_{-1.4}$\,kpc, respectively). The third galaxy,
GOODS-N\_6215, has a G141 \oiii\ radius of $3.0 \pm 1.0$\,kpc.
In the following, we show all twelve measurements in figures
(nine from MOSFIRE, three from HST), with lines connecting the
two independent measurements for UDS\_35673 and GOODS-S\_30274.


\subsection{Comparison of Inferred and Measured Sizes}
\begin{figure}[hbtp]
\epsfxsize=8.5cm
\epsffile{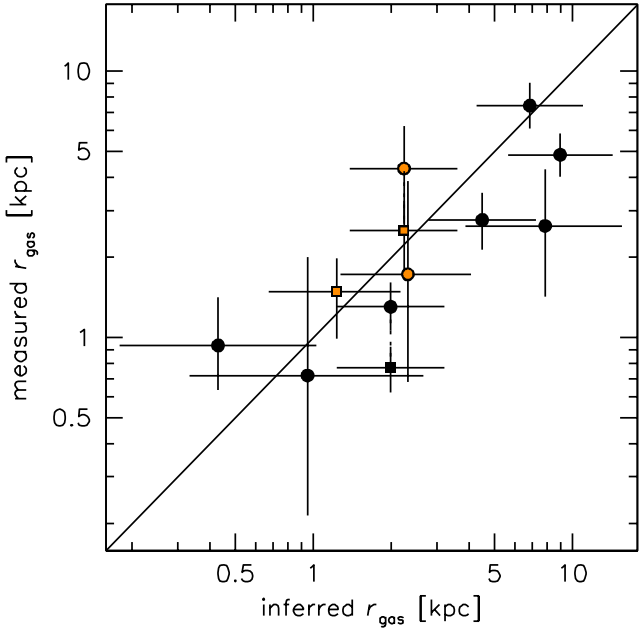}
\caption{\small
Relation between inferred and measured half-light radii of the gas
distribution in \sg{}s. Orange points are galaxies with an X-ray
AGN. Circles are Keck/MOSFIRE measurements of \ha;
squares are HST/WFC3 measurements of \oiii.
Points connected by dotted lines are measurements
for the same galaxy. The measured sizes were corrected
to account for the difference in orientation between
the slit and the galaxy's major axis.
The inferred sizes are based on the observed
velocity dispersions, axis ratios, and stellar masses of
the galaxies, and the measured sizes are determined directly from the spatial
extent of the emission lines. There is a strong
correlation, with no significant offset.
\label{comparesize.fig}}
\end{figure}

For the ten galaxies with gas size measurements
we can directly compare the
inferred sizes to the measured ones. The results are shown in
Fig.\ \ref{comparesize.fig}. There is a clear correlation, with a
significance of $>99$\,\%. Furthermore, the offset between the
two sets of radii is small. Giving equal weight to all twelve
measurements we find a difference of only $-0.09 \pm 0.07$ dex.
This excellent agreement between inferred and measured radii
provides support to our modeling of the observed kinematics of
\sg{}s.

\begin{figure*}[t]
\begin{center}
\epsfxsize=15.5cm
\epsffile{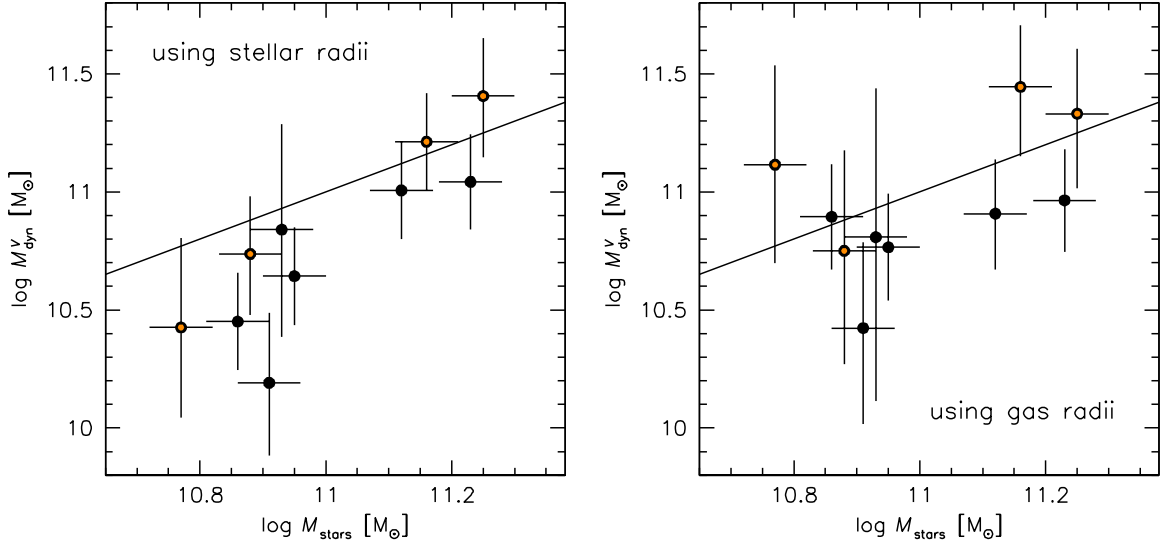}
\end{center}
\caption{\small
Dynamical mass versus stellar mass when using the stellar
half-light radii (left panel) or the H$\alpha$ half-light
radii (right panel) to calculate the dynamical mass.
The left panel shows the same information
as Fig.\ \ref{vdyn.fig}, but only for the ten galaxies with
measured H$\alpha$ radii. The dynamical masses derived from the
gas radii are self-consistent, as the rotation velocities
were measured at $r_{\rm gas}$, not $r_{\rm stars}$.
\label{massmass.fig}}
\end{figure*}

This result is presented in a different way in Fig.\
\ref{massmass.fig}, which shows the relation between
dynamical mass and stellar mass. The left panel is identical
to Fig.\ \ref{vdyn.fig}, but here we only show the ten galaxies
with measured H$\alpha$ effective radii. The dynamical
masses in the right panel were calculated using
\begin{equation}
M^V_{\rm dyn,\,gas} \equiv \frac{V_{\rm rot}^2 r_{\rm gas}}{f(r_{\rm gas})G},
\end{equation}
with $f(r_{\rm gas})$ accounting for the (small) fraction of the
mass that is outside $r_{\rm gas}$ (see Sect.\ \ref{sigcomp.sec}).
The dynamical masses in the right panel are consistent with the
stellar masses for all galaxies, although we note that the sample
is small. The mean offset is $\log M^V_{\rm dyn,\,gas} -
\log M_{\rm stars} = -0.07 \pm 0.08$, and the rms scatter is
0.25\,dex.

Summarizing the results from this and the previous Section,
we have inferred that \sg{}s have
rotating gas disks whose observed spatial extent is larger
by a factor of $\sim 2$ than their stellar
distribution. This is based on four related results:
1) Many of the galaxies have
very low galaxy-integrated velocity dispersions; this shows that the
gas does not have the same spatial distribution as the stars and that
galactic-scale winds do not dominate the kinematics for the majority
of the sample (Fig.\ \ref{sigsig.fig}a).
2) The observed dispersions display a significant
anti-correlation with the axis ratios of the galaxies; this is
consistent with disks under a range of viewing angles and difficult
to reconcile with M82-style galactic winds (Fig.\ \ref{sigq.fig}a).
3) Nearly all galaxies with spatially-resolved gas
distributions  show velocity gradients\footnote{There are
indications that
the presence of velocity gradients anti-correlates with the axis
ratio, as expected in the rotating disk interpretation, but larger
samples with higher spatial resolution are needed to confirm this.}
(Fig.\ \ref{2dspec.fig}).
4) Inferring the sizes
of the gas disks from the inclination-corrected rotation
velocities, we find good agreement between the inferred sizes and
the measured sizes (Fig.\ \ref{comparesize.fig}).

\subsection{Keplerian Rotation out to 7\,kpc}
\label{rotcurve.sec}

The measured kinematics can be used to constrain the total
mass within $\sim 7$\,kpc.
We can derive a spatially-resolved rotation curve by
making use of the fact that the measured spatial extent
of the gas varies by a factor of 10
(see Fig.\ \ref{comparesize.fig}), under the assumption that the galaxies
have similar inclination-corrected dynamics after scaling them
to a common mass. The validity of this approach is demonstrated in
Appendix C, where we calculate the relation between
the observed galaxy-integrated linewidths and the actual rotation
velocity at $r=r_{\rm gas}$. To bring all galaxies to the same
normalization, 
we first define the scaled rotation velocity as
\begin{equation}
V^{\rm s}_{\rm rot} =
\frac{V_{\rm rot}}{\sqrt{M_{\rm stars}/\langle M_{\rm stars}\rangle}},
\end{equation}
with $\langle M_{\rm stars}\rangle = 1.0\times 10^{11}$\,\msun\
the median stellar mass
of the full sample of 25 \sg{}s. 
We note that this scaling does not change the velocities by a large amount
as the galaxies in our sample span a small mass range.

\begin{figure}[hbtp]
\epsfxsize=8.5cm
\epsffile{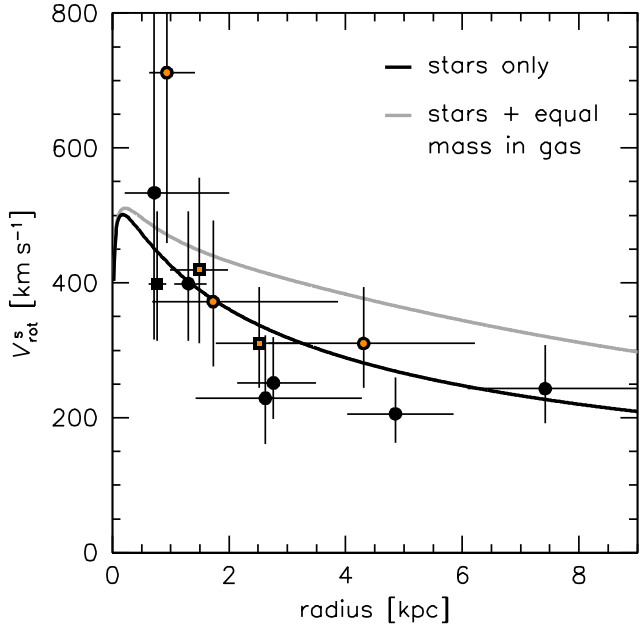}
\caption{\small
``Rotation curve'' for \sg{}s at $2.0<z<2.5$. Points with
errorbars are measured
inclination-corrected rotation velocities 
and measured gas effective radii
of ten different
galaxies. The quantities on the two axes are therefore
independent. The velocities were corrected to a common mass
of $10^{11}$\,\msun\  and the radii were multiplied
by a factor that
accounts for the slit alignment.
Galaxies with orange centers have an X-ray AGN. The rotation
curve declines, with $>99$\,\% significance. The black curve is not
a fit, but the
expected rotation curve if all the mass is in the compact stellar
component of the galaxies. This model is a good description
of the data. The grey curve assumes that 50\,\% of
the total mass is in the form of gas, with a spatial extent that
is a factor of 2.5 larger than that of the stars. This model
is inconsistent with the data.
\label{kepler.fig}}
\end{figure}

In Fig.\ \ref{kepler.fig} the scaled velocities
are plotted as a function of the measured gas
half-light radius $r_{\rm gas}$ (corrected for slit
orientation)
for the 10 galaxies that have this
measurement. 
The rotation curve declines: in galaxies where H$\alpha$
is measured at a larger distance from the center, the inclination-corrected
rotation velocity is lower. The decline has a formal significance
$>99$\,\%.
Falling rotation curves have been seen previously
in some individual (large, non-compact) $z\sim 2$ galaxies
(e.g., the galaxies D3a6397 and zC400690
in Genzel et al.\ 2014b).
The solid line is the predicted rotation curve for an
$M=10^{11}$\,\msun\ galaxy with the
median effective radius ($r_e = 1.3$\,kpc) and median Sersic
index ($n=4$) of the \sg{}s, calculated with Eq.\ \ref{rgas.eq}.
This model is a good description of the data: $\chi^2=6.5$
with 12 degrees
of freedom. The grey line is a model with two mass components: in addition
to the stellar component this model has a gas component with
the same mass as the stars (i.e., the gas fraction in this
model is $f_{\rm gas}\equiv M_{\rm gas}/(M_{\rm stars}+M_{\rm gas})=0.5$).
For consistency with the
previous Sections, the spatial distribution of the gas is
assumed to be exponential with $r_{\rm gas} = 2.5\times r_e$. The grey
line overpredicts the observed velocities: with $\chi^2=30.0$ this
model can be ruled out with $99$\,\% confidence.

We can derive an upper limit to the
gas mass within 7\,kpc
by assuming that the uncertainty in the stellar
mass is small and allowing the mass in the gas component to vary.
The 95\,\% confidence upper limit to the gas mass is
$M_{\rm gas}<0.6\times 10^{11}$\,\msun, corresponding to a limit on the
gas fraction of $f_{\rm gas} < 0.4$. It appears that
the gas is mostly a tracer of, rather than 
a contributor to, the kinematics.
Finally, we derive the best fitting  mass within
$r=7$\,kpc by assuming that
$f_{\rm gas}=0$ and allowing $M_{\rm stars}$ to vary: $M_{\rm fit}
= 0.8^{+0.6}_{-0.4} \times 10^{11}$\,\msun, where the errorbars are
95\,\% confidence limits. Although this estimate assumes that mass
follows light, we verified that
the results are very similar for more extended mass distributions.
We conclude that the dynamical mass within $r\sim 7$\,kpc
is fully consistent
with the stellar mass that is implied by the stellar population
fit;
and that there is little room for additional stars, gas, or 
dark matter inside this radius.

\section{Are Star Forming Compact Galaxies the Main
Progenitors of Quiescent
Compact Galaxies?}
\label{outer.sec}

In the previous Sections we have shown that 
a population of star forming galaxies exists
at $z\gtrsim 2$ whose dynamical
mass within $\sim 7$\,kpc is dominated
by a massive, compact, stellar component.
We now ask whether these galaxies can be
progenitors of the population of massive, compact, quiescent galaxies,
by considering their
number densities, morphologies, and star formation
rates.
This question has been discussed before,
by, e.g., Williams et al.\ (2014, 2015),
Bruce et al.\ (2014), Nelson et al.\ (2014), Dekel \& Burkert (2014),
Zolotov et al.\ (2015). Arguably
the most extensive observational study is
a series of papers
by Barro et al.\ ({Barro} {et~al.} 2013, 2014a, 2014b), using data
over two ({Barro} {et~al.} 2013, 2014b) or one ({Barro} {et~al.} 2014a) of the five
fields that we study here. Using our larger data set and more
restrictive selection we find broadly similar results. 

\subsection{Number Density Evolution}
\label{evoorder.sec}

\begin{figure}[hbtp]
\epsfxsize=8.5cm
\epsffile{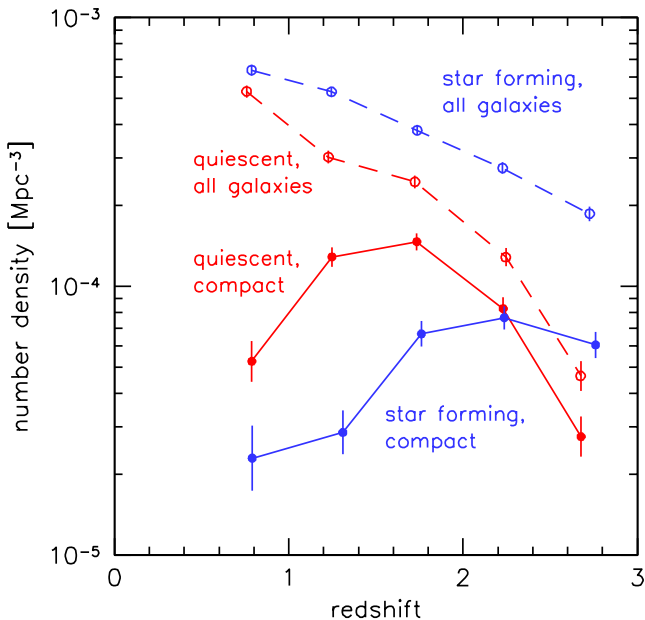}
\caption{\small
Evolution of the number density of \sg{}s (blue solid line) and of
\qg{}s (red solid line). The number density of all star forming
and quiescent galaxies with
$\log (M_{\rm stars})>10.6$ is also shown (dashed lines).
The data suggest that compact star forming galaxies continuously enter
the selection region from $z\gtrsim 2.8$ to $z\sim 1.8$ and quench,
leading to a strong increase in the number of compact quiescent galaxies.
When the number of \sg{}s begins to decrease at $z<1.8$,
the number of \qg{}s first plateaus and then drops,
as quiescent galaxies grow in size due to mergers at $0.5<z<1.5$.
\label{evo.fig}}
\end{figure}

A star forming compact massive
galaxy will resemble a quiescent compact massive galaxy if star formation
stops (quenching). However, the
opposite is also true: a quiescent compact galaxy that
starts forming stars due to the accretion of new gas (see,
e.g., {Zolotov} {et~al.} 2015; {Graham}, {Dullo}, \& {Savorgnan} 2015) could resemble a star forming compact galaxy
(rejuvenation). 
We can determine whether quenching or rejuvenation dominates by
measuring the number density of \sg{}s and \qg{}s as a function 
of redshift. The selection criteria of Sect.\ \ref{candidates.sec}
were applied in small redshift bins, and the number density
was determined by dividing the number of galaxies in the bin
by its volume. The result is shown in Fig.\ \ref{evo.fig}
(filled points and solid curves).

At $2.0<z<2.5$ the number densities of the two populations are very
similar, as already noted in Sect.\ \ref{predsig.sec}. However,
at higher and lower redshifts the number densities are different:
the \sg{}s have a roughly constant number density from $z\sim 2.8$
to $z\sim 1.8$, whereas the number density of \qg{}s increases by
an order of magnitude over that same redshift range.\footnote{The
evolution of compact quiescent galaxies may become
more gradual at $z>3$: {Straatman} {et~al.} (2015) recently reported
the existence of a sizeable population of compact, massive
quiescent galaxies at $z\sim 4$, based on medium-band near-IR
photometry.} A straightforward
interpretation is that star forming galaxies continuously enter the
``compact massive'' selection box (because of a decrease
in their size and/or an increase in their mass), and quench
shortly after. This continuous quenching then leads to a rapid build-up
of the number of quiescent galaxies in the compact/massive selection
region. We conclude that quenching likely
dominates over rejuvenation:
if rejuvenation dominated, we would have expected to see
quiescent galaxies disappear as their star formation (re-)started,
unless there are other channels to create quiescent compact
galaxies. We note that the evolution of the number densities of
the two populations is qualitatively similar to the simulations
of Zolotov et al.\ (2015).

It is difficult to determine how
long it takes before a compact star forming galaxy
turns into a quiescent galaxy, as this depends on the rate with
which new galaxies enter the sample. 
The
number density of \sg{}s is
constant from
$z\sim 2.8$ to $z\sim 1.8$, which means that new \sg{}s enter the
sample at approximately the same rate as existing ones quench.
We can obtain a very rough estimate of the ``compact life time''
of star forming galaxies $\tau_{\rm c}$ by adding
the number densities of the \sg{}s in the three redshift
bins that cover this period: if the average
quenching time is much shorter than
the time interval between redshift bins, all galaxies in each
bin are new arrivals and should be added to the sample of
progenitors of quiescent galaxies. The combined number density in these
bins (which are of nearly equal volume) is $2.0\times 10^{-4}$\,Mpc$^{-3}$,
higher than the increase in the number density of the \qg{}s
over this period
($1.3\times 10^{-4}$\,Mpc$^{-3}$). This implies that
only about half of the
star forming galaxies disappear from one redshift bin to the next,
and that the average
quenching timescale is roughly equal to the
time interval
between the redshift bins: $\tau_{\rm c} \sim 0.5$\,Gyr.
This is the average lifetime of star forming galaxies in
the ``compact massive'' selection box, assuming that they all turn
into quiescent galaxies. It is slightly lower than the
value of $\sim 0.8$\,Gyr
derived by Barro et al.\ (2013), but judging from their
Fig.\ 5 the two studies are broadly consistent.

Although  somewhat outside of the scope of this paper, we briefly
discuss
the number density evolution at lower redshift. The
number density of \sg{}s drops precipitously after $z\sim 1.8$. This
drop leads to a plateau in the number density of \qg{}s: as the number
of star forming progenitors decreases, no new quiescent galaxies are
added to the sample. At the lowest redshifts the number density of compact
quiescent galaxies decreases (as was also found
by {Taylor} {et~al.} 2010b,
{van der Wel} {et~al.} 2014b, and
{van Dokkum} {et~al.} 2014, among others),
while the number density
of {\em all} massive quiescent galaxies rises steeply (dashed red
curve). The likely explanation is that the compact galaxies accrete
extended envelopes through merging from $z\sim 1.5$ to the present
day
(e.g., {Bezanson} {et~al.} 2009; {Naab} {et~al.} 2009; {van Dokkum} {et~al.} 2010, 2014; {Newman} {et~al.} 2012; {Hilz} {et~al.} 2013).

Finally, we note that  Fig.\ \ref{evo.fig} is not new: 
the peak in the
number density of compact, massive quiescent galaxies was also shown
in Cassata et al.\ (2011, 2013), Barro et al.\ 2013,
and van der Wel et al.\ (2014b).
{Barro} {et~al.} (2013) derive a similar
lifetime for star forming galaxies in the compact selection region.
Although  uncertainties remain (particularly
at low redshift; see, e.g., Carollo et al.\ 2013), it is
encouraging that these largely independent samples give similar
results. 


\subsection{Morphologies and Radial Surface Brightness Profile}
\label{sb.sec}

The large spatial extent of the ionized gas raises the question whether
the stellar half-light radii and masses
of the compact star forming galaxies have been underestimated:
although it is difficult to bias
GALFIT measurements in this direction
(see, e.g., {Davari} {et~al.} 2014),
it is possible that the galaxies
have extended low surface brightness envelopes
(see, e.g., {Hopkins} {et~al.} 2009a).
If such envelopes
exists this would call into question whether \sg{}s
can be direct progenitors
of compact quiescent galaxies with the same apparent mass and half-light
radius.


Images of the galaxies are shown in Fig.\ \ref{hstim.fig} and
in Fig.\ A1 (see Sect.\
\ref{parent.sec}). Visually, most of the objects have a compact
luminosity distribution and no spiral arms, clumps,
star forming complexes, or other features outside of the dense center. 
Several of the reddest galaxies do not appear very compact: for
example, UDS\_42571 and, in particular, UDS\_16442 are faint and
fuzzy rather than bright and compact. The reason for their relatively
low surface brightness is that dust obscuration has dramatically
lowered their luminosity: as galaxies can have high $M/L$ ratios,
compact in mass does not necessarily imply compact in light.

Two objects show unambiguous
evidence for ongoing or past mergers: GOODS-S\_30274
has an asymmetric feature resembling a tidal tail, and COSMOS\_11363 is
one component of
a spectacular merger between two compact galaxies with a projected
separation of
$0\farcs 6$ (5\,kpc). The companion of
COSMOS\_11363
is COSMOS\_11337 in the {Skelton} {et~al.} (2014) catalog. Our Keck/NIRSPEC and
HST/WFC3 spectroscopy confirms that they are at the same redshift.
With $r_e = 1.0$\,kpc
and $M_{\rm stars} = 1.7\times 10^{11}$\,\msun\ COSMOS\_11337
is actually significantly
more compact than COSMOS\_11363. Its rest-frame $UVJ$ colors (just) give it a
quiescent classification.\footnote{We note that the rest-frame $J$ magnitudes
of these objects are somewhat uncertain as they rely on accurate deblending of
the IRAC fluxes; it may well be that both galaxies are \sg{}s.}
This merging pair seems to suggest that CMGs can
form in mergers ({Hopkins} {et~al.} 2009b), but that is not the right
interpretation: as both galaxies already fall in the ``compact massive''
selection region, this particular
type of merger actually {\em decreases} their number. Even if the result of
the merger falls in the selection region, there will be one less CMG.
Interestingly, several other galaxies show 
evidence for distorted
outer isophotes in Fig.\ A1. This could indicate interactions are
common for these galaxies, but the evidence is not conclusive at
the depth of the CANDELS imaging.

\begin{figure}[hbtp]
\epsfxsize=8.5cm
\epsffile{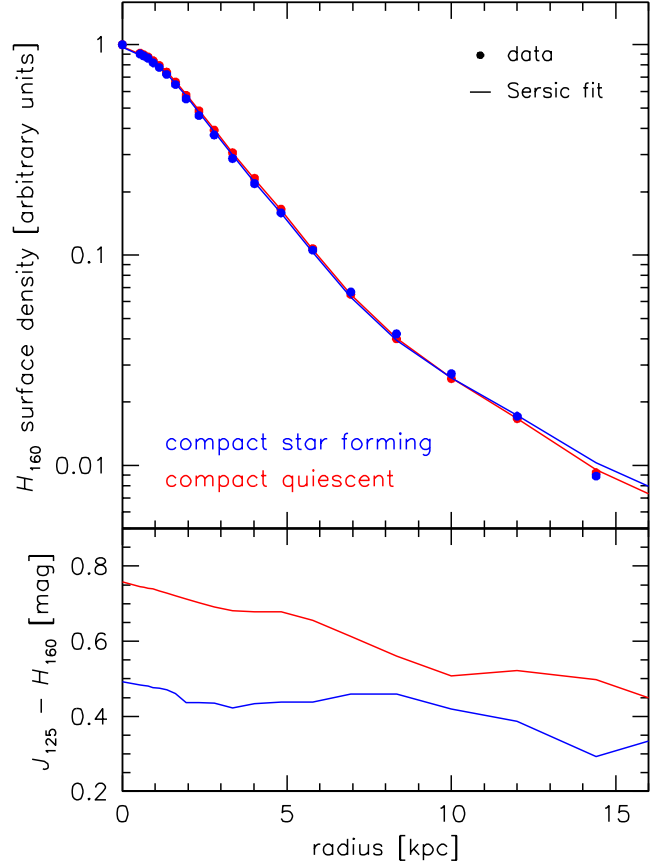}
\caption{\small
Radial surface brightness profile, measured from a stack of all 25
\sg{}s in our spectroscopic sample (blue points). The profile is
very well fit by a single Sersic profile, convolved with the PSF
(blue line). There is no excess emission at large radii. For
comparison, the red points and red line are for \qg{}s that
were selected to have the same median
size and mass as the \sg{}s.
Their profile is virtually identical to the star
forming galaxies. The bottom panel shows color profiles for both
samples. The galaxies have modest color gradients, with the outskirts
slightly bluer than the centers.
\label{sbprof.fig}}
\end{figure}

To quantify the stellar emission on scales $\gg 1$\,kpc we stacked the
$H_{160}$ images of the 25 \sg{}s and measured their averaged radial
surface brightness profile to faint levels. Each galaxy was normalized
by its total $H_{160}$ flux prior to stacking, so that the stack is not
dominated by a few bright objects. Neighboring objects, identified from
the SExtractor segmentation map (see {Bertin} \& {Arnouts} 1996; {Skelton} {et~al.} 2014),
were masked. The resulting surface brightness profile is shown in
the top panel of
Fig.\ \ref{sbprof.fig} (blue points). We fit the stack with a 
PSF-convolved
Sersic profile to determine whether there is
evidence for an additional component at large radii. This
fit, done with GALFIT ({Peng} {et~al.} 2002), is shown by the blue line.
It is an excellent description of the data out to 15\,kpc ($>10 r_e$): there
is no excess light beyond a single Sersic profile. Furthermore, the
best-fitting effective radius ($r_e = 1.3$\,kpc) and Sersic index
($n=3.6$) are similar to the median values of the 25 galaxies
that went into the stack: $\langle r_e\rangle = 1.4$\,kpc and $\langle
n\rangle =4.3$.

The stacked \sg{} profile is compared to a stacked \qg{} profile,
shown in red in Fig.\ \ref{sbprof.fig}.
The \qg{}s in this Figure are a subset of the full population:
they were selected in narrow bins of mass and effective
radius, centered on the median values of the 25 \sg{}s. This
ensures that any differences between the stacks are not caused by a
difference in the mean size or mass of the samples. The quiescent
profile is virtually indistinguishable from that
of the star forming galaxies. Finally, $J_{125}-H_{160}$
color profiles of both
stacks are shown in the bottom panel of Fig.\ \ref{sbprof.fig}.
Both stacks are bluer at larger radii and the gradients are small,
qualitatively consistent with previous work ({Szomoru}, {Franx}, \& {van  Dokkum} 2012).
The negative color
gradients imply that the galaxies are even more compact in
mass than in light, and that
any stellar emission at $r\gg r_e$ is not missed because it is
enshrouded in dust.

We conclude that the morphologies of the \sg{}s
are consistent with being direct progenitors of \qg{}s. When selected
to have the same mass and effective radius, their surface brightness
profiles are indistinguishable out to at least 15\,kpc. 
We find a relatively high Sersic index for both populations. 
Such high values (and the relatively round 3D morphologies;
see Sect.\ \ref{rot.sec})
are consistent with violent relaxation following
a merger, but also with composite structures, such as envelopes of
material around extremely compact exponential disks.

\subsection{Star Formation Rates and Gas Content}

\begin{figure}[hbtp]
\epsfxsize=8.5cm
\epsffile{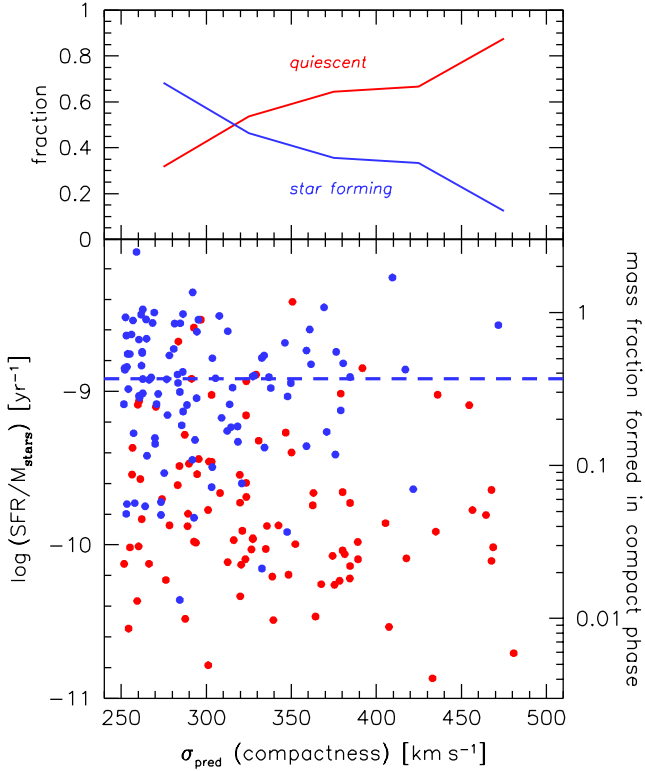}
\caption{\small
Relation between specific star formation rate and compactness ($\propto
M_{\rm stars}/r_{\rm e}$), for galaxies in the ``massive, compact''
selection box at $2<z<2.5$. Red points
are $UVJ$ quiescent galaxies; blue points are $UVJ$ star forming
galaxies. Within the sample of massive compact galaxies, the specific
star formation rate, and the fraction of $UVJ$ star forming
galaxies, declines with the degree of compactness.
The right axis is the fraction of mass that will be added to the galaxies
in 0.5\,Gyr, which is the estimated
average lifetime of star forming galaxies
in the massive, compact region. About $1/3$
of the mass of compact quiescent
galaxies was formed in the compact phase.
\label{ssfr.fig}}
\end{figure}

Accepting that the \sg{}s are direct progenitors of \qg{}s, an important
question is whether they are forming
a large fraction of the stars that are present in their
quiescent descendants. If the life times of the
\sg{}s are short, or the star formation rates are low,
they may account for only
a small fraction of the total stellar mass
in compact massive galaxies at $z\sim 2$. We address this question
in Fig.\ \ref{ssfr.fig}, which shows the relation between the
specific star formation rate and compactness within the sample
of compact, massive galaxies at $2<z<2.5$.

The right axis of this figures shows the fraction of the total stellar
mass that is formed in the compact phase:
\begin{equation}
\frac{M_{\rm c}}{M_{\rm stars}} \sim {\rm SSFR} \times
w\times
\tau_c,
\end{equation}
with {\rm SSFR} the specific star formation
rate, $w$
a correction for mass loss due to stellar winds, and $\tau_c$ the
average life time of star forming galaxies in the compact, massive
selection region.  The median specific star formation rate of
the \sg{}s is SSFR\,$= 1.2\times 10^{-9}$\,yr$^{-1}$, and for
$w\sim 0.6$ ({Chabrier} 2003) and
$\tau_c\sim 0.5$\,Gyr (Sect.\ \ref{evoorder.sec}) we find
$M_{\rm c} \sim 0.4 M_{\rm stars}$. As they are, on average,
observed halfway through their lifetime in the compact
selection region, their final mass before
quenching will be $M_{\rm stars,final} = M_{\rm stars}
+ 0.5 M_{\rm c} \sim 1.2 M_{\rm stars}$,
and the fraction of $M_{\rm stars,final}$
that is formed in the compact phase is then $\sim 1/3$.
We conclude
that \sg{}s are responsible for forming 
a significant fraction
of the stars that are present in compact quiescent galaxies. 

An implication of this result is that the spatial distribution
of the \ha\ emission in \sg{}s is probably more extended than the spatial
distribution of star formation in these galaxies. 
This is qualitatively similar to results for galaxies at $z\sim 1$
(Nelson et al.\ 2012, 2015), and may indicate that star formation
has ceased in the inner regions of the galaxies
(e.g., Genzel et al.\ 2014b; Tacchella et al.\ 2015).
However, as discussed in
Sect.\ \ref{sfr.sec} most of the star formation in \sg{}s is
obscured, 
and the observed \ha\ emission accounts for only
$\sim 10$\,\% of the total star formation. As the column
density is
a very strong function of radius in these compact
galaxies (see, e.g., {Gilli} {et~al.} 2014; {Nelson} {et~al.} 2014), the obscuration-corrected
distribution
of star formation is almost certainly much more compact than the
observed distribution of \ha\ emission -- at least for the galaxies
with low observed velocity dispersions.

A somewhat
puzzling aspect of the \sg{}s is that they have very high specific star
formation rates even though their observed kinematics
leave little room for a large
gas reservoir (see Sect.\ \ref{rotcurve.sec}). Many studies have found
that the molecular gas and dust
content of galaxies increases with redshift, and
reaches $>50$\,\% of the total baryonic mass for
$z\sim 2$ galaxies 
with the highest star formation rates
(e.g., {Tacconi} {et~al.} 2010; {Daddi} {et~al.} 2010; {Genzel} {et~al.} 2015; {Scoville} {et~al.} 2015).
Using the scaling relations derived in {Genzel} {et~al.} (2015), the
expected gas fraction for the galaxies in our sample
is $\sim 60$\,\%.
One possible explanation
for their relatively low gas fraction is
that the galaxies have nearly exhausted their reservoir
and are about to quench. If the galaxies typically build $\sim 40$\,\%
of their mass inside the compact, massive selection region, the average
\sg{} should have $\sim 30$\,\%
of their mass in gas (for $w\sim 0.6$); this is just consistent
with the 95\,\% confidence
upper limit on the gas fraction of 40\,\% that we
derived in Sect.\ \ref{rotcurve.sec}. Another explanation is that
newly accreted gas is continuously and efficiently funneled 
into the central regions, and the star formation
rates are ``accretion throttled'' ({Dekel} {et~al.} 2009); in that case the
gas depletion time can be shorter than the actual duration of
star formation (see, e.g., {Genzel} {et~al.} 2010).
Direct observations of the dust and molecular gas in
\sg{}s, at $\sim 1$\,kpc resolution, are needed to address these questions.

Finally, we note that
star forming
galaxies tend to be less compact than quiescent galaxies even
{\em within} the population of compact massive galaxies at $2<z<2.5$
(see Fig.\ \ref{ssfr.fig}).
As discussed earlier in the context of the sample
selection (Sect.\ \ref{predsig.sec}),
star forming galaxies are always less compact than quiescent galaxies,
irrespective of the precise criteria for their selection. In the
next Section we interpret the distribution of galaxies in the size-mass
plane in the context of a simple model, in which star forming galaxies
become gradually more compact and the probability of
quenching rises smoothly as their compactness increases.

\section{Formation of Star Forming Compact Galaxies}

\subsection{A Simple Model for Building Massive Galaxies}
\label{simplemod.sec}

In this Section we turn to the {\em formation} of compact, massive star
forming galaxies. Several distinct mechanisms have been discussed
in the literature,
including mergers of gas-rich galaxies
({Tacconi} {et~al.} 2008; {Hopkins} {et~al.} 2009b; Hammer et al.\ 2009;
{Wellons} {et~al.} 2015),
in-situ, inside-out growth of even more compact progenitors
({Oser} {et~al.} 2010; {Johansson}, {Naab}, \&  {Ostriker} 2012; {Williams} {et~al.} 2014; {Nelson} {et~al.} 2014; {Wellons} {et~al.} 2015),
``compaction'' of the gas in star forming galaxies due to disk
instabilities ({Dekel} \& {Burkert} 2014),
and hybrid models that include several of these
effects ({Zolotov} {et~al.} 2015).

\begin{figure}[hbtp]
\epsfxsize=8.5cm
\epsffile{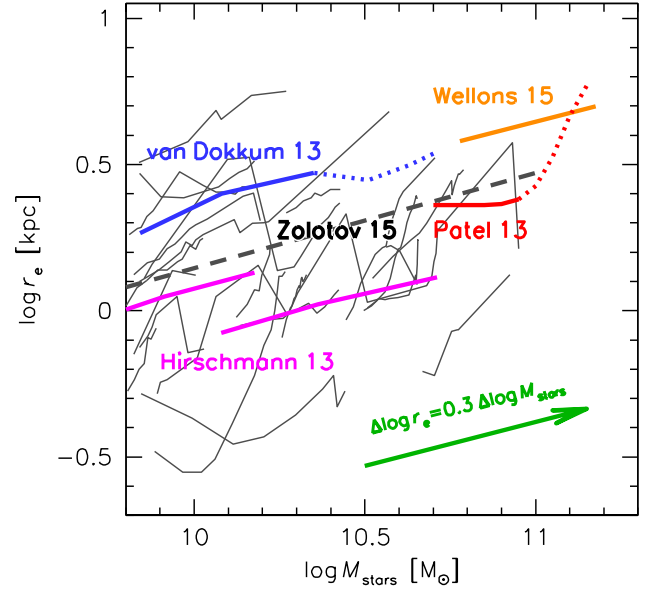}
\caption{\small
Tracks of galaxies in the size-mass
plane in different studies. The solid
blue
and red curves show the evolution from $z\sim 3$ to $z=1.5$
of number density-matched samples
of low mass ({van Dokkum} {et~al.} 2013) and high mass ({Patel} {et~al.} 2013) galaxies.
Broken curves show the evolution at $z<1.5$. 
Magenta tracks are the wind models of {Hirschmann} {et~al.} (2013),
for two different mass ranges and $1.5<z<2.5$.
The orange curve is
the evolution of the full sample of massive Illustris
galaxies from $z=3$ to $z=1.5$ in {Wellons} {et~al.} (2015). Thin black curves
are individual simulated galaxies in {Zolotov} {et~al.} (2015), from $z=3$
to $z=1.5$. The mean Zolotov
evolution is indicated by the thick black dashes. The green arrow is
a good match to the mean growth of galaxies
in all these studies: $\Delta \log r_{\rm e} \sim 0.3
\Delta \log M_{\rm stars}$.
\label{meangrowth.fig}}
\end{figure}

Although individual massive galaxies likely have complex formation histories,
including periods of compaction, mergers, and star bursts, the
{\em population} of massive galaxies should follow a particular
track in the size-mass plane that is determined by the 
dominant mode of growth
when the evolution of many galaxies is averaged. Tracks derived
from observations and simulations are shown in Fig.\ \ref{meangrowth.fig}.
The blue and red tracks show the evolution
of galaxies matched by their
cumulative number density, for (relatively) low mass galaxies
({van Dokkum} {et~al.} 2013, blue) and
high mass galaxies ({Patel} {et~al.} 2013, red). The solid parts of the curves
are for $1.5<z<3$ and the dotted parts for $0<z<1.5$. Low mass
galaxies evolve along a single track with a slope of
$\sim 0.3$. High mass galaxies evolve along a similar
track from $z\sim 3$ to $z\sim 1.5$ but then turn ``upward'',
around the time when star formation ceases and the growth becomes
dominated by dry mergers (see Sect.\
\ref{lowz.sec}).

Magenta, orange, and black curves are from simulations. 
The magenta tracks are the wind models shown in Fig.\ 10 of
{Hirschmann} {et~al.} (2013), for two different mass ranges. These
models are the same as those in {Genel} {et~al.} (2012), and are
updated versions of
the momentum-driven wind models of {Oppenheimer} \& {Dav{\'e}} (2006) in
cosmological simulations. They include both winds
and metal enrichment; as shown in {Hirschmann} {et~al.} (2013) models
without winds predict somewhat
steeper relations between size growth and mass growth.
The orange curve is the
track of galaxies in the Illustris project ({Vogelsberger} {et~al.} 2014), as
shown in Fig.\ 5 of {Wellons} {et~al.} (2015). This is the average track
of all galaxies with a stellar mass in the range
$1-3\times 10^{11}$\,\msun\ at $z=2$. The thin black curves show
the evolution from $z=3$ to $z=1.5$ of
individual galaxies
in the simulations of {Zolotov} {et~al.} (2015). We include all 34
simulations, irrespective of whether they have a ``compaction'' phase.
The thick dashed curve was created by averaging the evolution in
these simulations. The number density-matched observational
samples and the simulations all suggest that the ensemble-averaged
evolution of star forming galaxies
in the size-mass plane is well approximated by
\begin{equation}
\label{growth.eq}
\Delta \log r_{\rm e} = 0.3 \Delta \log M_{\rm stars},
\end{equation}
that is, galaxies increase their size by a factor of 2 for every
factor of 10 evolution in their mass. This simple inside-out growth
model is qualitatively consistent with a host of other data and
theory, including the expected growth of disks in $\Lambda$CDM
(e.g., {Mo}, {Mao}, \& {White} 1998) and the distributions of star formation and
existing stars in galaxies (e.g., {Nelson} {et~al.} 2012).
Interestingly, this track corresponds to an approximately
constant 3D density within the effective radius
(as $\rho(r_{\rm e}) \propto
M/r_{\rm e}^3$, it follows that $r_{\rm e} \propto M^{1/3}$ if
the density is constant).

\begin{figure}[hbtp]
\epsfxsize=8.5cm
\epsffile{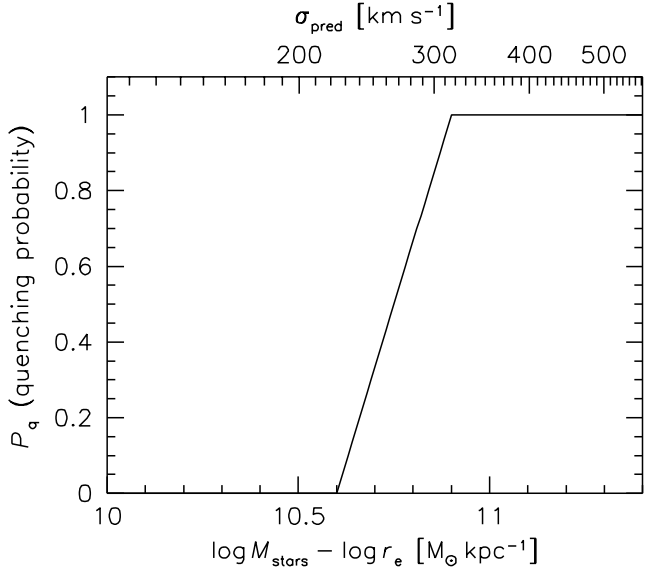}
\caption{\small
Parameterization of quenching. No galaxies with low velocity
dispersions are quenched, and
all galaxies with high velocity dispersions are quenched. The quenching
probability begins to increase at $\log M_{\rm stars} - \log r_{\rm e}
=10.6$. This threshold is held fixed in this paper, but is
in fact redshift dependent.
\label{q.fig}}
\end{figure}

Although the 3D density within the effective radius stays
constant,
a direct consequence of Eq.\ \ref{growth.eq} is that the
stellar density within a physical radius, the stellar
surface density, and the stellar
velocity dispersion all gradually increase as galaxies form
stars. We assume that
galaxies have
an increasing likelihood of quenching as their velocity dispersion
reaches a particular threshold. This is motivated by numerous studies
showing that the specific star formation rates of galaxies correlate
much better with compactness than with mass (e.g., {Kauffmann} {et~al.} 2003; {Franx} {et~al.} 2008).
We parameterize this process by a dispersion-dependent
quenching probability $P_{\rm q}$:
\begin{eqnarray}
\label{quench.eq}
P_q & = & 0\,\,\,\,\,\,\,\,(x<10.6)\nonumber\\
 & = & \frac{x - 10.6}{0.3}\,\,\,\,\,\,\,\,(10.6\leq x\leq 10.9)\nonumber\\
 & = & 1\,\,\,\,\,\,\,\,(x> 10.9),
\end{eqnarray}
with $x \equiv \log M_{\rm stars} - \log r_{\rm e}$
(see Fig.\ \ref{q.fig}).
Galaxies begin to quench at $\log M_{\rm stars}-\log r_{\rm e}>10.6$,
or $\sigma_{\rm q}=225$\,\kms\ (Eq.\ \ref{sigpred.eq}).
As we show below this particular choice of $\sigma_{\rm q}$ provides
a reasonably good fit to the data over the redshift range
$1.5<z<3.0$. We use a single value in this paper,
but we note that the threshold is a function of redshift:
low redshift galaxies quench at a lower density or dispersion than
high redshift galaxies ({Franx} {et~al.} 2008).

The average mass growth of the population
is assumed to be a simple function of the star
formation rate, modified by the quenching function:
\begin{equation}
\label{massgrowth.eq}
\Delta \log M_{\rm stars} = \beta \Delta t \times {\rm SFR} \times
(1-P_{\rm q}).
\end{equation}
The parameter $\beta$ encompasses mass loss due to stellar winds,
possible effects of mergers, and the well-documented offset between
the evolution of the star forming sequence and the evolution of the
stellar mass function
(see {Leja} {et~al.} 2015; {Papovich} {et~al.} 2015, and references therein). We adopt
$\beta = 0.45$; values of $0.4<\beta<0.5$ produce very similar
results. A pure mass loss model would have $\beta = w
\approx 0.6$ for
a {Chabrier} (2003) IMF. The star formation rate is
given by the star forming ``main
sequence''. We adopt the mass-dependent 
parameterization of {Whitaker} {et~al.} (2014):
\begin{equation}
\label{ms.eq}
\log({\rm SFR}) = a + b \log M_{\rm stars} + c (\log M_{\rm stars})^2,
\end{equation}
with  $a=-19.99$, $b=3.44$, and $c=-0.13$ for the redshift range of interest. As shown in Fig.\ \ref{select.fig}c the actual
star formation rates of \sg{}s are broadly
consistent with this relation.
 
\begin{figure}[hbtp]
\epsfxsize=8.5cm
\epsffile{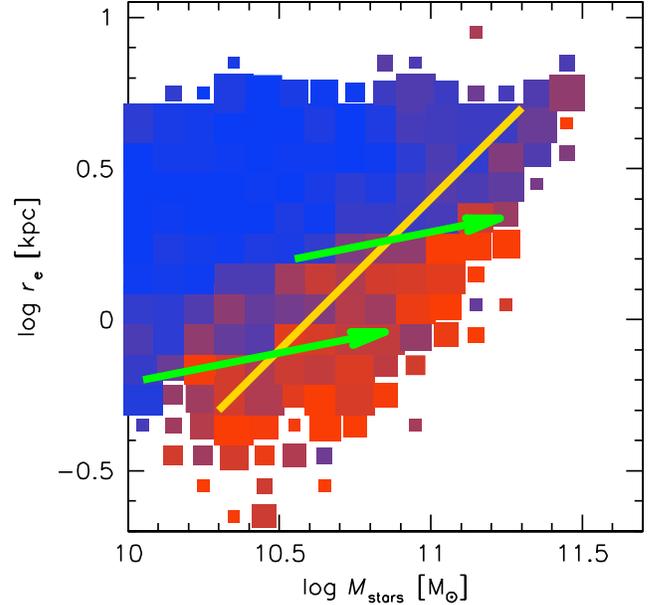}
\caption{\small
Illustration of the ``parallel track'' model of massive galaxy evolution.
The blue and red squares show the distribution of galaxies in the
size-mass plane at $1.5<z<2.25$, with
the size of the square proportional to the
number of galaxies and the color indicating the fraction
of quiescent galaxies. Galaxies move along parallel tracks in the
size-mass plane, with $\Delta \log r_{\rm e} \sim 0.3 \Delta \log
M_{\rm stars}$, until they cross the yellow quenching line of constant
$\sigma_{\rm q}\sim 225$\,\kms. 
\label{moddemo.fig}}
\end{figure}

The model is illustrated in Fig.\ \ref{moddemo.fig}, which shows
galaxies in the size-mass plane at $1.5<z<2.25$. The color indicates
the fraction of galaxies that are quiescent according to the $UVJ$
criteria. Galaxies move along the green curves until they cross the
yellow line, when their quenching probability rises steeply.
In  this model
galaxies follow parallel tracks in the 
size-mass plane, which means that large galaxies and small galaxies
at fixed mass have different formation histories. However, we emphasize
that individual galaxies likely have complex histories,
involving excursions above and below these mean tracks
(see, e.g., {Zolotov} {et~al.} 2015).
Our description is qualitatively
similar to the work of {Williams} {et~al.} (2014, 2015), who
identified low mass Lyman break galaxies with small sizes as possible
progenitors of quiescent compact massive galaxies. 

\begin{figure*}[hbtp]
\begin{center}
\epsfxsize=15.5cm
\epsffile{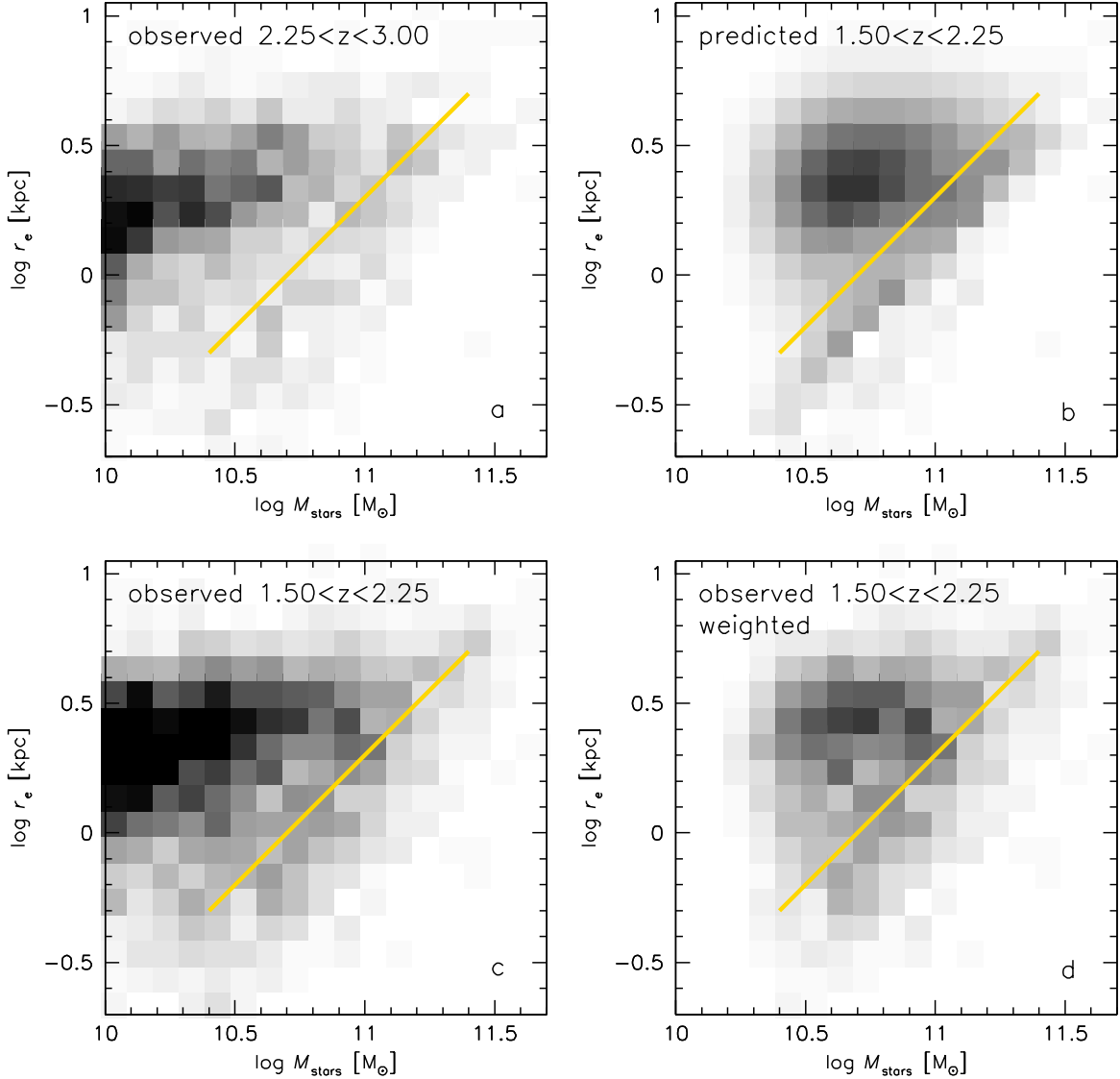}
\end{center}
\vspace{-0.3cm}
\caption{\small
Testing the ``parallel track'' model for the creation of compact
massive galaxies. Panel (a) shows
the observed number density of galaxies in the size-mass
plane at $2.25<z<3.00$,
with the grey scale proportional to the number of galaxies.
In panel (b) the distribution is evolved forward in time by
1.0\,Gyr to  $1.50<z<2.25$, by assuming that galaxies grow
along lines of $\Delta \log r_e = 0.3\Delta \log M_{\rm stars}$
and quench after they pass the yellow line. Panel (c) shows the
{\em observed} number density of galaxies at $1.50<z<2.25$. 
Panel (d) is identical to panel (c), but weighted to account for
the edge effect at low masses in the model prediction of panel (b).
The distribution of galaxies in panel (d) is remarkably similar to
that in panel (b), demonstrating that compact massive galaxies
at $z\sim 2$
can be formed by simple mass growth of galaxies at higher redshift.
\label{modtest1.fig}}
\end{figure*}

\begin{figure*}[hbtp]
\begin{center}
\epsfxsize=15.5cm
\epsffile{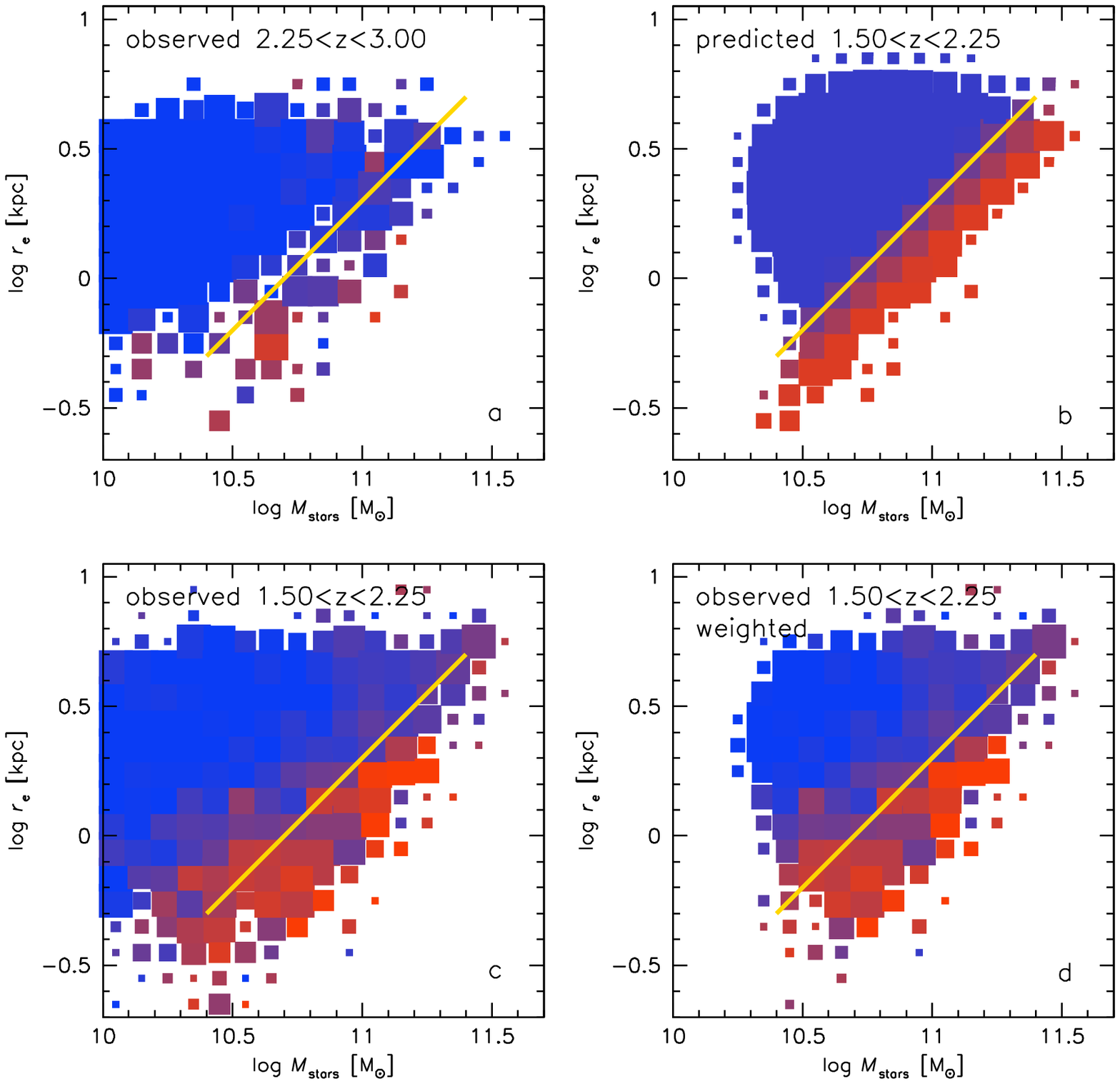}
\end{center}
\vspace{-0.3cm}
\caption{\small
Same as Fig.\ \ref{modtest1.fig}, but with color coding indicating
the median specific star formation rate of the galaxies.
Our simple model naturally
produces a
population of massive, compact quiescent galaxies at $M_{\rm stars}
\sim 10^{11}$\,\msun\ and $r_e\sim 1$\,kpc. The model overpredicts the
quiescent fractions at the largest masses and sizes.
\label{modtest2.fig}}
\end{figure*}

\subsection{Testing the Model}

We test the model in the following way. We first quantify the distribution
of galaxies in the size-mass plane at $2.25<z<3.0$, by measuring the
number of galaxies in bins of 0.1\,dex\,$\times$\,0.1\,dex (see
Fig.\ \ref{modtest1.fig}a). Next, we evolve this distribution forward
in time, using timesteps of $\Delta t=100$\,Myr.
For each combination of $(M_{\rm stars}, r_{\rm e})$
we can calculate the SFR from Eq.\ \ref{ms.eq}, $P_{\rm q}$ from
Eq.\ \ref{quench.eq}, the change in mass from
Eq.\ \ref{massgrowth.eq}, and the corresponding change in size from
Eq.\ \ref{growth.eq}.

The evolved distribution
after 10 timesteps (i.e., 1 Gyr) is shown in Fig.\ \ref{modtest1.fig}b,
with a small (4\,\%) correction to account for the volume difference
between $2.25<z<3.00$ and $1.50<z<2.25$.
As expected, the galaxies have shifted to larger masses and to slightly
larger radii in the size-mass plane. The distribution artificially
falls off at low masses
due to the $M_{\rm stars} = 10^{10}$\,\msun\ limit in Fig.\
\ref{modtest1.fig}a. This limit was chosen to ensure that the
galaxies with the lowest masses and highest redshifts
have robust size measurements: the median brightness
of the 28 galaxies with $10.0<\log M_{\rm stars}<10.1$ and $2.9<z<3.0$
is $\langle H_{160}\rangle = 23.9$, well within the regime where
size measurements are reliable (see {van der Wel} {et~al.} 2014b).

The {\em observed} distribution of galaxies at $1.50<z<2.25$ is shown
in Fig.\ \ref{modtest1.fig}c. In panel (d) this observed distribution
is multiplied by a weight mask, to account for the artificial fall-off
at low masses in panel (b). The weight mask was constructed by evolving
a galaxy population with a uniform density distribution
in the size-mass
plane and a cutoff at $M_{\rm stars}
<10^{10}$\,\msun\ forward in time (in the same way as described
above). The distribution in Fig.\
\ref{modtest1.fig}d is remarkably similar to that in Fig.\
\ref{modtest1.fig}b. Furthermore,
the total number density of galaxies in the
two panels is almost identical; panel (d) has 7\,\% less galaxies
than panel (b).

In Fig.\ \ref{modtest2.fig} the color-coding reflects the specific
star formation rates of the
galaxies, with redder squares indicating a lower SSFR.
The figure looks very similar when the fraction of quiescent galaxies
is used for the color coding instead of the SSFR.
The sizes of the squares are proportional to the number of galaxies.
The  model naturally produces a population of
quiescent galaxies with $M_{\rm stars}\sim 10^{11}$\,\msun\
and $r_e \sim 1$\,kpc. In our model,
the progenitors of these galaxies
have masses of $\sim 3\times 10^{10}$\,\msun\
and sizes of $\sim 0.7$\,kpc at $z\sim 3$.
The model does not produce the right fraction of quiescent galaxies
at the highest masses and largest sizes: many of
these galaxies are forming stars at $z\sim 1.9$ even though they
have high
galaxy-averaged velocity dispersions. This suggests that our quenching
prescription is too simplistic in this regime
(see Sect.\ \ref{modelsum.sec}).

\begin{figure}[hbtp]
\epsfxsize=8.5cm
\epsffile{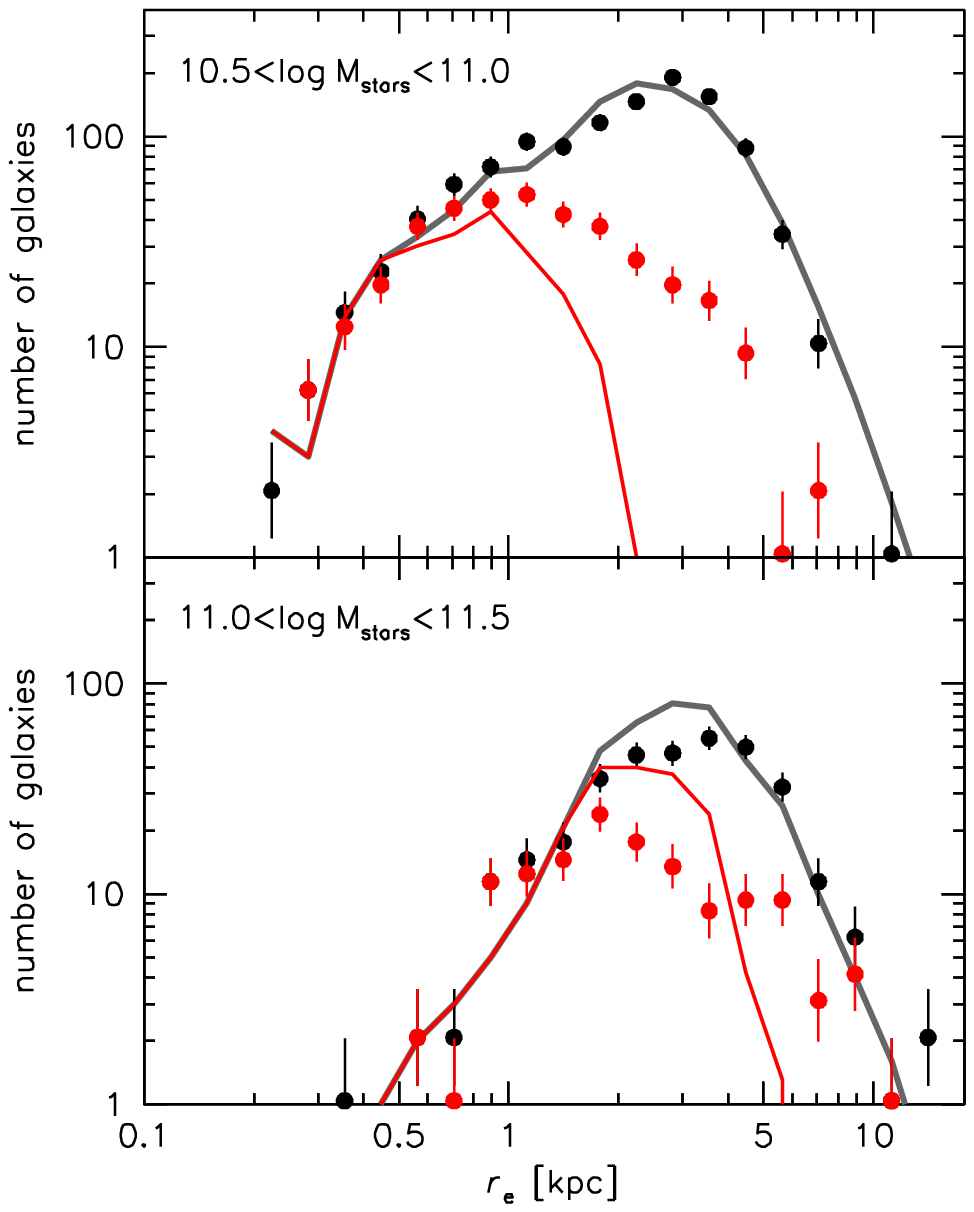}
\caption{\small
The number density of galaxies as a function of size at $1.50<z<2.25$,
in two mass bins. Points with errorbars are the observed values; black
points show all galaxies and red points show quiescent galaxies only.
The lines are the predicted distributions in our model, that is, the
observed distribution at $2.25<z<3.00$ evolved forward in time by
1.0 Gyr. The size distributions are well reproduced in this
model, in both mass bins (black lines). The match to
the subset of quiescent galaxies is very good at the smallest sizes but
shows systematic differences at intermediate and large sizes.
\label{compmod.fig}}
\end{figure}

We compare the predicted to the observed number densities explicitly
in Fig.\ \ref{compmod.fig}. This Figure highlights the excellent match
of our model to the size distribution of all galaxies over the entire mass
range $10.5<\log M_{\rm stars}<11.5$: it not only reproduces the peak
in the distribution at $r_e \sim 2.5$\,kpc but also the ``shoulder''
of compact quiescent galaxies. It also demonstrates that the
modeling of quenching is too simplistic for large
galaxies, as was already clear
from the comparison of panels (b) and (d) of Fig.\
\ref{modtest2.fig}. In particular, nearly 100\,\% of galaxies
with $r_e>2$\,kpc are forming stars in the model, whereas the
observed star forming fraction is only $\sim 85$\,\%.

\subsection{Summary of the Modeling}
\label{modelsum.sec}

In summary, we have shown that the population of compact, massive galaxies
at $z\sim 2$ can be explained by a model in which galaxies form stars
at a rate that is dictated by the star forming sequence, experience
a modest increase in size for a given increase in mass, and quench
after passing a velocity dispersion threshold. This was demonstrated
by evolving the observed galaxy population at $z\sim 2.6$ forward by
1\,Gyr to $z\sim 1.9$. This is a critical period as the number
density of \qg{}s increases by an order of magnitude
over that redshift range.

Although it is beyond the scope of this (already somewhat unwieldy)
paper, we note that
the modeling can easily be extended. In particular,
it would be straightforward to
fit for the two tunable parameters (the quenching dispersion
$\sigma_q$ and the parameter $\beta$, which relates the mass growth
to the star formation rate).  Furthermore, our quenching description
is inadequate in the high mass / large size regime; the yellow
line in Fig.\ \ref{moddemo.fig} is somewhat too steep.
A possible explanation is that quenching depends on the
galaxy properties in the central $\sim 1$\,kpc, and the simple
$M_{\rm stars}/r_e$ criterion no longer ``works'' in a regime
where $r_e \gg 1$\,kpc. Some evidence for this comes from a study
of the mass in the central $r_{\rm 3D}<1$\,kpc of galaxies
({van Dokkum} {et~al.} 2014): as we showed in Fig.\ 9 of that paper the
mass inside of 1\,kpc is an excellent predictor of quiescence at
all redshifts.
Finally, the modeling can be extended to lower redshifts, taking
evolution in $\sigma_q$ into account (see Sect.\ \ref{lowz.sec}).

\section{Discussion}

\subsection{The Formation of Today's Massive Galaxies}
\label{lowz.sec}

In the preceding sections we discussed a simple model for the evolution
of massive galaxies at $2<z<3$:
they grow inside-out with $\Delta \log r_{\rm e}
\sim 0.3 \Delta \log M_{\rm stars}$ (Eq.\ \ref{growth.eq})
while they are forming stars, and
quench when they reach a density or velocity dispersion threshold.
This model provides an explanation for the fact that large
galaxies have younger stellar populations than small galaxies
at fixed mass (e.g., {Franx} {et~al.} 2008), as
only the smallest galaxies have reached the quenching
threshold.
Galaxies enter the massive, compact selection
region in the size-mass plane ``from the left'', that is, by increasing
their masses.  This seems different from models in which large,
massive galaxies enter this region ``from above'', that is, by
decreasing their sizes through mergers (e.g., {Hopkins} {et~al.} 2009b)
or by 
gas ``compaction'' followed by star formation
({Dekel} \& {Burkert} 2014).  This apparent difference may
reflect a difference in approach: in this paper we are concerned with the
average evolution of the population of massive galaxies, whereas
simulations such as those of {Zolotov} {et~al.} (2015) are able to
follow the tracks of individual galaxies in the size-mass plane.
Judging from the {Zolotov} {et~al.} (2015) tracks,
Eq.\ \ref{growth.eq} may simply
be the time- and population average 
of periods of proportional size and mass growth
($\Delta \log r_{\rm e} \sim \Delta \log M_{\rm stars}$),
periods of compaction, and the effects of
mergers.\footnote{Note that the term ``compaction'' refers
to the gas, not the stars; in the Zolotov et al.\ models
the (indirect) effect on the stellar
effective radius is generally
much smaller than that on the gas radius.}

At lower redshifts
massive galaxies evolve along a markedly different track in the size-mass
plane: {van Dokkum} {et~al.} (2010), {Patel} {et~al.} (2013), and others find
that the size  and mass
evolution of massive galaxies are related through
$\Delta \log r_{\rm e} \sim 2 \Delta \log M_{\rm stars}$ at $0<z<2$
(as indicated by the dotted section of the red curve in
Fig.\ \ref{meangrowth.fig}). This evolution can be explained by
minor, gas-poor mergers building up the outer envelopes of galaxies
({Bezanson} {et~al.} 2009; {Naab} {et~al.} 2009; {Hopkins} {et~al.} 2010; {Hilz} {et~al.} 2013). In {van Dokkum} {et~al.} (2010)
we showed that {\em any} physical process that deposits mass
at $r>r_{\rm e}$ leads to a steep track in the size-mass plane, due
to the definition of the effective radius.

\begin{figure}[hbtp]
\epsfxsize=9cm
\epsffile{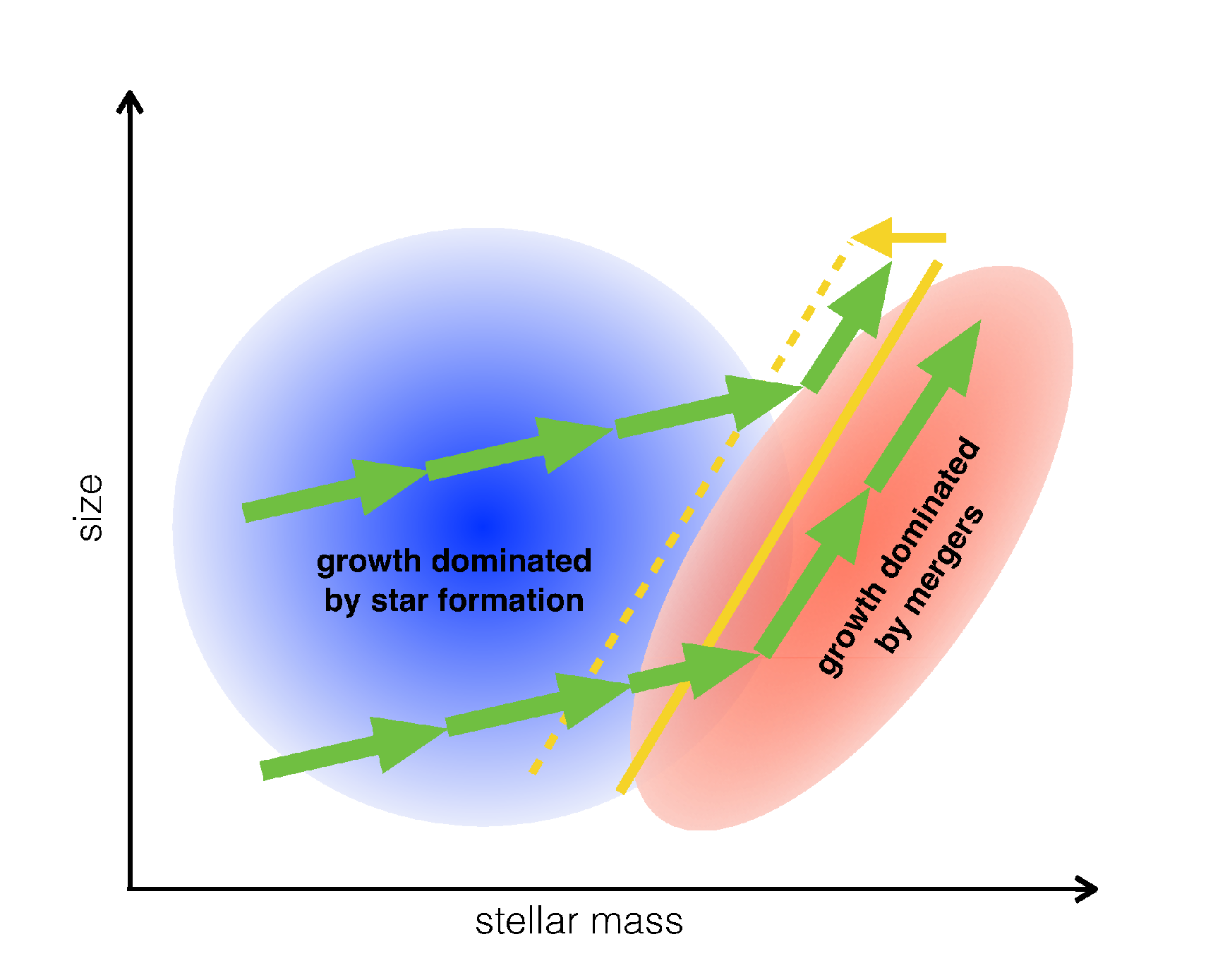}
\caption{\small
Illustration of possible average
tracks of galaxies in the size-mass plane
from $z\sim 3$ to $z\sim 0$. While they are forming stars, galaxies
grow mostly in mass and gradually increase their density. After
reaching a velocity dispersion or stellar
density threshold (the yellow
line, whose location is redshift dependent)
they quench, due to AGN feedback or other processes that correlate
with stellar density. The dominant mode of growth after quenching
is dry merging, which takes galaxies on a steep track in the size-mass
plane.
\label{cartoon.fig}}
\end{figure}

A schematic of the growth of massive galaxies from $z\sim 3$ to $z\sim 0$
is shown in Fig.\ \ref{cartoon.fig}. After galaxies quench, their
mass growth per unit time is reduced, but their effective radii
continue to increase. This Figure suggests that there are multiple paths
leading to large, massive, quiescent galaxies in the local Universe,
as was also noted in, e.g.,
Cappellari et al.\ (2013) and {Barro} {et~al.} (2014a).
Their $z\sim 2$ progenitors can be large star forming (disk) galaxies, such as
those studied extensively by, e.g., {Genzel} {et~al.} (2008)
and {F{\"o}rster Schreiber} {et~al.} (2011), or compact, massive, quiescent galaxies that
have grown through mergers (e.g., {Trujillo} {et~al.} 2011;
{Patel} {et~al.} 2013; Ownsworth et al.\ 2014).
As shown in Fig.\ 2 of {van Dokkum} {et~al.} (2014) massive $z=0$ galaxies
have a large range of central densities at fixed total mass, as
expected in such scenarios.
It is possible that
massive S0 galaxies formed from large star forming galaxies and
massive elliptical galaxies formed from compact star forming galaxies,
although it remains to be seen whether the stellar populations of
massive early-type galaxies are sufficiently diverse to accommodate a
large range in formation histories ({Gallazzi} {et~al.} 2005; {van Dokkum} \& {van der Marel} 2007).

\subsection{Winds, Shocks, and AGN}
\label{complex.sec}

In this paper we mostly ignored the effects of AGN, despite the fact
that nearly half of the 25 galaxies with Keck spectra
have X-ray luminosities above the canonical AGN
limit of $L_{\rm X}>10^{43}$\,ergs\,s$^{-1}$.\footnote{The number
of galaxies with active nuclei could be even higher, as the X-ray
selection is biased against Compton-thick AGN (see, e.g.,
Fiore et al.\ 2008).} The reason is that these
effects are difficult to constrain and quantify. {Barro} {et~al.} (2013) discuss
the high occurrence rate of
AGN in compact star forming galaxies extensively, and argue that
they are the agent of quenching. This may be true: in many
galaxy formation models AGNs play a crucial role in quenching
star formation precisely in this mass and redshift range
(e.g., {Croton} {et~al.} 2006; {Hopkins} {et~al.} 2008). However, the star
formation rates of the \sg{}s are (still) high and consistent with the
$z\sim 2.3$ star forming sequence ({Whitaker} {et~al.} 2014), and there is no
evidence for a direct effect of the AGNs on star formation. 
Turning this around,
it is obviously the case that the black holes are growing in these
galaxies, and that they are growing at a time when the dense stellar
centers are also growing. This is not surprising, as it is difficult
to see how to {\em avoid} a high accretion rate onto the central
object in these extremely dense, highly star forming galaxies.

An obvious point of concern is that the presence of
AGNs causes errors in the derived physical parameters of the galaxies.
In principle, an AGN in a relatively low mass, relatively large,
and relatively quiescent galaxy could push the galaxy in the \sg{}
category: the extra light of the AGN could be mistaken for star light,
increasing the mass; the combination of a point source with a normal
galaxy could be mistaken for a compact bulge-dominated object; and
the hot IR flux from the AGN could be mistaken for PAH features from
star formation. This can only be addressed properly with data of much
higher spatial resolution than is available today, but we note here
that the galaxies with AGNs do not stand out in any of the figures.
The only exception is that the four galaxies with the highest measured
velocity dispersions all have X-ray AGN, and also \niiha\ ratios
of $\sim 1$. We have treated these four galaxies in the same way
as the others.

A related issue is the almost-certain presence of galactic-scale winds
and outflows. Such winds can be driven by star formation 
(e.g., {Heckman}, {Armus}, \& {Miley} 1987) and/or AGNs (e.g., {Proga}, {Stone}, \& {Kallman} 2000)
and are ubiquitous in star forming galaxies at high redshifts
({Franx} {et~al.} 1997; {Pettini} {et~al.} 1998; {F{\"o}rster Schreiber} {et~al.} 2014; {Genzel} {et~al.} 2014a).
Galactic superwinds can create bubbles
and shock fronts whose kinematics, spatial extent, and emission line
ratios are very similar to what we observe. 
In at least one of the galaxies
in our sample, COSMOS\_1014,
there is evidence for a broad H$\alpha$
line in addition to a narrow component, similar to 
IRAS\,11095--0238 ({Soto} \& {Martin} 2012) and galaxies in
{F{\"o}rster Schreiber} {et~al.} (2014).
Furthermore,
four of the galaxies in our sample are part of the sample
of massive galaxies of {Genzel} {et~al.} (2014a) (COSMOS\_11363,
GOODS-S\_30274, GOODS-S\_37745, and GOODS-S\_45068), and
they find broad
nuclear velocity components in two of them
(COSMOS\_11363 and GOODS-S\_30274). A detailed study of
the kinematics and line ratios of GOODS-S\_30274 was 
also done by {van Dokkum} {et~al.} (2005).

Although winds are almost certainly {\em present},
two results suggest that they are not 
dominating the galaxy-integrated
emission line widths. First, winds
tend to escape in a direction perpendicular to the plane of the
galaxy ({Heckman}, {Armus}, \& {Miley} 1990),
which is difficult to reconcile with the observed anti-correlation
between velocity dispersion and axis ratio (Fig.\ \ref{sigq.fig}).
Second, the observed kinematics are fully explained by the stellar
mass, leaving little room for additional broadening due to winds.
In fact, broad components in the velocity profiles are
expected just from rotating gas at small radii: as shown in
Fig.\ \ref{kepler.fig} gas at $\sim 1$\,kpc should have
FWHM\,$\approx 1000$\,\kms\ even in the absense of winds.
Judging from other $z\sim 2$ galaxies the disks are also likely
to be highly turbulent, with a relatively high internal
dispersion (see, e.g., Cresci et al.\ 2009;
F\"orster Schreiber et al.\ 2009).
The gaseous environments of \sg{}s may be similar to those of ULIRGs,
which are highly complex: as shown in {Soto} \& {Martin} (2012)
they can have rotating, large-scale disks
in addition to outflows and shocks.

Finally, we note that the presence of
spatially-extended gas disks in these galaxies had been predicted
by {Zolotov} {et~al.} (2015). They also predicted that the gas dispersions
are, on average, lower than the stellar dispersions (Fig.\ \ref{sigsig.fig}a),
as the gas is in disks which are sometimes seen face-on.
Interestingly, {Zolotov} {et~al.} (2015) also find that the gas
constitutes only a small fraction of the total baryonic mass of
the simulated compact massive star forming galaxies, although
they note that this result is sensitive to the feedback
prescription. Similarly,
{Johansson} {et~al.} (2012) predicted that compact, massive galaxies
are stellar mass-dominated and have
Keplerian rotation curves; the model rotation curves in their
Fig.\ 7 are remarkably similar to the inferred
rotation curve shown in our Fig.\ \ref{kepler.fig}.

\subsection{Submm-Galaxies, Far-IR Selected Galaxies, and Quasars}
\label{extreme.sec}

This study begins with an HST/WFC3-selected sample in a total area of
$\sim 0.25$ square degrees. Many other studies have found extreme
star forming galaxies by selecting them on the basis of their
far-infrared, submm, or radio  emission instead
(e.g., {Kormendy} \& {Sanders} 1992; {Sanders} \& {Mirabel} 1996; {Barger} {et~al.} 1998; {Smail} {et~al.} 2000; {Barger} {et~al.} 2001; {Casey} {et~al.} 2012).
These extreme galaxies are plausible ancestors of early-type
galaxies; as an example, {Tacconi} {et~al.} (2008),
{Toft} {et~al.} (2014), and {Simpson} {et~al.} (2015) have suggested that many
submm galaxies could be direct progenitors of compact quiescent
galaxies at $z\sim 2$.

We do not select {\em against} such objects, and our sample
should include the proper number of submm galaxies, radio galaxies,
and other extreme objects.
However, there are (at least)
two possible reasons
why galaxies selected at other wavelenghts could
be underrepresented in our sample:
some fraction may be
too faint in the near-IR to be included (or to be properly
characterized) in the {Skelton} {et~al.} (2014) catalogs, and
some may be too
rare to be represented in the 3D-HST/CANDELS area. 
\sg{}s have
such high column densities in the central regions that some may be
entirely obscured at rest-frame optical wavelengths ({Gilli} {et~al.} 2014; {Nelson} {et~al.} 2014). {Wang}, {Barger}, \& {Cowie} (2012) and {Caputi} {et~al.} (2014)
show that objects exist that are
relatively bright in the IRAC bands but that are undetected in deep
near-IR data. It is obviously difficult to measure the redshifts
and masses of these objects with traditional means, but
it may be possible using molecular lines (see {Walter} {et~al.} 2012; {Riechers} {et~al.} 2013).
In the context of the study presented here the question is not
whether {\em any} massive, compact,
``optically-dark'' galaxies were missed, but what
fraction of mass and star formation is in such objects.

The second class of potentially missed objects are extremely rare,
extremely luminous galaxies. The median star formation rates of
\sg{}s in our study is $\langle {\rm SFR}\rangle = 134$\,\msun\,yr$^{-1}$,
and we have 112 such objects at $2<z<2.5$. Therefore,
objects that are so rare
that there are only a few (or zero) in our survey volume must have
star formation rates $\gtrsim 5000$\,\msun\,yr$^{-1}$ to have a significant
impact on our results. This seems extreme, but such objects probably
exist: the most extreme Herschel-selected galaxies at $2<z<5$ have
estimated star formation rates up to $\sim 9000$\,\msun\,yr$^{-1}$
(Casey et al.\ 2012). Furthermore,
recently identified highly obscured quasars have
bolometric luminosities of $L_{\rm bol} \sim 10^{47}$\,ergs\,s$^{-1}$
({Banerji} {et~al.} 2012, 2015), and it seems likely that the growth
of the black holes in these objects is accompanied by prodigious
star formation. 
It remains to be seen whether such objects are sufficiently
common (or rather, long-lived)
to impact results derived from CANDELS-sized areas.

Finally, we note that we do not find a correlation between size and
IR luminosity at fixed stellar mass, that is, an IR selection does not
preferentially select compact galaxies but objects with a wide range
of rest-frame optical sizes (see also {Wiklind} {et~al.} 2014;
{Simpson} {et~al.} 2015).
As an IR selection is effectively
a star formation selection at high masses
(see, e.g., {Whitaker} {et~al.} 2012; {Rodighiero} {et~al.} 2014), this is
perhaps not surprising.

%
%



\section{Summary and Conclusions}
\label{summary.sec}

In this paper we have identified a population of star forming, compact,
massive galaxies  in the five fields of the CANDELS and
3D-HST surveys. Such objects have been studied previously by
{Barro} {et~al.} (2013, 2014b, 2014a) and
{Nelson} {et~al.} (2014), and we build on their
results. Compared to the Barro et al.\ studies,
our selection is more restrictive, focusing only on the most
massive and most compact galaxies;
we study an area that is $\sim 2.5$
times larger; and our redshift catalogs make use of the 3D-HST grism
spectra for all objects brighter than $H_{160}<24$.

We first confirm the redshifts and masses of the galaxies using Keck
MOSFIRE and NIRSPEC spectroscopy of 25 compact massive star forming
galaxies at $2<z<2.5$. The gas dynamics suggest that the galaxies are
embedded in spatially-extended rotating disks; this explains the low
measured dispersions of a large fraction of the sample and the
observed anti-correlation between the disperion and the axis ratio of
the galaxies. Support for this interpretation comes from direct
measurements of the sizes of the H$\alpha$ disks for 10 galaxies; the
fact that this is possible at all from ground-based, seeing-limited
data already shows that the gas extends to scales $\gg$\,1\,kpc.  The
derived sizes of the gas disks, and the fall-off of the
rotation curve that we construct for the galaxies, are in very good
agreement with recent models for the formation of massive galaxies
({Johansson} {et~al.} 2012; {Zolotov} {et~al.} 2015).

It is important to note that, in our interpretation, the measured
gas velocity dispersions of the galaxies generally do not reflect
the true $V_{\rm rot}$ in the stellar body. We predict that the
(inclination-corrected)
velocities at $r\lesssim 1$\,kpc are $400-500$\,\kms for all
galaxies. This can be tested with adaptive optics-assisted 
observations of the H$\alpha$ line. There is evidence for
broad components in several of the velocity profiles (see
Sect.\ \ref{complex.sec}), and these complex profiles
may reflect the combined effect of high rotation velocities
at small radii and lower velocities at larger radii. A more
direct measurement could come from CO line widths, as these
likely probe much smaller radii than the H$\alpha$ emission
(see, e.g., {Downes} \& {Solomon} 1998).

Next,
we interpret the existence of star forming, compact galaxies at
$2<z<2.5$ in the context of a simple model for the evolution
of galaxies in the size-mass plane.
We describe the average evolution of star-forming
galaxies by
the simple relation $\Delta \log r_{\rm e}
\sim 0.3\Delta \log M_{\rm stars}$, with the mass evolution
proportional to the main sequence star formation rate.
We show that this evolution
is a consistent feature in galaxy formation models of
{Hirschmann} {et~al.} (2013), {Wellons} {et~al.} (2015), and {Zolotov} {et~al.} (2015),
and is also seen in observations of number density-matched samples
of galaxies ({van Dokkum} {et~al.} 2013; {Patel} {et~al.} 2013;
Ownsworth et al.\ 2014).

As galaxies move
along this track their average 3D density within $r_{\rm e}$
remains approximately constant (as $\rho(r_{\rm e}) \propto
M/r_{\rm e}^3$, it follows that $r_{\rm e} \propto M^{1/3}$ if
the density is constant). However, their density within a
fixed physical radius increases, as does their projected (2D)
density and their velocity dispersion.
Following
many other studies (e.g., {Franx} {et~al.} 2008; {Bell} {et~al.} 2012), we assume that
quenching occurs when galaxies
reach a threshold in either velocity dispersion or physical
density.
We show  that this model explains the evolution of
the distribution of galaxies in the size-mass plane from $z\sim 2.6$
to $z\sim 1.9$, the redshift range when the number density of
massive compact quiescent galaxies increases by nearly an order
of magnitude. In the
context of this straightforward model,
the progenitors of compact massive star forming galaxies at
$z\sim 2.5$ were simply somewhat less
massive and slightly smaller galaxies at $z\gtrsim 3$.

Our study has several important systematic uncertainties.
First, the stellar masses of the galaxies are derived from
fitting stellar population synthesis models to the photometry,
and these models have not been tested for the extreme galaxies
that are under discussion in this paper.
Such tests are urgently needed but they are
difficult, even for quiescent galaxies and
for ``normal'' star forming
galaxies in the local Universe ({Muzzin} {et~al.} 2009b; {Conroy} 2013). 
One interpretation of Fig.\ \ref{sigsig.fig}b is that the stellar
masses are off by factors up to $\sim 10$; however, as we
show in the remainder of Sect.\ 5 the dynamical masses and
stellar masses are consistent with each other once orientation
effects and the spatial extent of the gas are taken into account.
Our final dynamical
result ($M_{\rm fit} = 0.8^{0.6}_{-0.4} \times M_{\rm stars}$;
Sect.\ \ref{rotcurve.sec}) suggests that the contributions
of dark matter and gas to the mass within $\sim 7$\,kpc are
small. We have assumed a relatively bottom-light
{Chabrier} (2003) IMF when deriving stellar masses;
if we assume a {Salpeter} (1955) IMF instead
(see, e.g., {van Dokkum} \& {Conroy} 2010; {Conroy} \& {van Dokkum} 2012; {Cappellari} {et~al.} 2012)
we find
$M_{\rm fit} = 0.5^{+0.4}_{-0.2} \times M_{\rm stars}$, and
even tighter constraints on the amount
of gas and dark matter. We emphasize, however,
that the conversion of light to stellar mass for these dusty,
compact star forming galaxies is highly uncertain. We also
note here that the stellar masses are not corrected for the
contribution of emission lines to the SEDs. These corrections
are generally small ($\sim 10$\,\%).

Second, the role of winds and active nuclei in these galaxies
is not well understood (Sect.\ \ref{complex.sec}).
They almost certainly influence the
measured dynamics and line ratios, but without spatially-resolved
data  it is very difficult
to disentangle the
effects of winds, a falling rotation curve, and the spatial
distribution of the ionized gas.
Third, the fact that the galaxies are all very dusty may
imply that we are missing part of the population due to
selection effects (Sect.\ \ref{extreme.sec}). We could
be missing galaxies outright (see Fig.\ 3
in {Nelson} {et~al.} 2014), or they could be misclassified as less
compact, lower mass galaxies if only their outer edges
are detected in the currently available data. Another
potential effect of the dust is that the stellar population
modeling may produce incorrect stellar masses: the modeling
uses a screen approximation for dust, whereas in reality
the dust and stars are almost certainly mixed.

Fortunately, the prospects for addressing these uncertainties
are excellent. Adaptive optics-assisted spectroscopic
observations with
integral field units on 8\,m -- 10\,m telescopes can be
used to measure kinematics and line ratios on $\lesssim 1$\,kpc
scales (e.g., {Newman} {et~al.} 2013). The morphology of
the dust and molecular gas emission can be studied with
interferometers such as the Very Large Array, the
Plateau de Bure Interferometer, and the Atacama Large Millimeter
Array (see, e.g., {Simpson} {et~al.} 2015, for impressive early ALMA results
on submm-selected galaxies).
These instruments can also measure the kinematics of
the molecular gas (e.g., {Tacconi} {et~al.} 2008). On a longer
timescale, the James Webb Space Telescope can measure the
stellar kinematics of the galaxies, as well as identify and
characterize compact galaxies that are entirely obscured
in the $K$ band ({Wang} {et~al.} 2012). Finally, the upcoming generation
of extremely large ground-based optical/near-IR telescopes is
needed to spatially
resolve these compact, massive galaxies within their effective
radius. 

\begin{acknowledgements}
We thank Adi Zolotov for providing model tracks of galaxies in
digital form, and Thorsten Naab and Peter Johansson for pointing
us to the key figures that show model tracks in their simulations.
The comments from the anonymous referee improved the manuscript,
and prompted us to add the three appendices.
Support from STScI grant GO-12177
is gratefully acknowledged.
The data presented herein were obtained at the
W.\ M.\ Keck Observatory, which is operated as a scientific
partnership among the California Institute of Technology, the
University of California and the National Aeronautics and Space
Administration. The Observatory was made possible by the
generous financial support of the W.\ M.\ Keck Foundation. The
authors recognize and acknowledge the very significant cultural
role and reverence that the summit of Mauna Kea has always
had within the indigenous Hawaiian community. We are most
fortunate to have the opportunity to conduct observations from
this mountain.
\end{acknowledgements}


\newpage

\begin{appendix}

\section{A.\ $H_{160}$ Images}

\begin{figure*}[hbtp]
\begin{center}
\epsfxsize=17cm
\epsffile{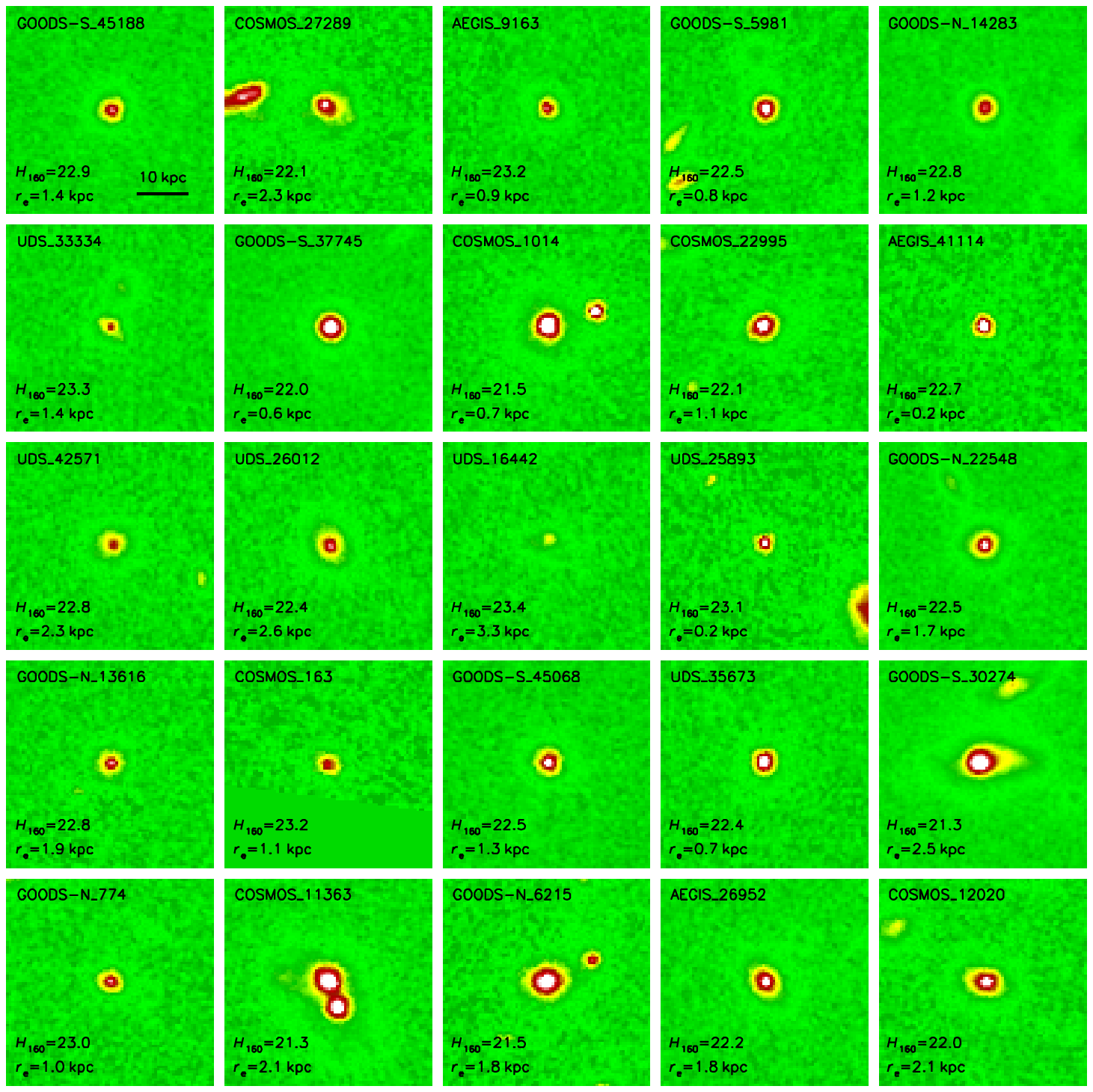}
\end{center}
\vspace{-0.3cm}
\caption{\small
HST images of the galaxies in Figs.\ 5, 6, 7, and 15,
in the $H_{160}$ band. The galaxies are displayed with a high 
dynamic range, so that
faint structures around bright cores can be seen more clearly
than in Fig.\ 6 of the main text.
GOODS-S\_3027
4 and COSMOS\_11363 show clear
tidal features.
\label{Hband_color.fig}}
\end{figure*}

In the main text we show color images of the 25 star forming compact
massive galaxies, created from the $J_{125}$ and $H_{160}$ CANDELS
data
(Fig.\ 6). In Fig.\ \ref{Hband_color.fig}
we show the $H_{160}$ images separately, with a higher dynamic
range than in Fig.\ 6. The tidal features around GOODS-S\_30274 and
COSMOS\_11363 are very clear, and several other galaxies 
also show structure at faint surface brightness.
We fit all galaxies with a single Sersic profile, which
is an excellent approximation of the average surface brightness
profile of the full sample (see Sect.\ \ref{sb.sec}); however, it is
clear that these fits do not capture the full information in the
HST images.

\section{B.\ Expected and Observed Uncertainties in the Spectra}
\label{noise.sec}

As described in Sect.\ 3.4.1 we fit Gaussian models to the
emission lines. The fits are done with the {\tt emcee}
code, with the observed 1D spectrum and a noise model as inputs
for each galaxy. Here we briefly analyze the residuals from these
fits to determine the accuracy of the noise models.

\begin{figure*}[hbtp]
\begin{center}
\epsfxsize=17cm
\epsffile{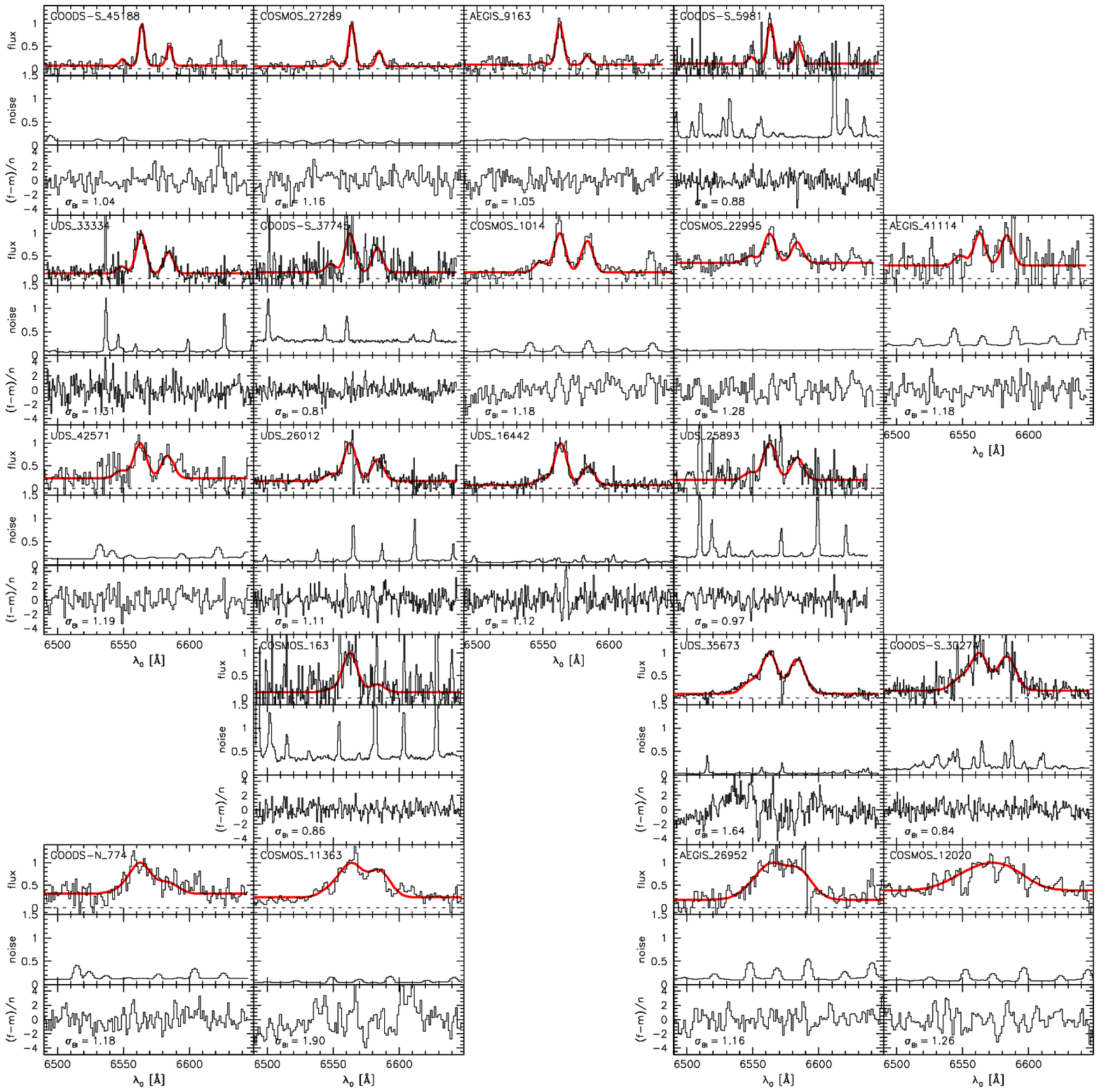}
\end{center}
\vspace{-0.3cm}
\caption{\small
Analysis of the noise in the NIRSPEC and MOSFIRE spectra.
The galaxies have the same order as in Fig.\ \ref{spectra.fig};
panels for objects taken from Barro et al.\ (2014) are left blank.
For each galaxy, the top panel shows the spectrum and the
best-fitting model; the middle panel shows the expected noise
(see Sect.\ 3.1 and Sect.\ 3.2); and the bottom panel shows
the difference between the observed spectrum and the best-fitting
model divided by the expected noise. The width of the
distribution of these residuals
is $\sim 1$ in nearly all cases.
\label{spectra_noise.fig}}
\end{figure*}

In Fig.\ \ref{spectra_noise.fig} we show the spectra of the 20
galaxies that were observed by us. For convenience,
the figure has the
same format as Figs.\ 5, 6, 7, and 15 in the main text,
except that the five galaxies from
Barro et al.\ (2014) are left blank. For each galaxy three
subpanels are shown. The top subpanel is identical to
the main panel of Fig.\ 5, and shows the observed spectrum
in black along with the best-fitting model in red.
The middle subpanel shows the noise model (empirical in the
case of MOSFIRE and theoretical in the case of NIRSPEC;
see Sect.\ 3.1 and Sect.\ 3.2). The bottom subpanel is
the residual from the fit divided by the noise model.

The residuals are well-behaved, and generally exhibit no
indications of poorly subtracted sky lines or other irregularities.
We quantified this by calculating the biweight
scatter $\sigma_{\rm BI}$ (see Beers et al.\ 1990)
in the distribution of residuals. The value of $\sigma_{\rm BI}$
deviates by more than $\sim 30$\,\% from unity in only two cases,
UDS\_35673 and COSMOS\_11363. Both galaxies have very high
S/N ratio spectra, and the higher than expected
residuals are not caused by errors in the noise spectra but
by the fact that the velocity distributions are not exactly
Gaussian. The average scatter of the remaining 18 galaxies
is $\langle \sigma_{\rm BI}\rangle = 1.09$, which means that
the noise models that we use are accurate to $\sim 10$\,\%.

\section{C.\ Converting Galaxy-Averaged Velocity Dispersions to
a Rotation Curve}

\subsection{Motivation}

In Sect.\ \ref{rotcurve.sec} we construct the average rotation curve for
star forming compact massive galaxies. This is done by combining
information for 10 different galaxies: all galaxies have approximately
the same stellar masses and $H_{160}$ half-light radii, but they have
a wide range of H$\alpha$ effective radii. For each
galaxy we measure the galaxy-integrated
velocity dispersion and the inclination, and convert these
to an inclination-corrected rotation velocity at $r=r_{\rm gas}$, where
$r_{\rm gas}$ is the half-light radius of the H$\alpha$ emission.
The rotation velocities of the galaxies
are then plotted versus $r_{\rm gas}$ in Fig.\ \ref{kepler.fig},
and the resulting relation is interpreted as
a rotation curve.

\begin{figure*}[hbtp]
\begin{center}
\epsfxsize=15.5cm
\epsffile{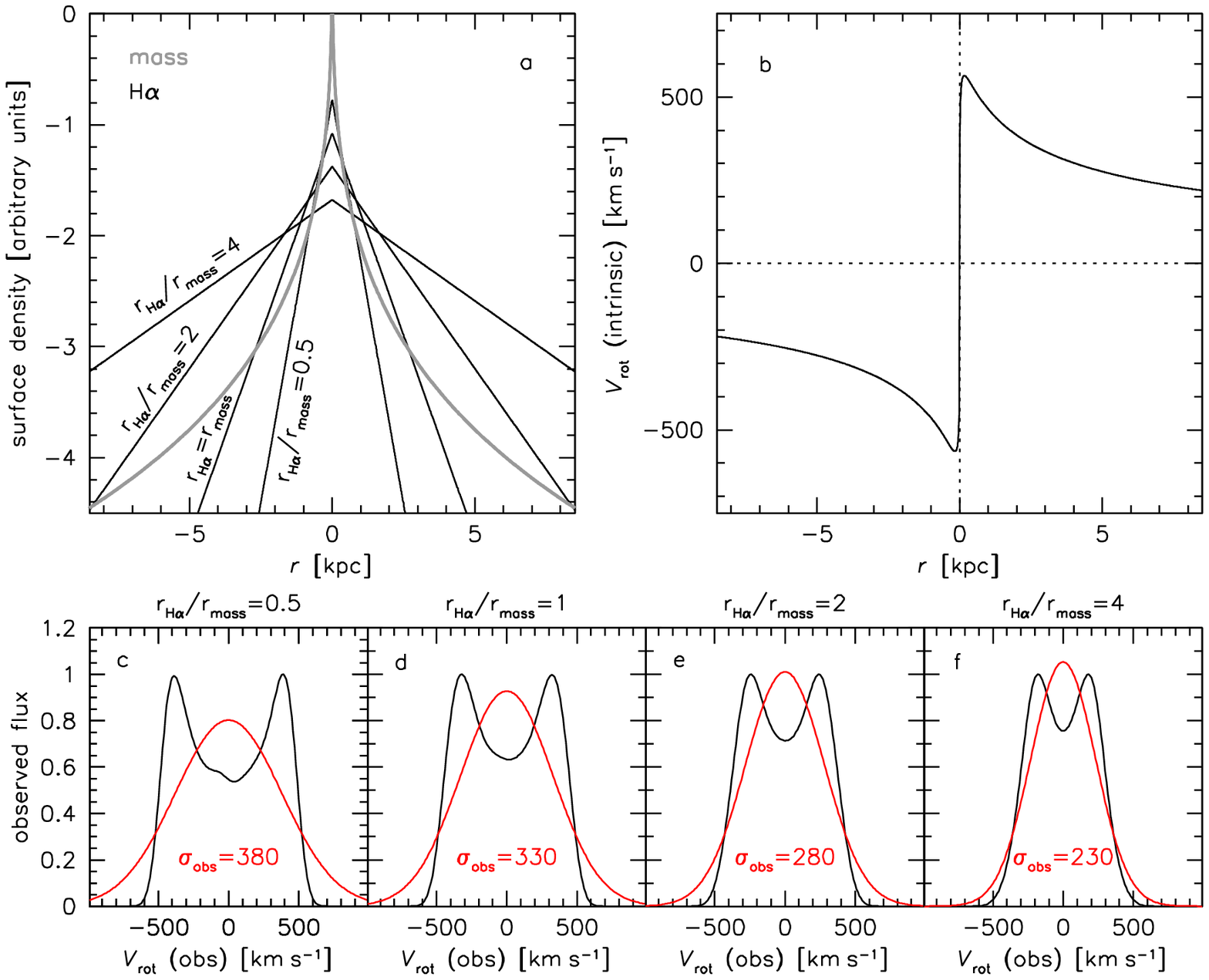}
\end{center}
\vspace{-0.3cm}
\caption{\small
a) Surface density profile of a
model galaxy with a stellar mass of $10^{11}$\,\msun,
Sersic index $n=4$, and
an effective radius $r_{\rm mass}=1$\,kpc (grey). Black lines show
four different 
H$\alpha$ $n=1$ surface brightness profiles, with effective
radii ranging from 0.5\,kpc to 4\,kpc.
b) Rotation curve of the H$\alpha$-emitting gas disks in the
model galaxy. The H$\alpha$ emission is
assumed to be a tracer, not a contributor, to the mass, and the
rotation curve is identical in all four models.
c-f) Observed galaxy-integrated H$\alpha$ velocity profiles for the four
surface brightness profiles shown in panel a, assuming an inclination
of 60$\arcdeg$ and an instrumental resolution of 60\,\kms.
The red curves are Gaussian fits to the observed profiles.
The measured dispersion is lower for higher values of
$r_{{\rm H}\alpha}/r_{\rm mass}$, as the profile is weighted
toward larger radii.
\label{alpha_1.fig}}
\end{figure*}

Here we test whether this method is viable, that is, whether the
actual rotation curve of a model galaxy can be reconstructed in this way.
We also test whether we are using the correct
conversion constant to go from a galaxy-integrated
velocity dispersion to a rotation velocity at the half-light
radius of H$\alpha$. This constant, 
together with
an inclination correction, relates the velocity dispersion $\sigma$
to the rotation velocity $V_{\rm rot}$:
\begin{equation}
\alpha = \frac{\sigma_{\rm gas}}{V_{\rm rot} \sin^{-1}(i)}
\end{equation}
(see Eq.\ \ref{siggen.eq} and Eq.\ \ref{correcti.eq}). In the main text
we use $\alpha=0.8\pm 0.2$, based on previous studies (see Sect.\
\ref{gasstars.sec}). However, these studies did not consider the specific
model of a compact, $r^{1/4}$-law mass distribution combined with
an extended, exponential gas distribution.

\subsection{Modeling Velocity Profiles}

We simulated the observations in the following way. We constructed a
model mass distribution that follows a Sersic surface density profile.
This mass distribution is characterized by three parameters: the Sersic
index $n$, the effective radius $r_{\rm mass}$ (this parameter is
equivalent to both $r_{\rm stars}$ and $r_e$ in the main text), and
the total mass $M$. We fixed $r_{\rm mass}=1$\,kpc and $M
=1.0\times 10^{11}$\,\msun, and for the initial model we set
$n=4$. Apart from a slight rescaling of the effective radius,
this model closely matches the actual average stellar
mass distribution of the sCMGs, if mass traces the $H_{160}$ light.
The model is shown in Fig.\ \ref{alpha_1.fig}a by the grey line.

Next, we constructed 10 model galaxies, each with the same mass distribution
but with different distributions of the H$\alpha$ emission. The
ionized gas is in thin exponential disks, with effective radii ranging
from $r_{{\rm H}\alpha} = 0.5$\,kpc (and hence $r_{{\rm H}\alpha} = 0.5
r_{\rm mass}$) to $r_{{\rm H}\alpha} = 5$\,kpc. Four of these model
gas distributions are shown by the black lines in Fig.\ \ref{alpha_1.fig}a.
The gas disks mimic
the derived extended ionized gas of sCMGs, with $r_{{\rm H}\alpha}$
equivalent to the parameter $r_{\rm gas}$ in the main text. 
Galaxy-integrated velocity profiles were created by integrating the
projected velocities along the line of sight and
over the full spatial extent
of the model galaxies. The velocities were calculated from the mass
profile shown in Fig.\ \ref{alpha_1.fig}a and  weighted by
the H$\alpha$ flux. In order to model the observed profiles
as closely as possible, we used
an inclination of 60$\arcdeg$
(where $90\arcdeg$ is edge-on) and an instrumental resolution of
60\,\kms\ (in between the MOSFIRE and NIRSPEC resolution).

The velocity profiles of the four model galaxies
are shown in panels c-f of Fig.\ \ref{alpha_1.fig}. As expected
they have the
classic ``double-horned'' form that is characteristic of rotating disks.
The profile is not the same for all four models even though the mass
distribution, and hence the underlying velocity field, is identical in
all cases. The more extended
the H$\alpha$ distribution is with respect
to the mass, the narrower the profile becomes,
and the more closely it resembles
a Gaussian. The reason for this behavior is that the H$\alpha$ emission
is more weighted toward larger radii, where the rotation velocity is
lower. Velocities in excess of $\sim 350$\,\kms\ are still sampled, but
they have relatively low weight and are responsible for the high velocity
tails of the profile.

\subsection{Relation Between Global Dispersion and Rotation Velocity
at $r=r_{{\rm H}\alpha}$}

We fitted Gaussian models to the line profiles, just as we do in the
data analysis described in the main text. These Gaussian fits are
shown by the red curves in panels c-f of Fig.\ \ref{alpha_1.fig}.
The width of these Gaussians decreases with increasing $r_{{\rm H}\alpha}/
r_{\rm mass}$, as discussed above. 
We note here that the actual profile shape is not
very well approximated by a Gaussian, particular in panels c and d.
Interestingly, we see hints of double-horned profiles in the data for
some of the galaxies (e.g., UDS\_16442 and, particularly, GOODS-N\_774,
which was published in Nelson et al.\ 2014), although the S/N ratio
is  not high enough to quantify this.

\begin{figure*}[hbtp]
\begin{center}
\epsfxsize=15.5cm
\epsffile{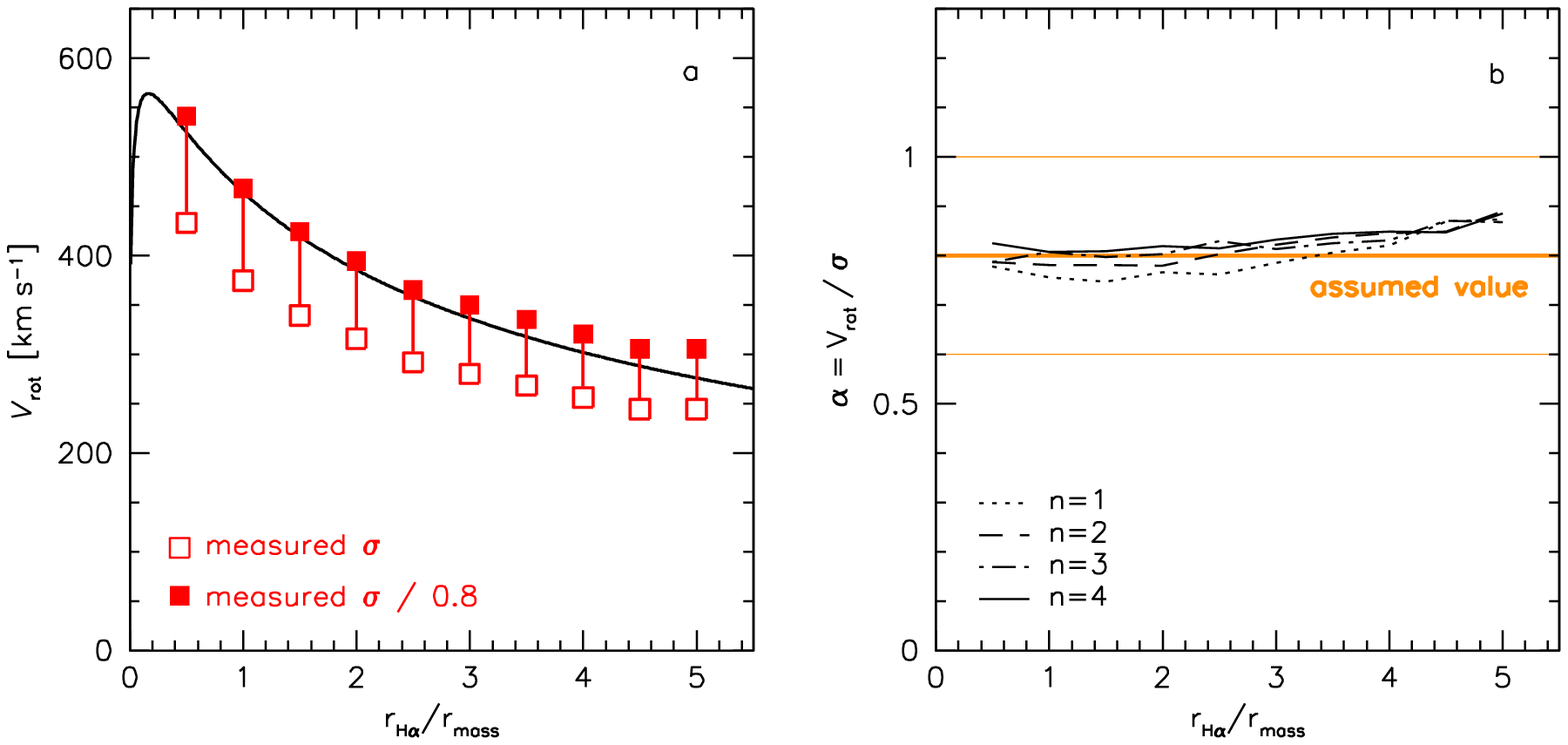}
\end{center}
\vspace{-0.3cm}
\caption{\small
a) Rotation curve of the model in Fig.\ \ref{alpha_1.fig} (black line),
compared to the inclination-corrected,
galaxy-integrated velocity dispersion $\sigma$
for 10 different H$\alpha$ distributions (open red squares).
The half-light radius
of the H$\alpha$ emission ranges from $0.5\times r_{\rm mass}$ to
$5\times r_{\rm mass}$. Solid red squares are corrected for the
parameter $\alpha = \sigma/V_{\rm rot} = 0.8$.
b) Derived values of $\alpha$ from our model (black lines). The
value $\alpha=0.8\pm 0.2$ that is used in the main text is shown by
the orange line. Different line types indicate results for different
Sersic indices $n$ of the mass distribution; the value of $\alpha$ is nearly
independent of $n$.
\label{alpha_2.fig}}
\end{figure*}

In Fig.\ \ref{alpha_2.fig}a these measured galaxy-integrated velocity
dispersions are plotted versus the half-light radii of the H$\alpha$
disks, after correcting for inclination and instrumental broadening
(open squares). All ten galaxy models are shown, with H$\alpha$
effective radii ranging from $0.5 \times
r_{\rm mass}$ to $5\times r_{\rm mass}$. For comparison, the black curve
shows the actual rotation curve of the galaxies. The squares show the
same fall-off as the actual rotation curve, with a roughly constant
multiplicitative offset. The solid squares are obtained by dividing the
measured dispersions by 0.8, which is the value of $\alpha = \sigma/V_{\rm
rot}$ that we used in the analysis of Sect.\ \ref{rotcurve.sec}.
They are in almost perfect agreement with the black curve, demonstrating
that it is possible to reconstruct the average
rotation curve of sCMGs with our method.

The analysis is generalized in Fig.\ \ref{alpha_2.fig}b, where we show
the value of $\alpha$ as a function of the ratio of the effective radius
of H$\alpha$ and the effective radius of the mass. We repeated the
analysis for different assumed mass profiles, ranging from exponential
($n=1$; dotted) to an $r^{1/4}$ law ($n=4$; solid). The ratio between
dispersion and rotation velocity at $r=r_{{\rm H}\alpha}/r_{\rm mass}$
is remarkably constant: it does not vary appreciably either
with $r$ or with $n$. We conclude that the assumed value of $\alpha=0.8\pm
0.2$ is reasonable for the mass and H$\alpha$ profiles discussed in this
paper.

\end{appendix}

\end{document}